\documentclass[useAMS,usenatbib]{mn2e}

\usepackage{longtable}
\usepackage{lscape}
\usepackage{natbib}
\usepackage{psfig}
\usepackage{graphicx}
\usepackage{multicol}

\def\degree{\mbox{$^{\circ}$}}
\def\farcsn{\hbox{$\ \!\!^{\prime\prime}$}}

\def\kms{$\mathrm{km\;s}^{-1}$}
\def\kmsmpc{$\mathrm{km\;s^{-1}\;Mpc^{-1}}$}
\def\lsun{L$_{\odot}$}
\def\msun{M$_{\odot}$}
\def\mas{mag arcsec$^{-2}$}

\def\hst{{\it HST\/}}

\def\mbh{$M_\bullet$}
\def\vc{$V_{\rm c}$}
\def\re{$r_{\rm e}$}
\def\reb{$r_{\rm e, bulge}$}
\def\reg{$r_{\rm e, gal}$}
\def\co{$C_{28}$}
\def\n{$n$}
\def\sigmas{$\sigma$}

\def\sigmae{$\sigma_{\rm e}$}
\def\mstar{$M_{\star, \rm gal}$}

\def\li{$L_{i, {\rm gal}}$}
\def\ligal{$L_{\rm gal}$}
\def\Lbui{$L_{i, {\rm bulge}}$}
\def\Lbu{$L_{\rm bulge}$}
\def\Mbu{$M_{\rm bulge}$}
\def\Mbuvir{$M_{\rm bulge}$}
\def\megal{$M_{\rm e, gal}$}
\def\mdyn{$M_{\rm dyn, gal}$}

\def\muemean{$\langle\mu_{\rm e, bulge}\rangle$}

\def\mlbu{$M_\bullet-L_{\rm bulge}$}
\def\mlbui{$M_\bullet-L_{i, {\rm bulge}}$}
\def\mlbug{$M_\bullet-L_{\rm bulge}$}
\def\mlbu{$M_\bullet-L_{\rm bulge}$}

\def\mli{$M_\bullet-L_{i, {\rm gal}}$}
\def\ms{$M_\bullet-\sigma$}
\def\mse{$M_\bullet-\sigma_{\rm e}$}

\def\vs{$V_{\rm c}-\sigma_{\rm e}$}
\def\mv{$M_\bullet-V_{\rm c}$}
\def\mmstar{$M_\bullet-M_{\star, \rm gal}$}
\def\mmegal{$M_\bullet-M_{\rm e, gal}$}
\def\mmbuvir{$M_\bullet-M_{\rm bulge}$}

\def\msres{$\log(M_\bullet/M_\bullet[\sigma_{\rm e}])$}
\def\mlibures{$\log(M_\bullet/M_\bullet[L_{i, {\rm bulge}}])$}
\def\mmassbures{$\log(M_\bullet/M_\bullet[\rm M_{bulge}])$}
\def\mlires{$\log(M_\bullet/M_\bullet[L_{i, {\rm gal}}])$}
\def\mstarres{$\log(M_\bullet/M_\bullet[M_{\star, \rm gal}])$}

\def\niipg {[{\sc N$\,$ii}]$\,\lambda\lambda6548,6583$}
\def\ha{H$\alpha$}

\def\siipg {[{\sc S$\,$ii}]$\,\lambda\lambda6716,6731$}

\def\hiire {{\sc H$\,$ii}}
\def\hire  {{\sc H$\,$i}}

\def\aj{AJ}                   
\def\araa{ARA\&A}             
\def\apj{ApJ}                 
\def\apjl{ApJ}                
\def\apjs{ApJS}               
\def\aap{A\&A}                
\def\aaps{A\&AS}              
\def\mnras{MNRAS}             
\def\nat{Nature}
\def\nat{Nature}              

\defcitealias{Gultekin2009}{G09}
\defcitealias{Beifiori2009}{B09}

\def\Eq#1{Eq.~(\ref{eq:#1})}
\newcommand{\be}{\begin{equation}}
\newcommand{\ee}{\end{equation}}
\def\se#1{\S\ref{sec:#1}}

\def\Fig#1{Figure~\ref{fig:#1}}

\newcommand{\sersic}{S\'{e}rsic\ }

\newcommand{\placefigone}{  
\begin{figure}
\begin{center}
\includegraphics[width=\columnwidth]{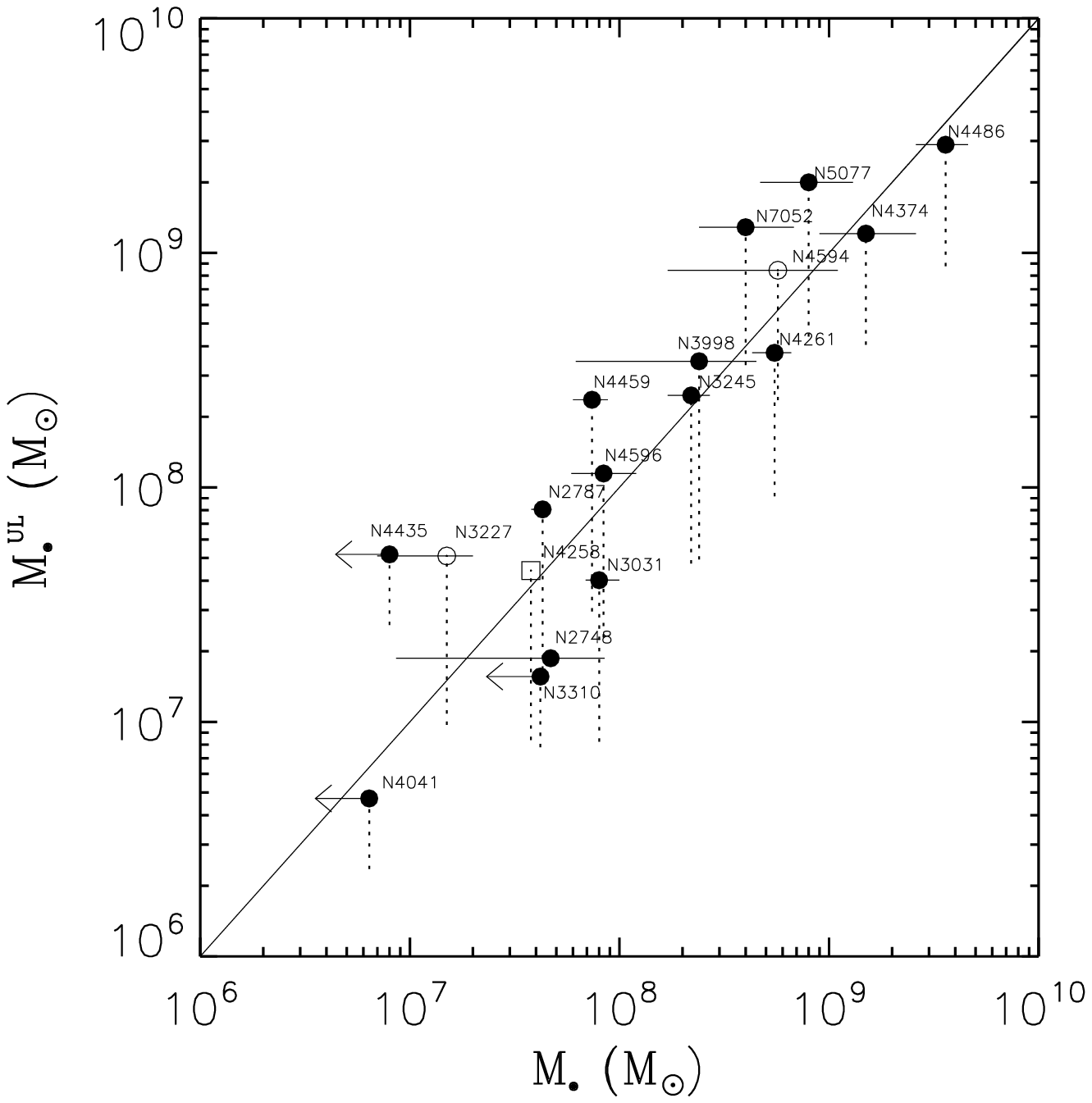}
\end{center}
\caption{Comparison between the \mbh\ upper limits by
  \citetalias{Beifiori2009} and accurate \mbh\ measurements (symbols)
  and upper limits (leftward arrows) by \citetalias{Gultekin2009} and
  based on the resolved kinematics of gas (filled circles), stars
  (open circles), and water masers (open square). The upper and lower
  edges of the dotted lines correspond to \citetalias{Beifiori2009}'s
  \mbh\ values estimated assuming an inclination of $i = 33^\circ$ and
  $81^\circ$ for the unresolved Keplerian disc, respectively.}
\label{fig:comparison_b09_g09}
\end{figure}
}

\newcommand{\placefigtwo}{  
\begin{figure*}
\begin{center}
\includegraphics[width=0.45\textwidth]{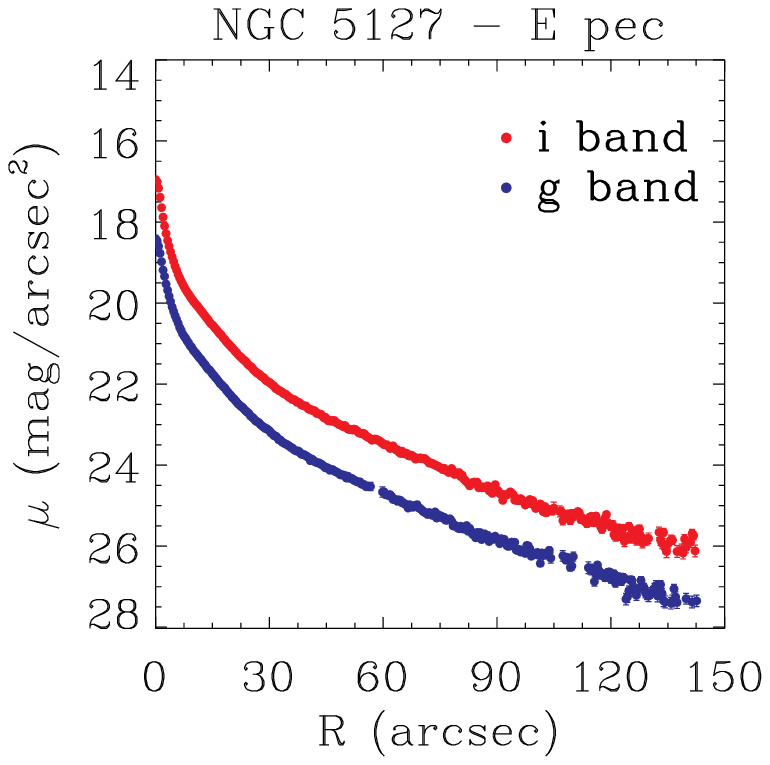}
\includegraphics[width=0.45\textwidth]{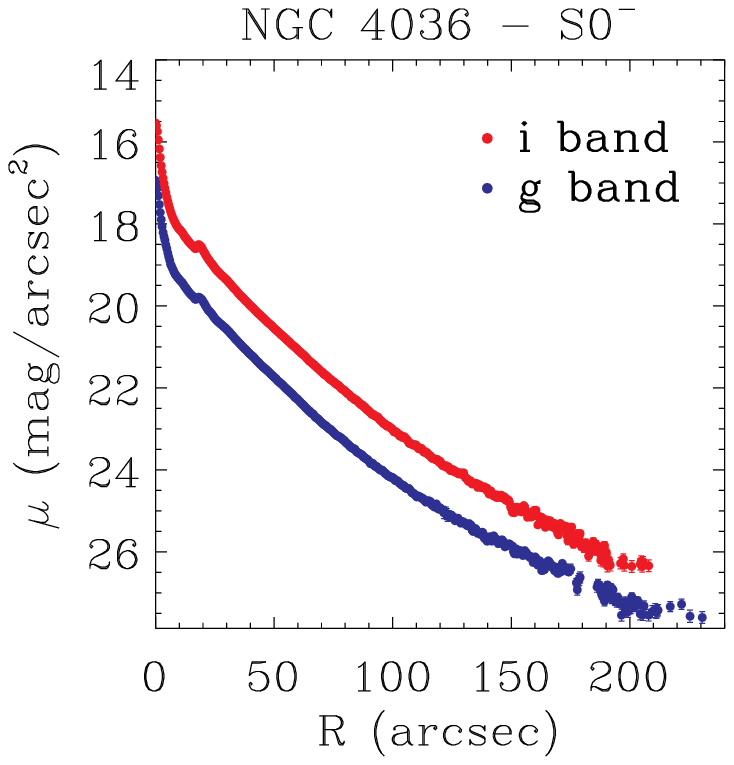}\\
\includegraphics[width=0.45\textwidth]{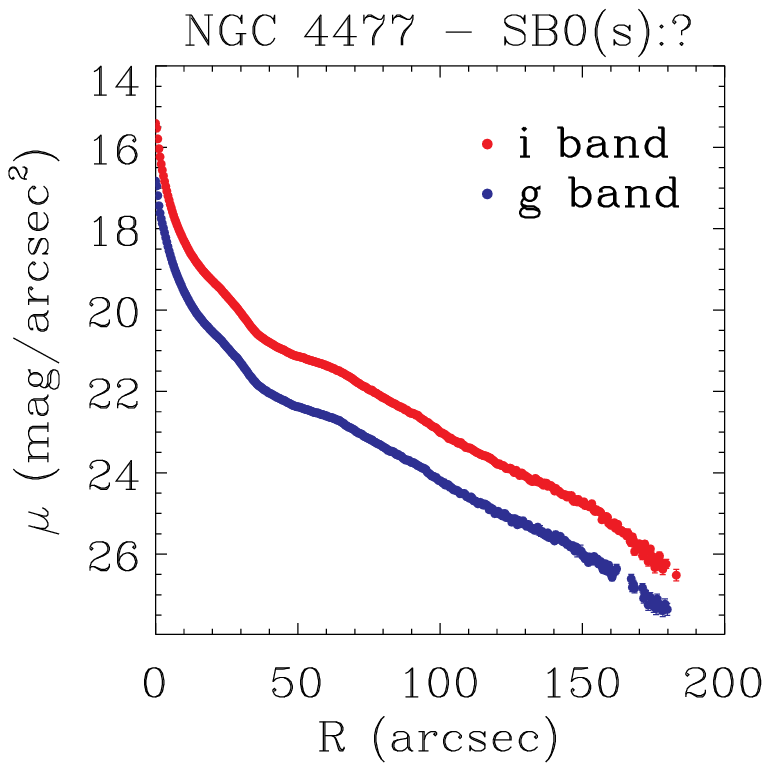}
\includegraphics[width=0.45\textwidth]{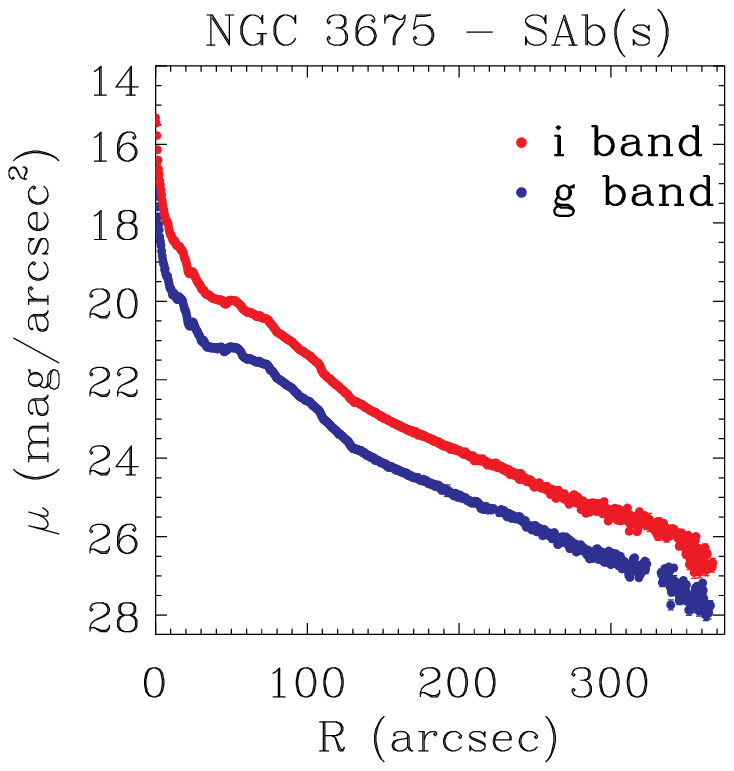}
\end{center}
\caption{Typical $g$ (red points) and $i$-band (blue points) surface
  brightness profiles extracted along the major axis for a few sample
  galaxies.}
\label{fig:profiles}
\end{figure*}
}

\newcommand{\placefigthree}{  
\begin{figure*}
\begin{center}
\includegraphics[angle=90.0,width=0.45\textwidth]{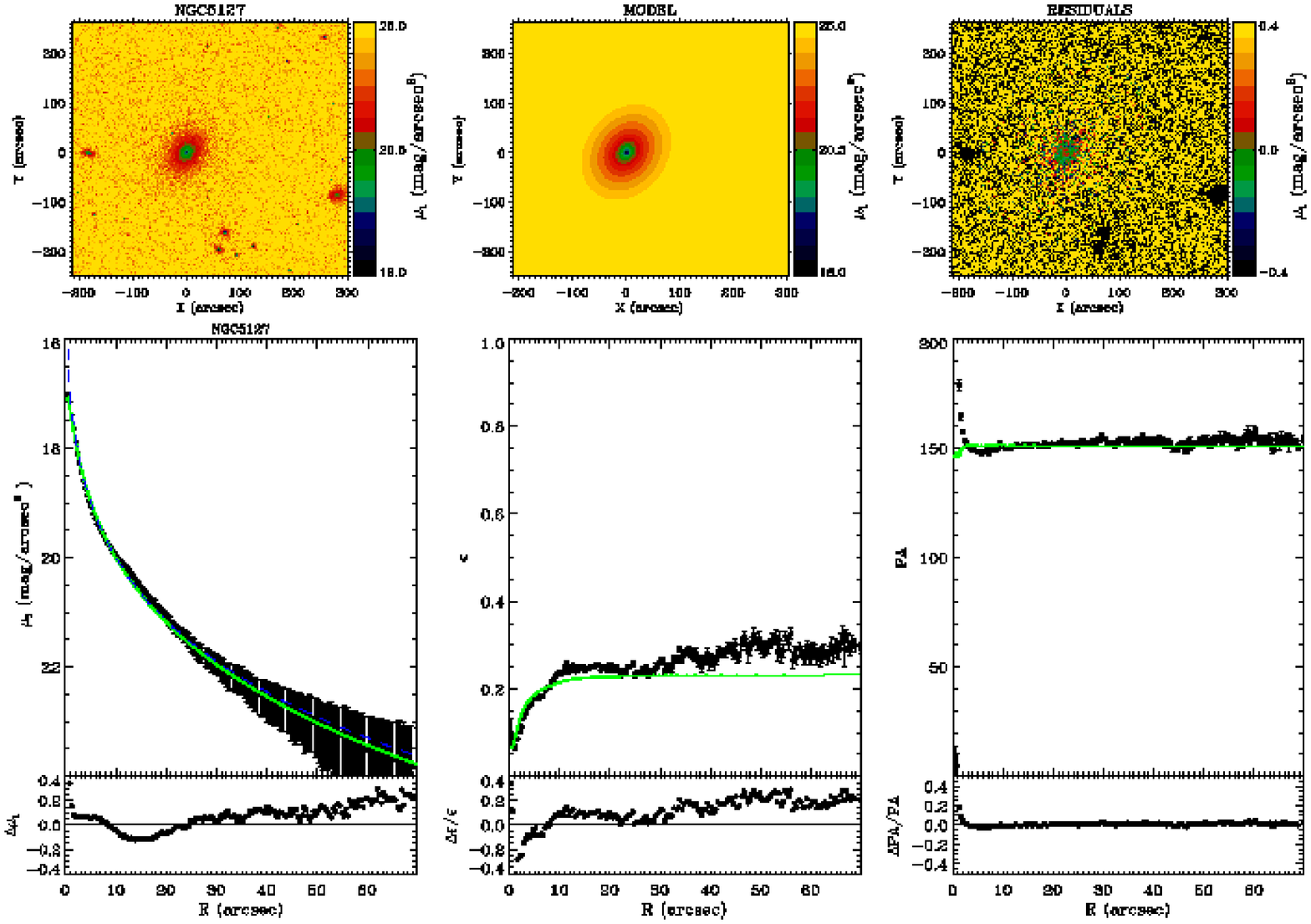}
\hspace{1cm}
\includegraphics[angle=90.0,width=0.45\textwidth]{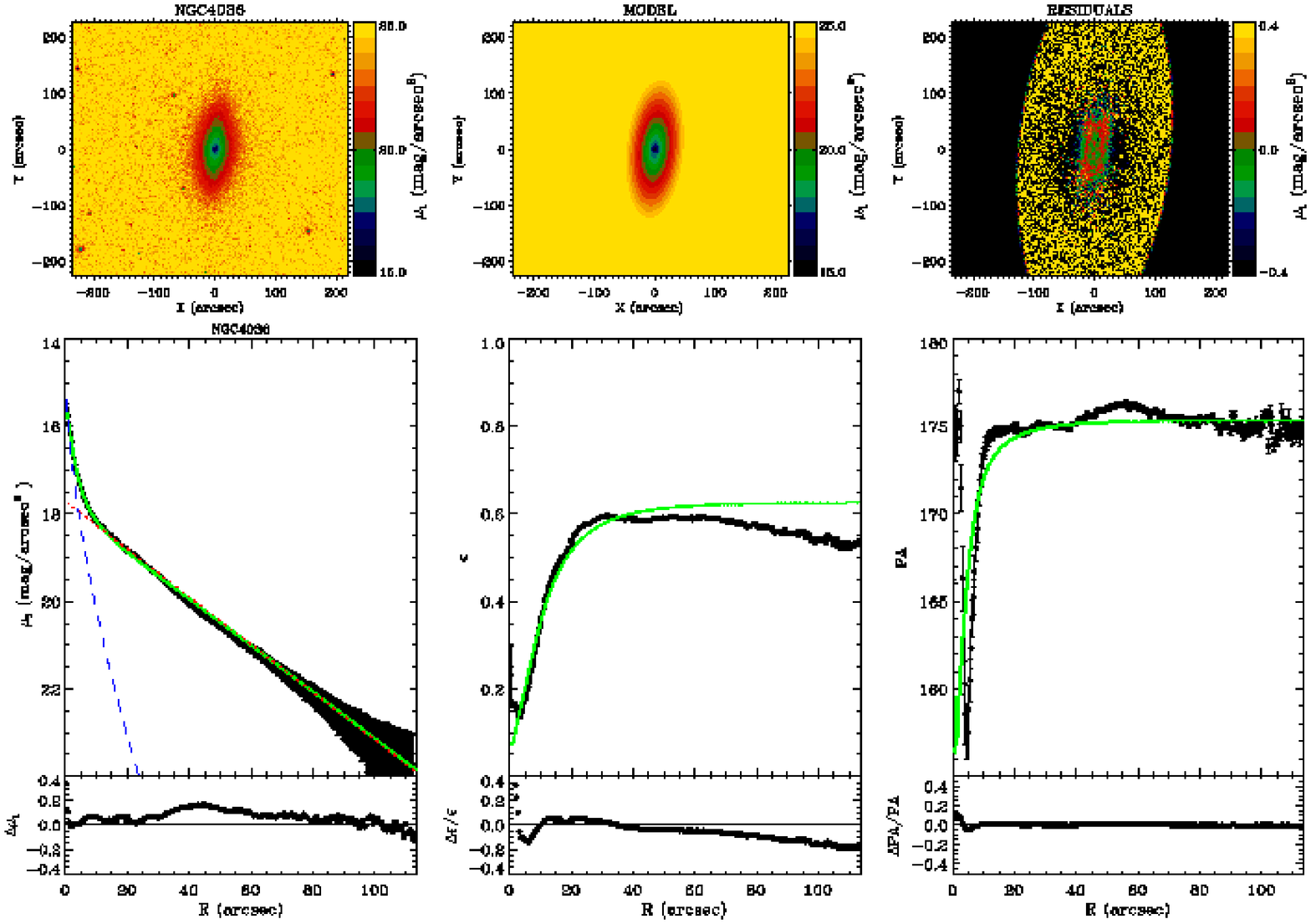}\\
\vspace{1cm}
\includegraphics[angle=90.0,width=0.45\textwidth]{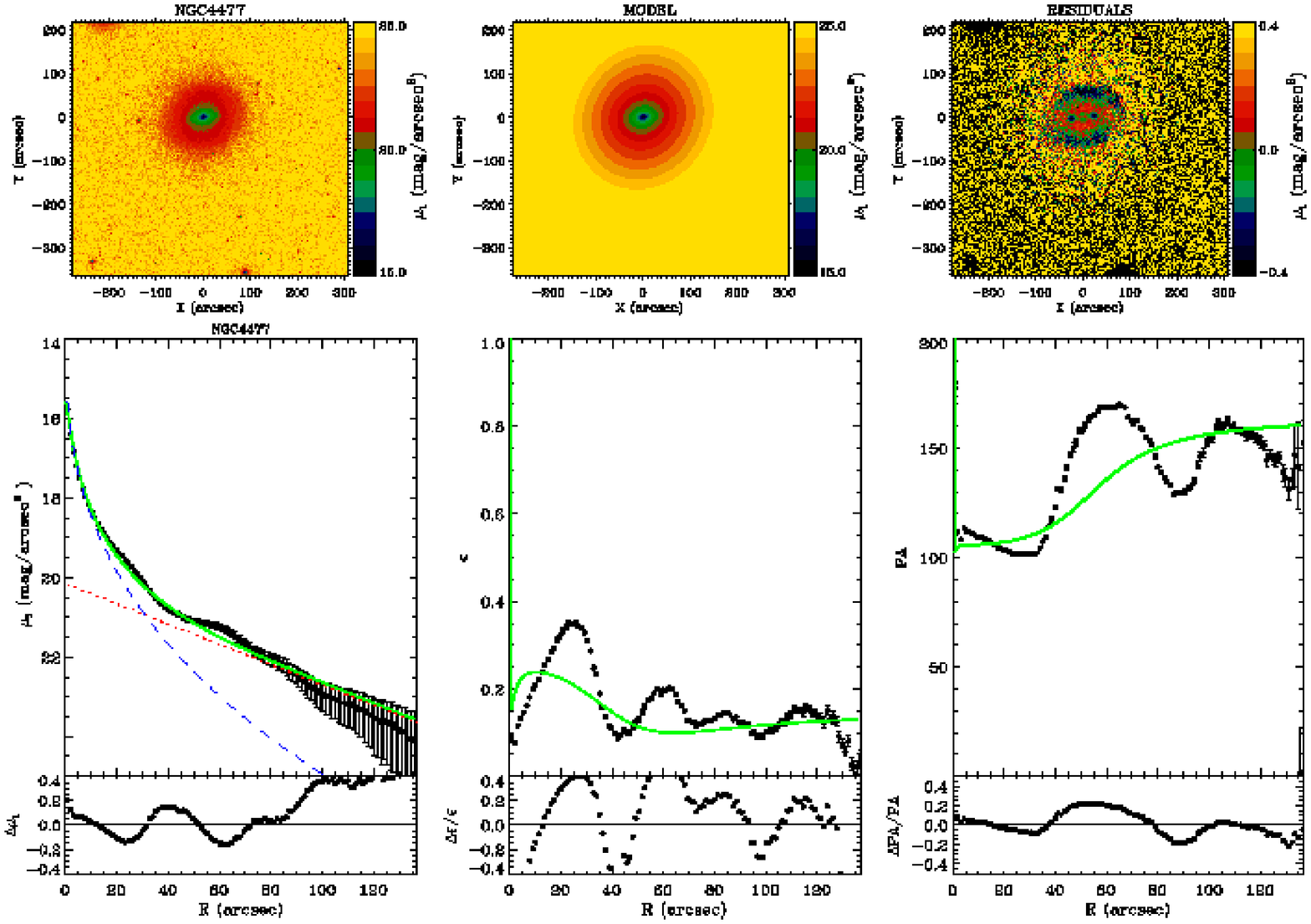}
\hspace{1cm}
\includegraphics[angle=90.0,width=0.45\textwidth]{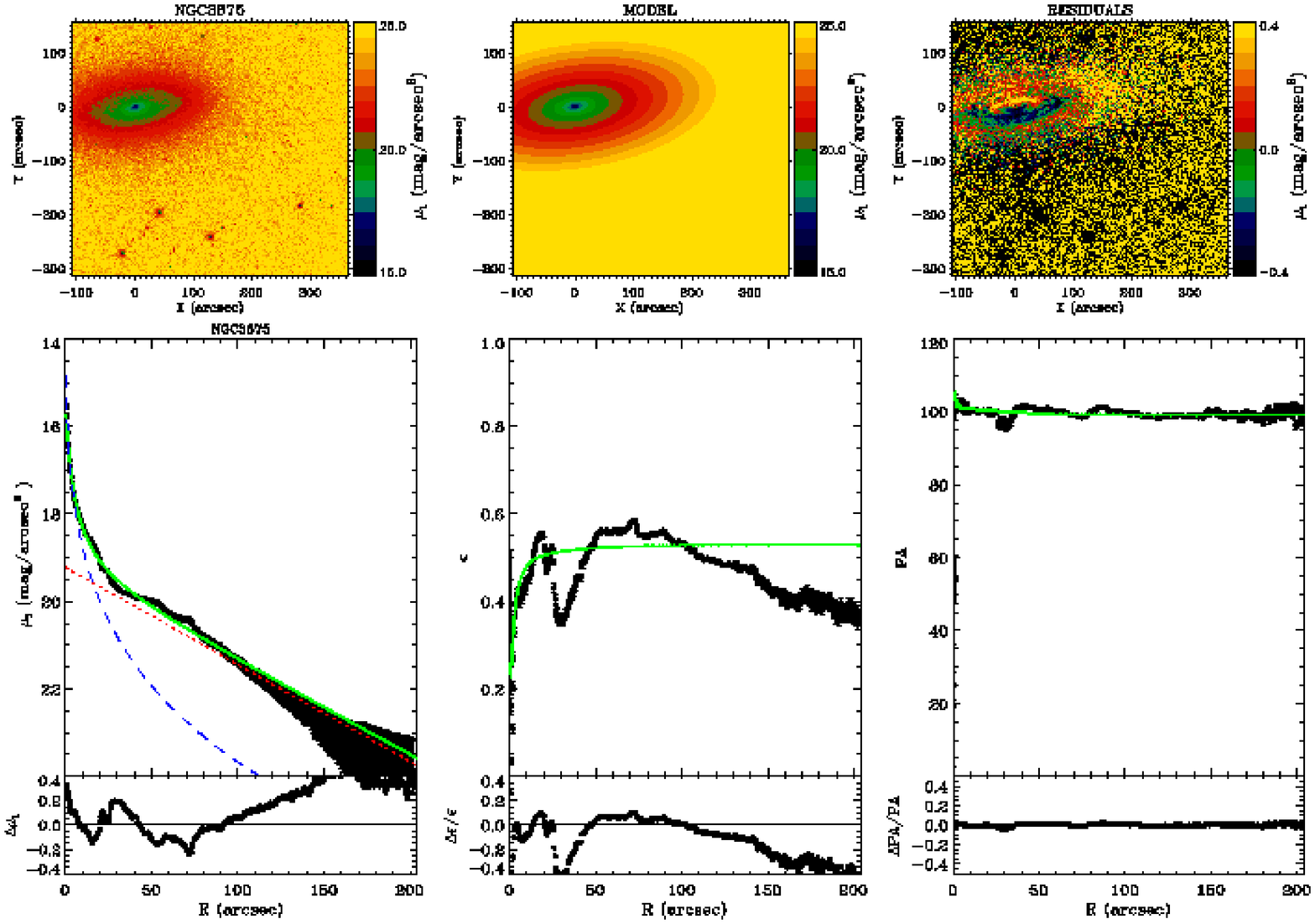}
\end{center}
\caption{Two-dimensional photometric decomposition of the sample
  galaxies shown in Figure~\ref{fig:profiles} illustrating the various
  fitting strategies adopted with {\sc GASP2D}. For each galaxy, we show in
  the upper panels the SDSS $i$-band image (left), the best-fit image
  (middle), and the residual (i.e., observed-model) image (right).
  The lower panels show the ellipse-averaged radial profiles of the
  surface brightness (left), ellipticity (middle), and position angle
  (right) measured in the SDSS (dots) and model image (green
  continuous line). The difference between the ellipse-averaged radial
  profiles from the observed and model images are also shown. The
  dashed blue and dotted red lines represent the intrinsic surface
  brightness radial profiles of the bulge and disc, respectively. No
  disc contribution is assumed for the elliptical galaxy NGC~5127.}
\label{fig:profiles_gasp2d}
\end{figure*}
}

\newcommand{\placefigfour}{ 
\begin{figure}
\begin{center}
\includegraphics[width=0.9\columnwidth]{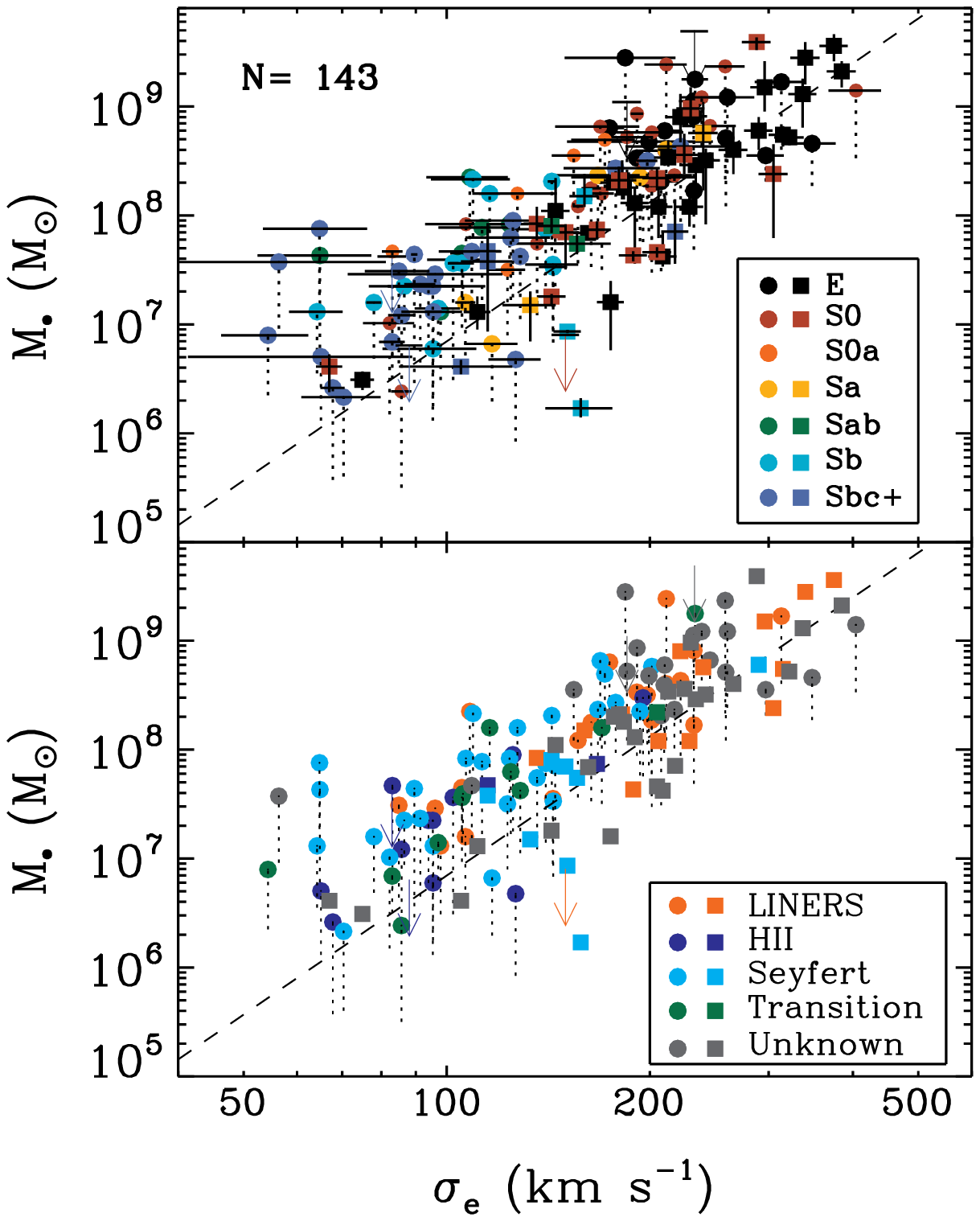}
\end{center}
\caption{\mbh\ as a function of \sigmae\ for 94 Sample A galaxies (89
  upper limits from nebular line widths, circles; 5 upper limits from
  resolved kinematics, arrows) and 49 Sample B galaxies (squares).
  The total number of galaxies is $N=143$.  The error bars for
  \sigmae\ are shown only in the upper panel for clarity.  The lower
  and upper ends of the dotted lines correspond to \mbh\ upper limits
  estimated assuming an inclination of $i=33$\degree\ and 81\degree\
  for the unresolved Keplerian disc, respectively. Galaxies are
  plotted according to morphological type (upper panel) and nuclear
  activity (lower panel). The dashed line is the
  \citetalias{Gultekin2009} \mse\ relation.  The Sbc$+$ bin includes
  all galaxies classified as Sbc or later.}
\label{fig:m_s}
\end{figure}
}

\newcommand{\placefigfive}{ 
\begin{figure}
\begin{center}
\includegraphics[width=0.9\columnwidth]{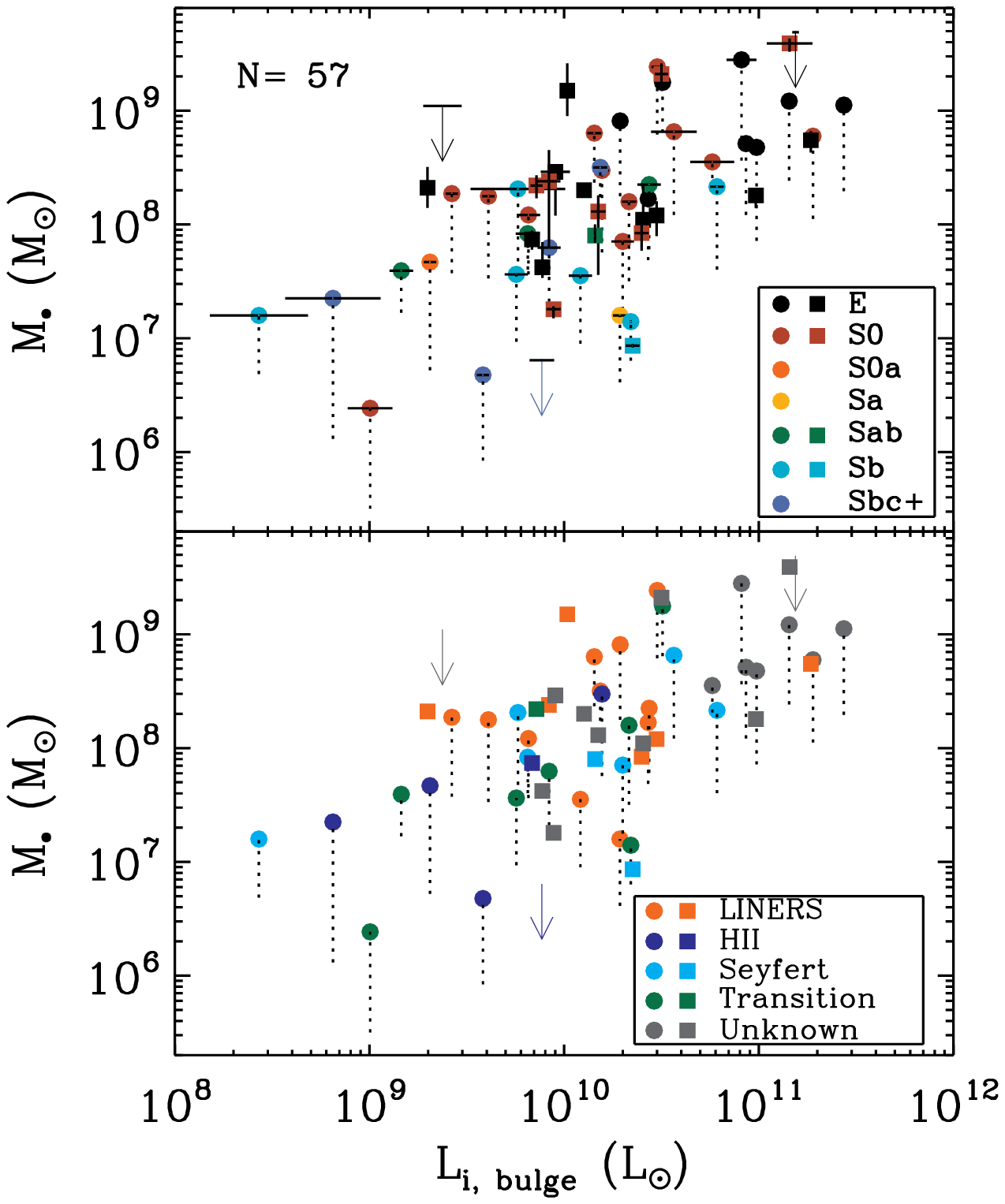}
\end{center}
\caption{\mbh\ as a function of \Lbui\ for 38 Sample A galaxies (35
  upper limits from nebular line widths, circles; 3 upper limits from
  resolved kinematics, arrows) and 19 Sample B galaxies (squares) for
  which two-dimensional bulge-to-disc decompositions of the SDSS
  $i$-band images were performed.  Symbols and panels are as in
  \Fig{m_s}.}
\label{fig:ml_bulge}
\end{figure}
}

\newcommand{\placefigsix}{ 
\begin{figure}
\begin{center}
\includegraphics[width=0.9\columnwidth]{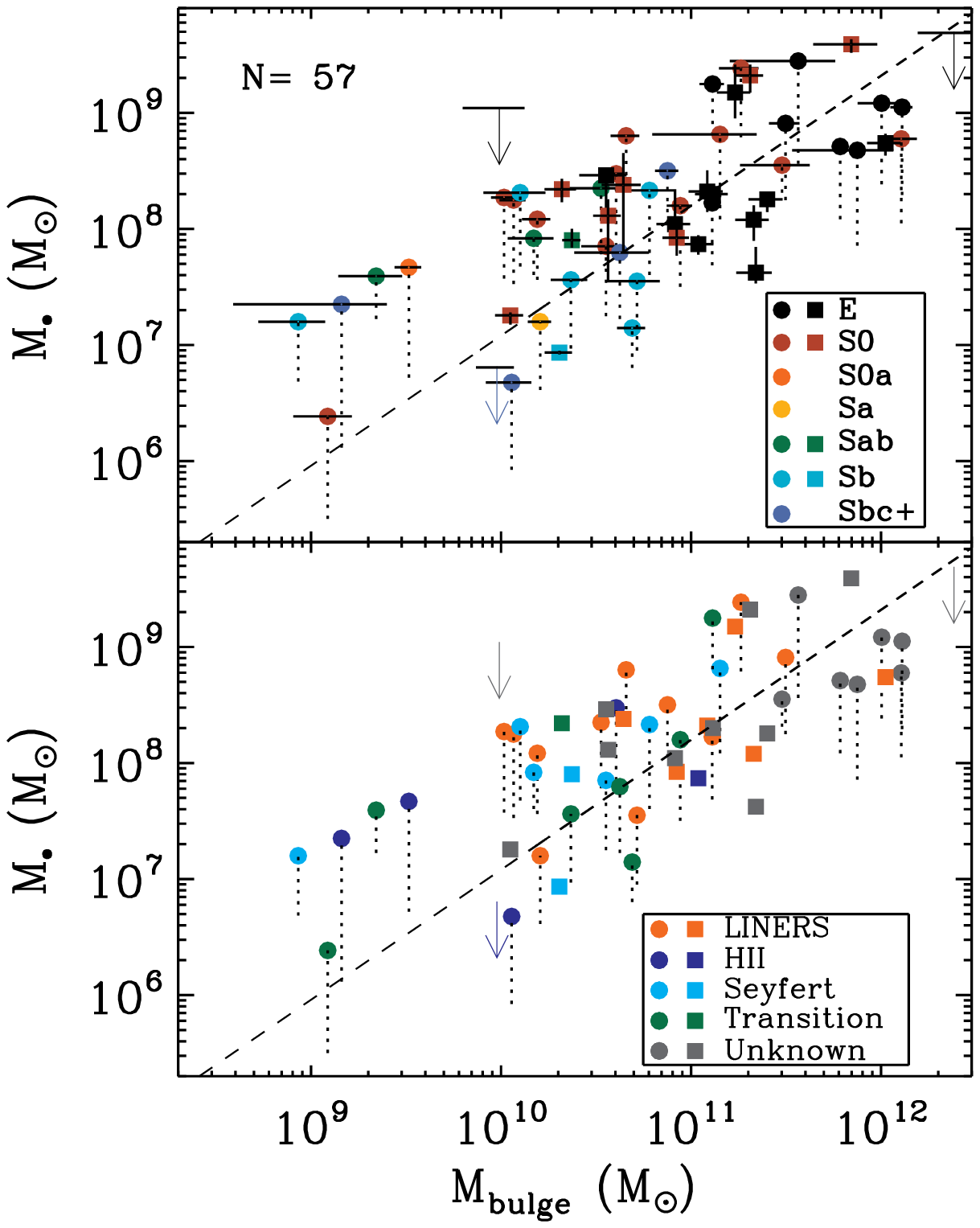}
\end{center}
\caption{\mbh\ as a function of \Mbuvir\ for the same sample as in
  \Fig{ml_bulge}.  Symbols and panels are as in \Fig{m_s}.  The dashed
  line is the \citet{Haering2004} \mbh$-$\Mbu\ relation.}
\label{fig:mmass_bulge}
\end{figure}
}

\newcommand{\placefigseven}{ 
\begin{figure}
\begin{center}
\includegraphics[width=0.9\columnwidth]{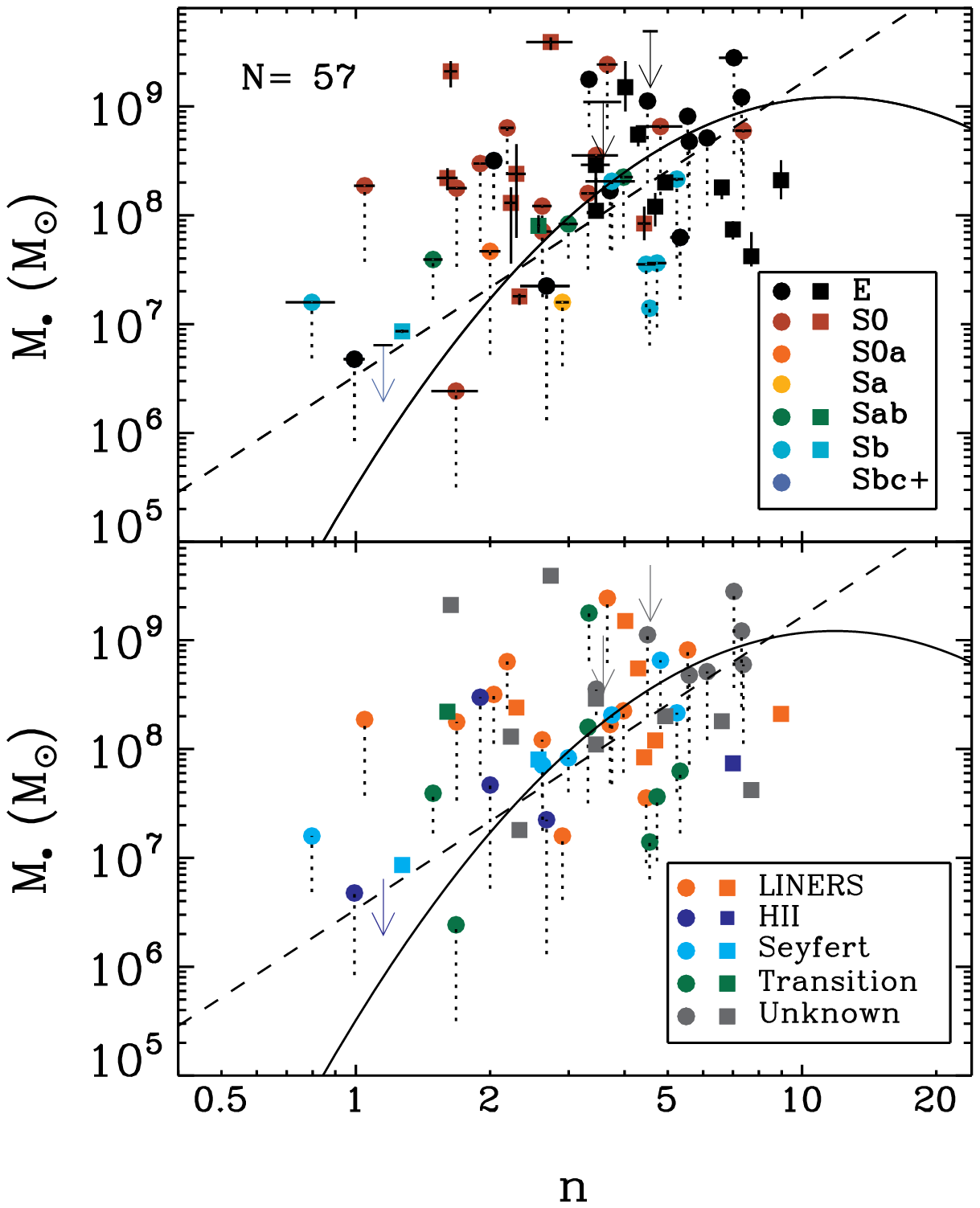}
\end{center}
\caption{\mbh\ as a function of \n\ for the same sample as in
  \Fig{ml_bulge} for which two-dimensional bulge-to-disc
  decompositions of the SDSS $i$-band images were performed.  Symbols
  and panels are as in \Fig{m_s}.  The \mbh$-$\n\ relations by
  \citet{Graham2001, Graham2003a} and \citet{Graham2007} are shown as
  the dashed and continuous lines, respectively.}
\label{fig:mbh_n}
\end{figure}
}

\newcommand{\placefigeight}{ 
\begin{figure}
\begin{center}
\includegraphics[width=0.9\columnwidth]{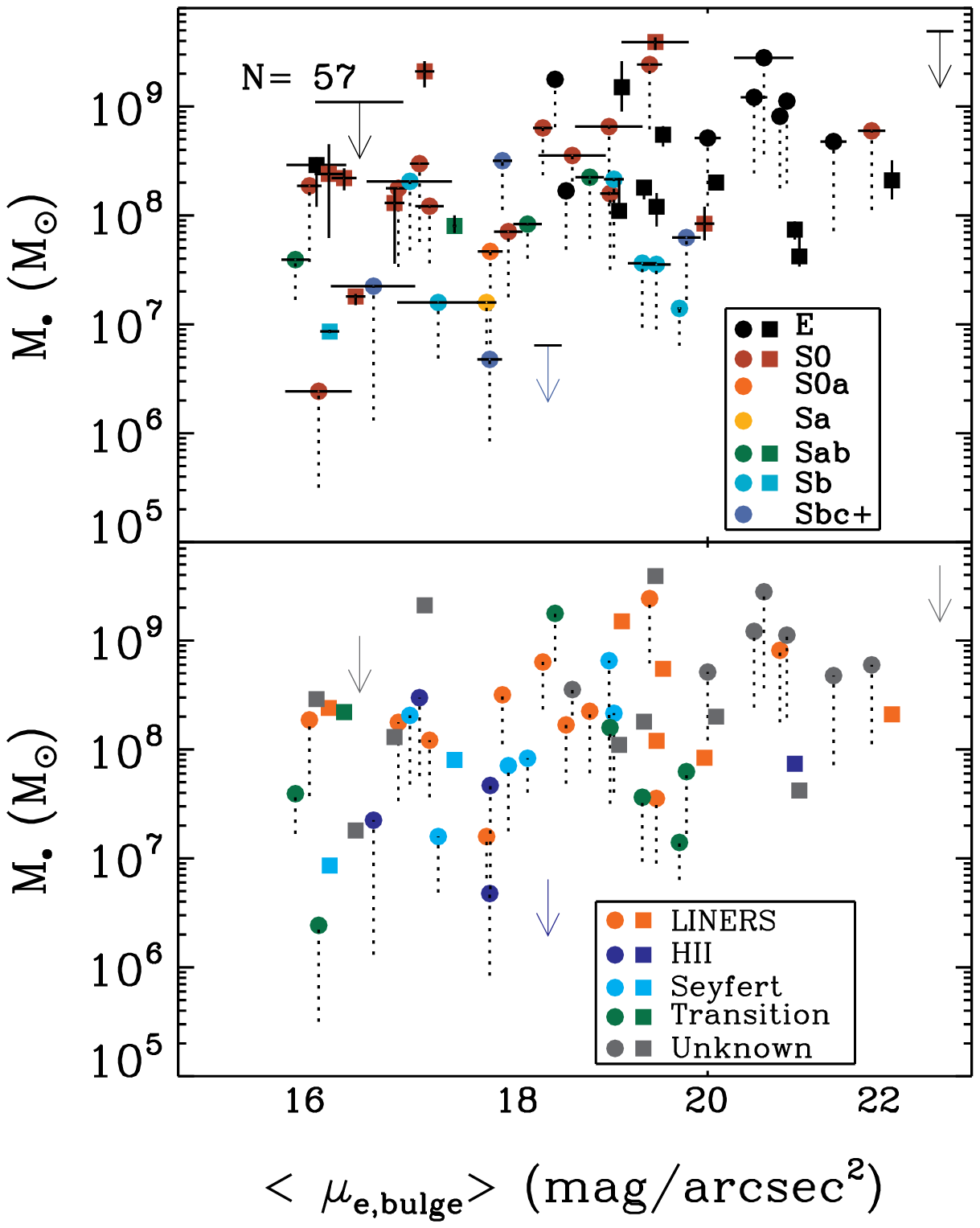}
\end{center}
\caption{\mbh\ as a function of \muemean\ for the same sample as in
  \Fig{mbh_n}.  Symbols and panels are as in \Fig{m_s}.}
\label{fig:mbh_mue}
\end{figure}
}

\newcommand{\placefignine}{ 
\begin{figure}
\begin{center}
\includegraphics[width=0.9\columnwidth]{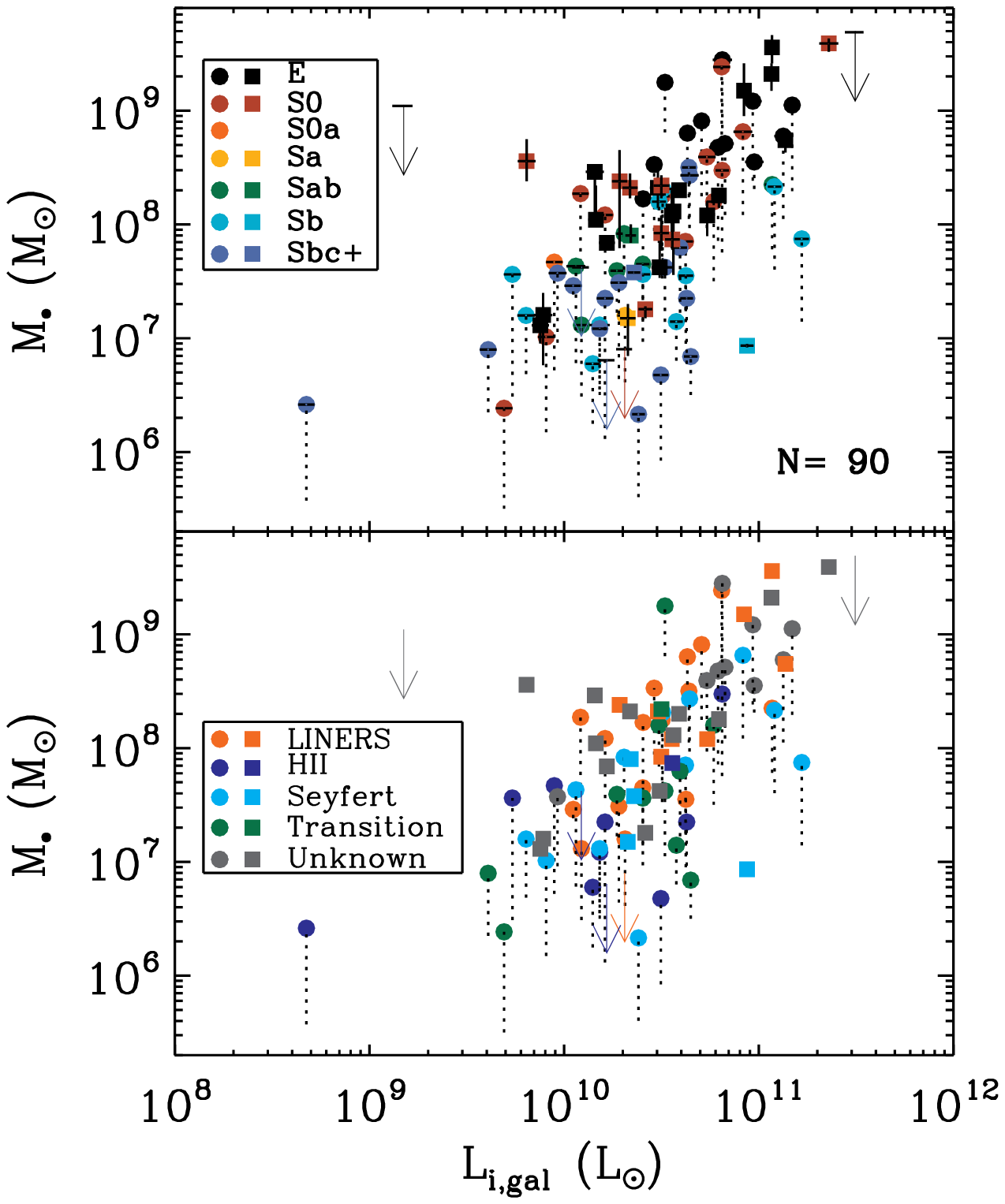}
\end{center}
\caption{\mbh\ as a function of \li\ for 62 Sample A galaxies (57
  upper limits from nebular line widths, circles; 5 upper limits from
  resolved kinematics, arrows) and 28 Sample B galaxies (squares) for
  which azimuthally-averaged luminosity profiles from SDSS $i$-band
  images were extracted.  Symbols and panels are as in \Fig{m_s}.}
\label{fig:mltot}
\end{figure}
}

\newcommand{\placefigten}{ 
\begin{figure}
\begin{center}
\includegraphics[width=0.9\columnwidth]{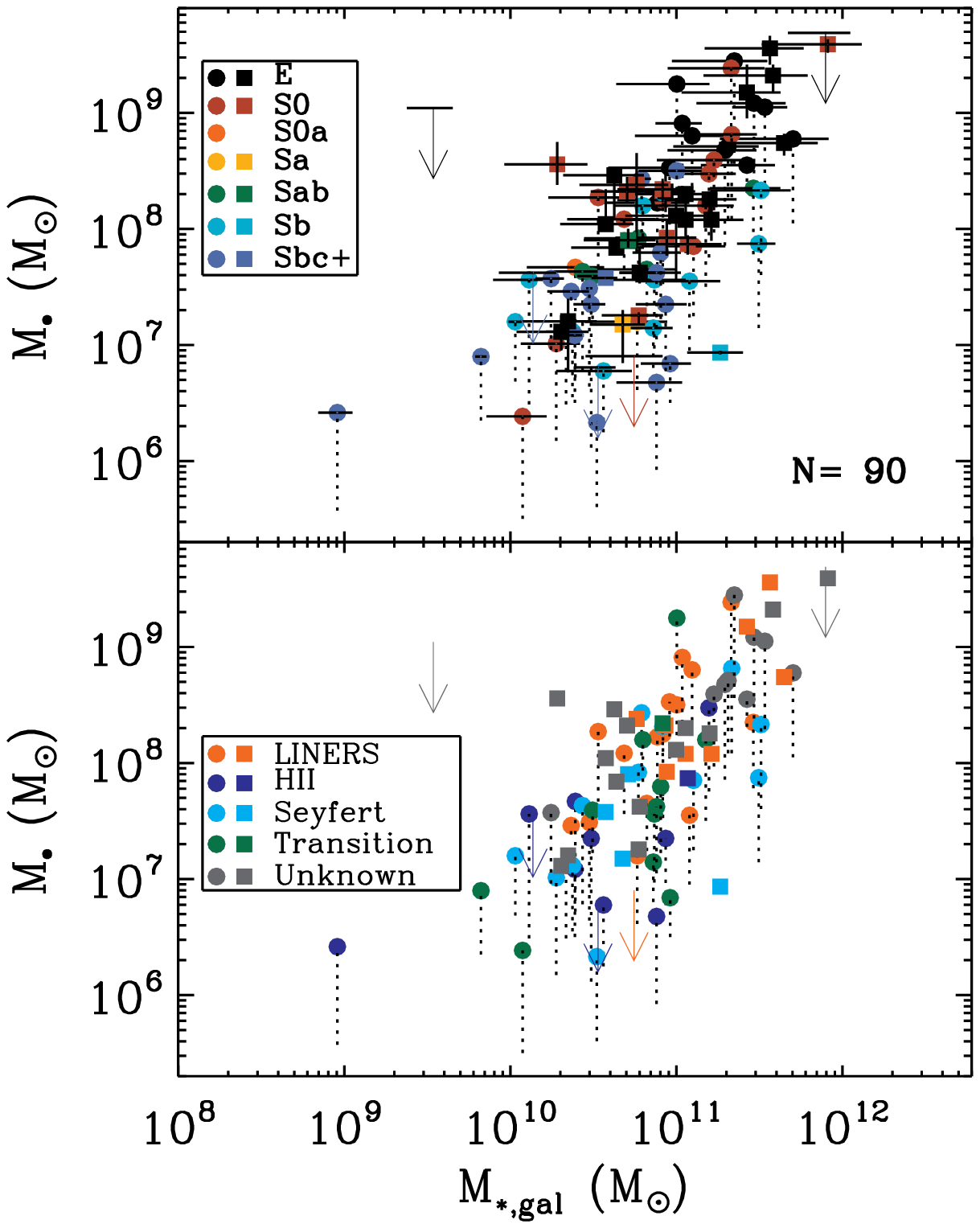}
\end{center}
\caption{\mbh\ as a function of \mstar\ for the same sample as in
  \Fig{mltot} for which $(g-i)$ colours from SDSS $g$ and $i$-band
  images were derived.  Symbols and panels are as in \Fig{m_s}.}
\label{fig:m_stellar_mass}
\end{figure}
}

\newcommand{\placefigeleven}{ 
\begin{figure}
\begin{center}
\includegraphics[width=0.9\columnwidth]{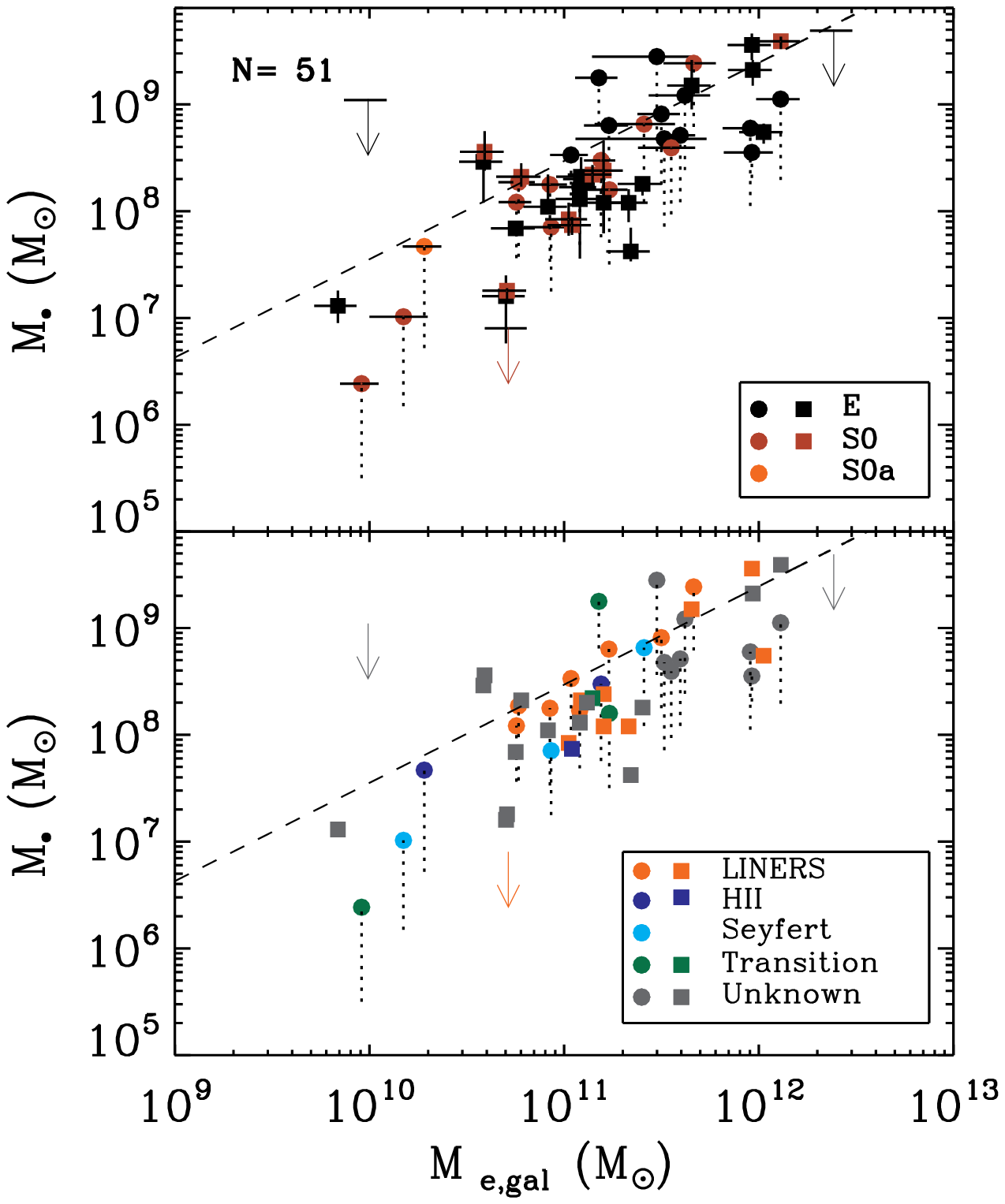}
\end{center}
\caption{\mbh\ as a function of \megal\ for 27 Sample A galaxies (24
  upper limits from nebular line widths, circles; 3 upper limits from
  resolved kinematics, arrows) and 24 Sample B galaxies (squares)
  ranging from E to S0a.  Symbols and panels are as in \Fig{m_s}. The
  dashed line is the \citet{Ferrarese2006} \mbh$-$\megal\ relation.}
\label{fig:mbh_vir}
\end{figure}
}

\newcommand{\placefigtwelve}{ 
\begin{figure}
\begin{center}
\includegraphics[width=0.9\columnwidth]{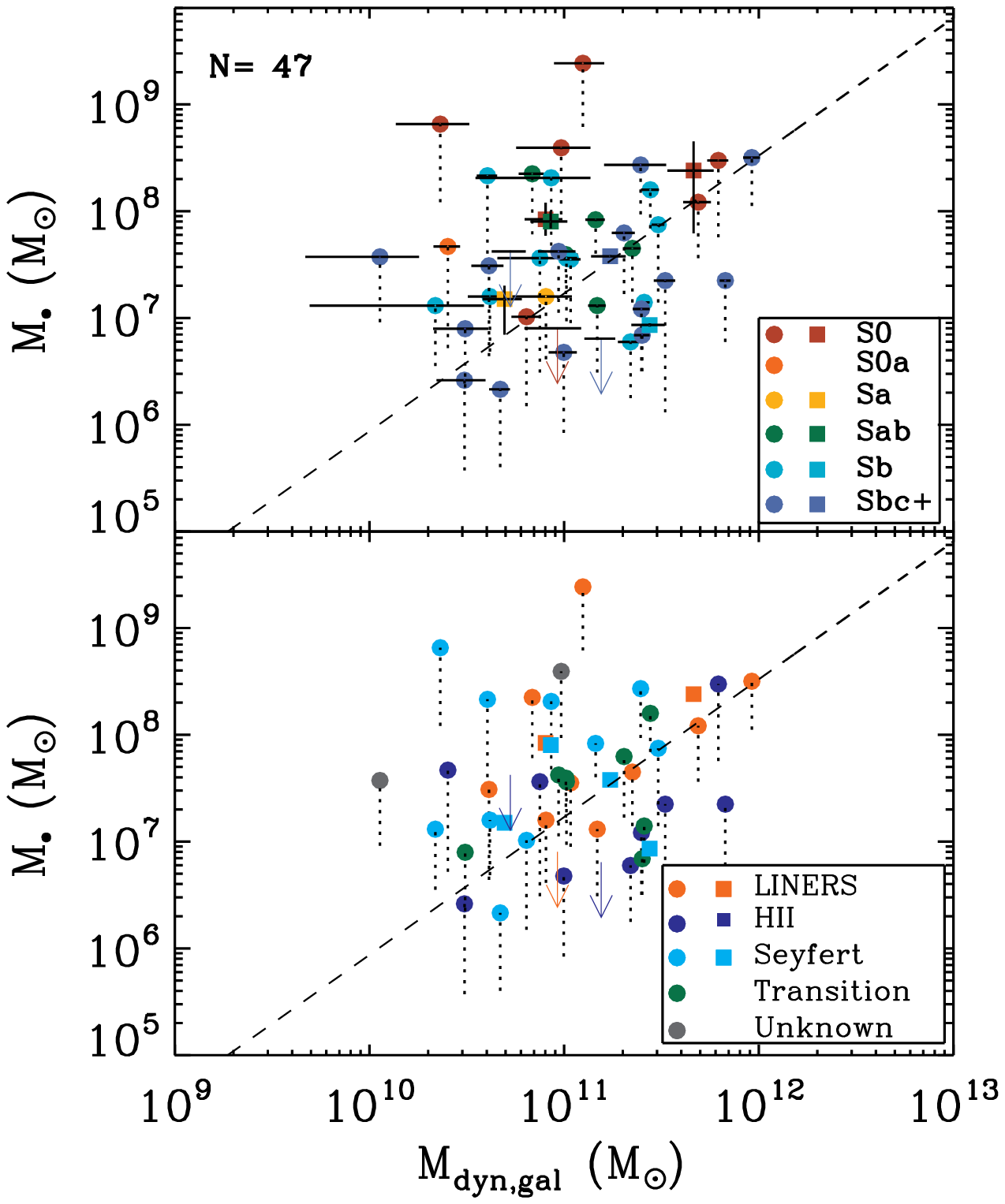}
\end{center}
\caption{\mbh\ as a function of \mdyn\ for 41 Sample A disc galaxies
  (38 upper limits from nebular line widths, circles; 3 upper limits
  from resolved kinematics, arrows) and 6 Sample B (squares).  Symbols
  and panels are as in \Fig{m_s}. The dashed line is the
  \citet{Ho2008b} \mbh$-$\mdyn\ relation.}
\label{fig:MBH_Mdyn}
\end{figure}
}

\newcommand{\placefigthirteen}{ 
\begin{figure}
\begin{center}
\includegraphics[width=0.9\columnwidth]{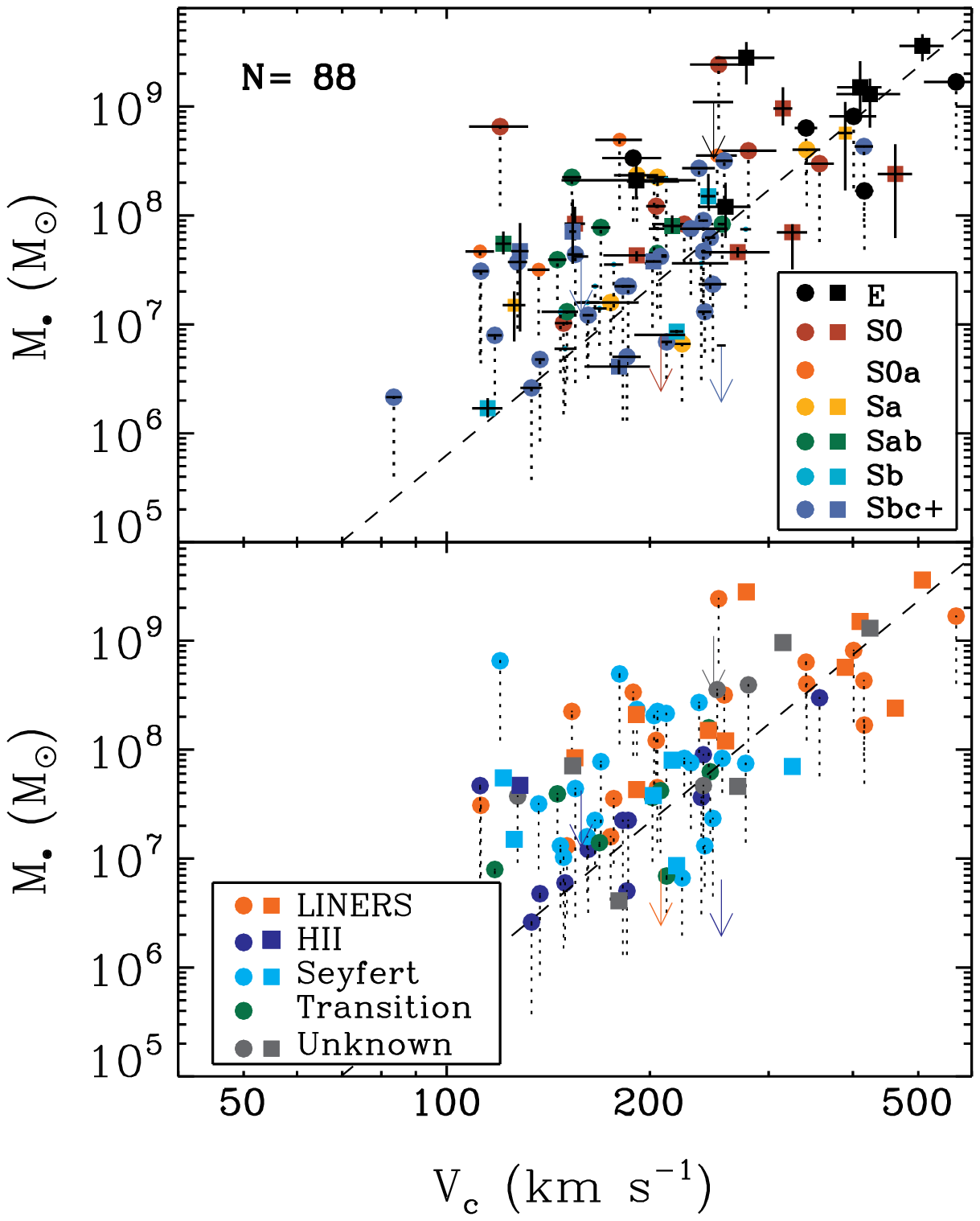}
\end{center}
\caption{\mbh\ as a function of \vc\ for 65 Sample A disc galaxies (61
  upper limits from nebular line widths, circles; 4 upper limits from
  resolved kinematics, arrows) and 23 Sample B galaxies (squares).
  Symbols and panels are as in \Fig{m_s}.  The dashed line is the
  \citet{Ho2007} \mbh$-$\vc\ relation.}
\label{fig:m_v}
\end{figure}
}

\newcommand{\placefigfourteen}{ 
\begin{figure}
\begin{center}
\includegraphics[width=0.9\columnwidth]{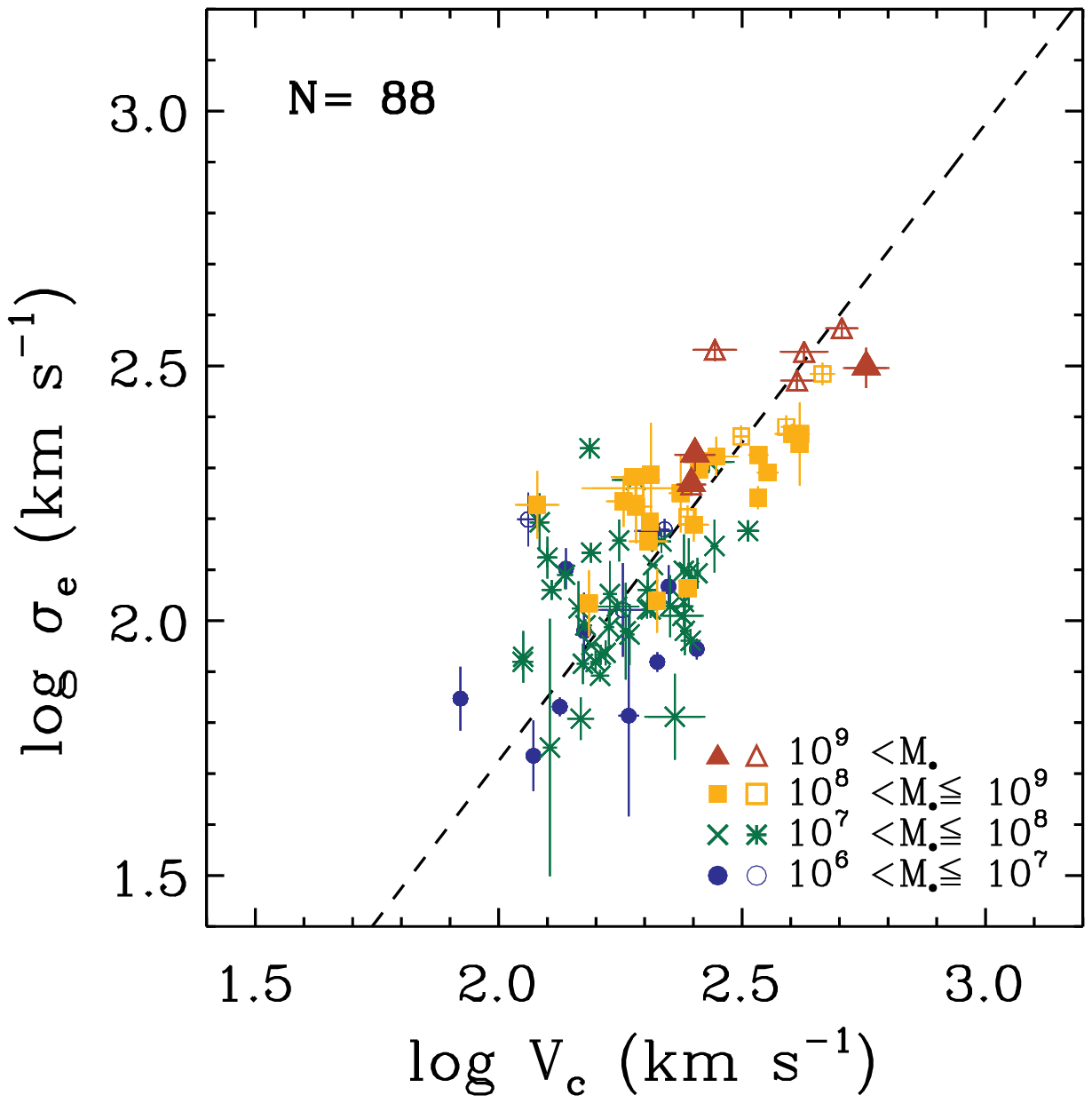}
\end{center}
\caption{\vs\ relation as a function of \mbh\ for the same sample as
  in \Fig{m_v}.  The dashed line is the \citet{Ho2007} \vc$-$\sigmae\
  relation.}
\label{fig:v_sigma_MBH}
\end{figure}
}

\newcommand{\placefigfifteen}{ 
\begin{figure*}
\begin{center}
\includegraphics[width=0.8\textwidth]{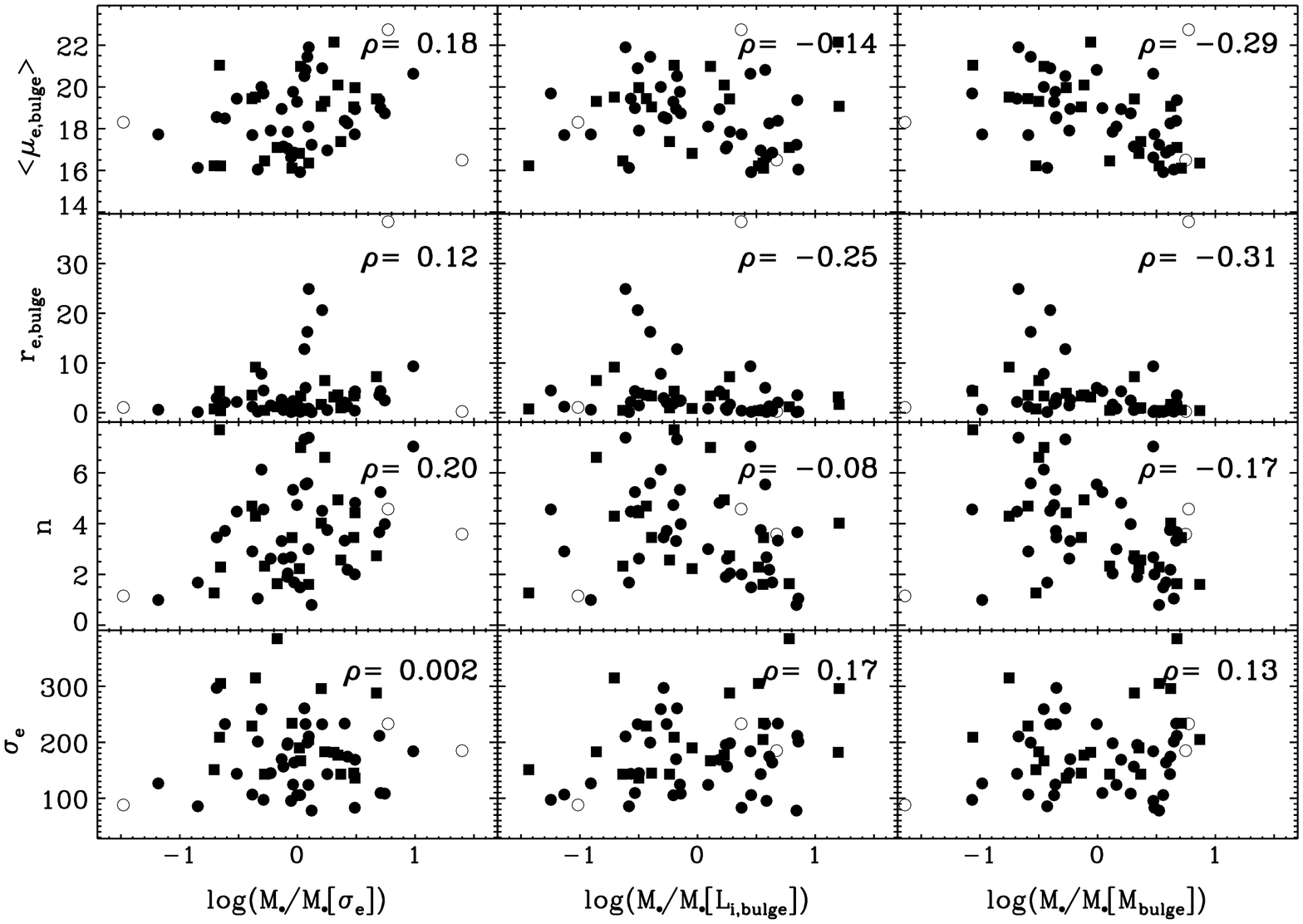}
\end{center}
\caption{Residuals of the \mse, \mlbui, \mbh$-$\Mbuvir\ relations
  versus \muemean, \reb, \n, \sigmae\ for 38 Sample A galaxies (35
  upper limits from nebular line widths, filled circles; 3 upper
  limits from resolved kinematics, open circles) and 19 Sample B
  galaxies (squares). The total number of galaxies is $N=57$.  The
  Spearman rank correlation coefficient $\rho$ is given in each
  panel.}
\label{fig:residual_mlbule}
\end{figure*}
}

\newcommand{\placefigsixteen}{ 
\begin{figure*}
\begin{center}
\includegraphics[width=\textwidth]{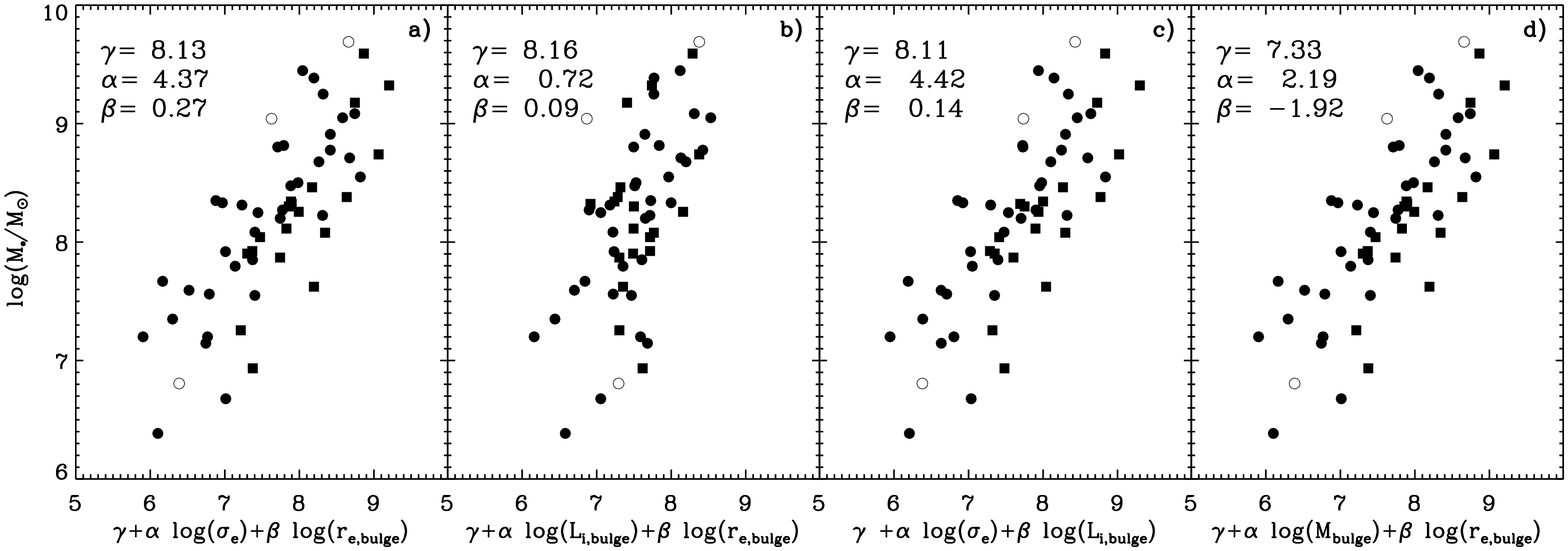}
\end{center}
\caption{\mbh\ as a function of \sigmae\ and \reb\ (a), \Lbui\ and
  \reb\ (b), \sigmae\ and \Lbui\ (c), \Mbuvir\ and \reb\ (d) for the
  same sample as in \Fig{residual_mlbule}.  The logarithmic slopes
  $\alpha$ and $\beta$ and offset $\gamma$ of the fitted relation are
  given in each panel.}
\label{fig:FP}
\end{figure*}
}

\newcommand{\placefigseventeen}{ 
\begin{figure*}
\begin{center}
\includegraphics[width=0.8\textwidth]{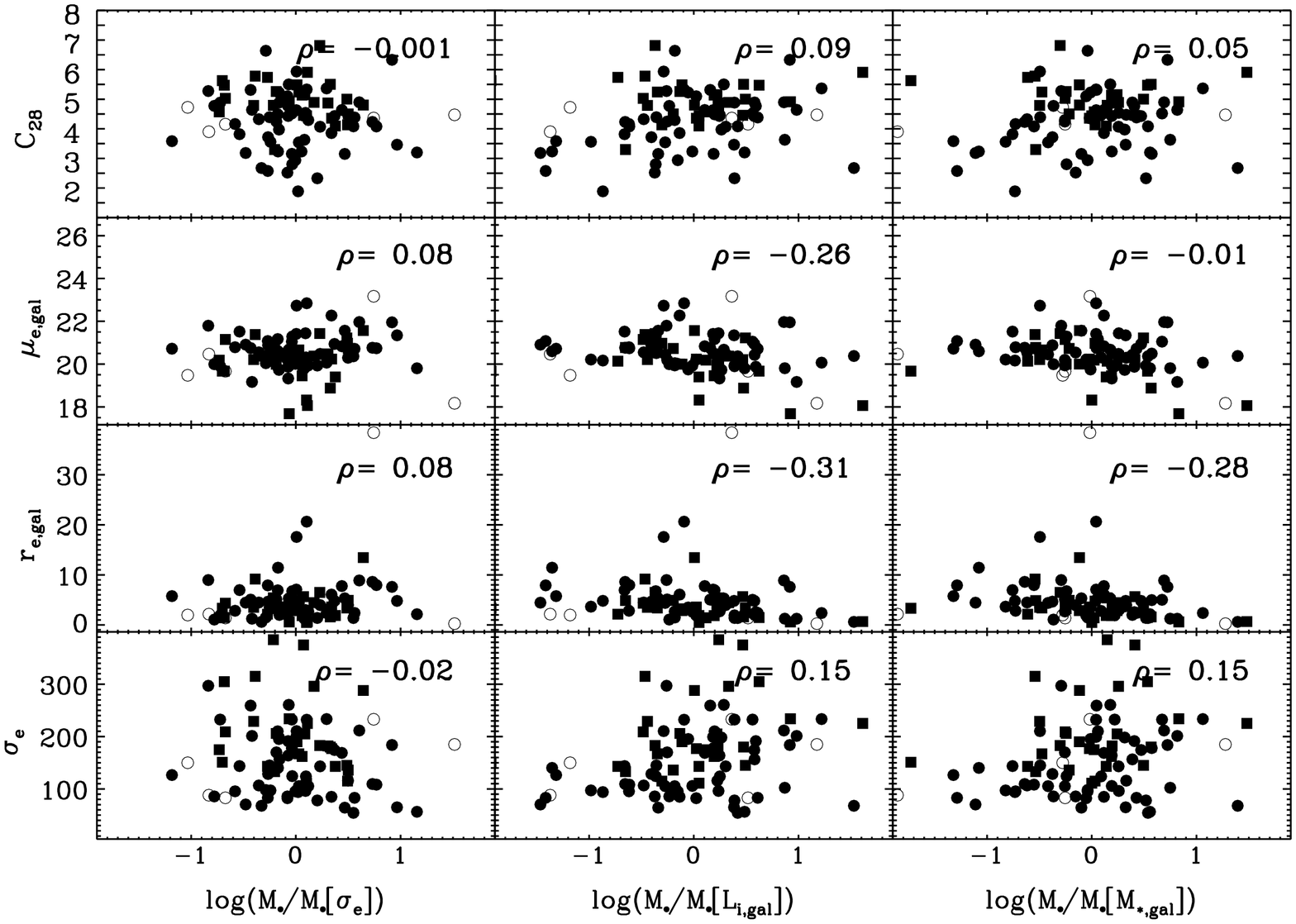}
\end{center}
\caption{Residuals of the \mse, \mli, \mbh$-$\mstar\ relations versus
  \co, $I_{\rm e, gal}$, \reg, \sigmae\ for 62 Sample A galaxies (57
  upper limits from nebular line widths, filled circles; 5 upper
  limits from resolved kinematics, open circles) and 28 Sample B
  galaxies (squares).  The total number of galaxies is $N=90$.  The
  Spearman rank correlation coefficient $\rho$ is given in each
  panel.}
\label{fig:mbh_residual_ms}
\end{figure*}
}

\newcommand{\placefigeighteen}{ 
\begin{figure*}
\begin{center}
\includegraphics[width=\textwidth]{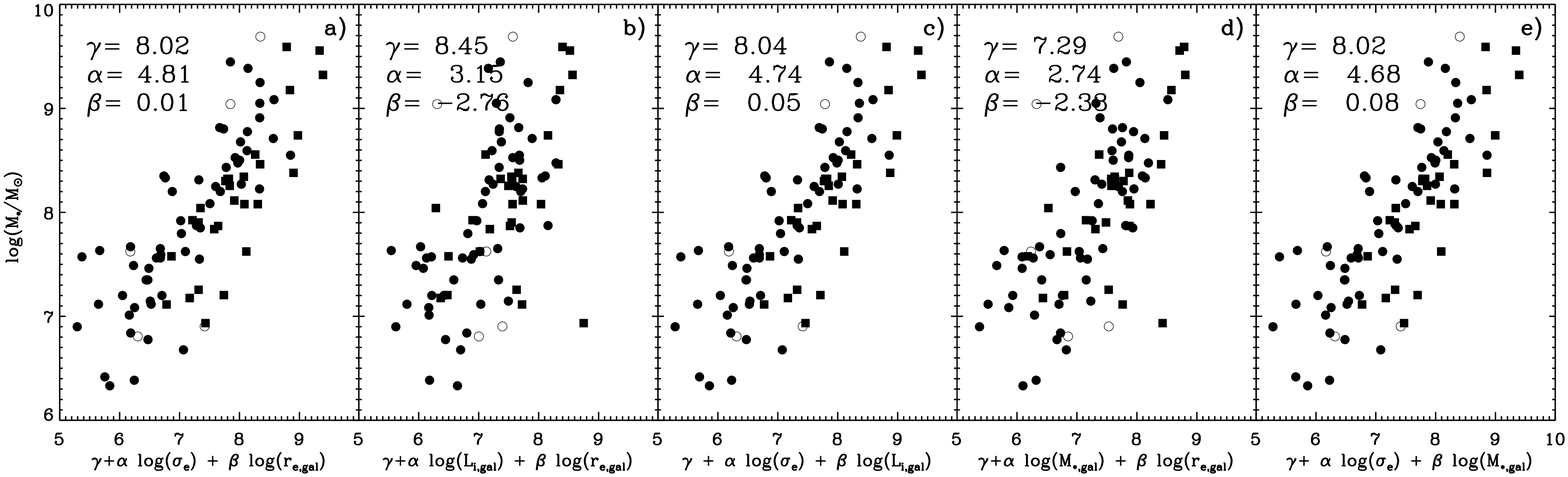}
\end{center}
\caption{\mbh\ as a function of \sigmae\ and \reg\ (a), \li and \reg\
  (b), \sigmae\ and \li\ (c), \mstar\ and \reg\ (d), and \sigmae\ and
  \mstar\ (e) for the same sample as in \Fig{mbh_residual_ms}.  The
  logarithmic slopes $\alpha$ and $\beta$ and offset $\gamma$ of the
  fitted relation are given in each panel.}
\label{fig:FP_li}
\end{figure*}
}

\begin{document}

\title[SMBH mass and galaxy properties]{On
the Correlations between Galaxy Properties and Supermassive Black Hole Mass}

\author[A. Beifiori et al.]{A. ~Beifiori$^{1,2}$\thanks{E-mail:beifiori@mpe.mpg.de}, S. Courteau$^{3}$, E.~M. Corsini$^{4}$, and Y. Zhu$^{3}$\\
$^1$Max-Planck-Institut f\"{u}r Extraterrestrische Physik,
Giessenbachstra\ss e, D-85748 Garching, Germany\\
$^2$Institute of Cosmology and Gravitation, Dennis Sciama Building,
      Burnaby Road, Portsmouth, PO1 3FX, United Kingdom\\
$^3$Queen's University, Department of Physics, Engineering Physics
and Astronomy, Kingston, ON, K7L 3N6, Canada\\
$^4$Dipartimento di Astronomia, Universit\`a di Padova,
  vicolo dell'Osservatorio 3, I-35122 Padova, Italy\\
}

\date{Accepted 2011 September 27.  Received 2011 September 13; in original form 2011 January 5}

\pagerange{\pageref{firstpage}--\pageref{lastpage}} \pubyear{2011}

\maketitle

\label{firstpage}

\begin{abstract}
  We use a large sample of upper limits and accurate estimates of
  supermassive black holes (SMBHs) masses coupled with libraries of
  host galaxy velocity dispersions, rotational velocities and
  photometric parameters extracted from Sloan Digital Sky Survey
  $i$-band images to establish correlations between the SMBH and host
  galaxy parameters.  We test whether the mass of the black hole,
  \mbh, is fundamentally driven by either local or global galaxy
  properties.
We explore correlations between \mbh\ and stellar velocity dispersion
\sigmae, $i$-band bulge luminosity \Lbui, bulge mass \Mbuvir, bulge
\sersic index \n, bulge mean effective surface brightness \muemean,
$i$-band luminosity of the galaxy \li, galaxy stellar mass \mstar,
maximum circular velocity \vc, galaxy dynamical and effective masses
\mdyn\ and \megal.
We verify the tightness of the \mse\ relation and find that correlations
with other galaxy parameters do not yield tighter trends.
We do not find differences in the \mse\ relation of barred and
unbarred galaxies.  The \mse\ relation of pseudo-bulges is also
coarser and has a different slope than that involving classical
bulges. The \mbh$-$\Mbuvir\ is not as tight as the \mse\ relation,
despite the bulge mass proving to be a better proxy of \mbh\ than
bulge luminosity, and despite adding the bulge effective radius as an
additional fitting parameter.  Contrary to various published reports,
we find a rather poor correlation between \mbh\ and \n\ (or \muemean)
suggesting that \mbh\ is not related to the bulge light concentration.
The correlations between \mbh\ and galaxy luminosity or mass are not a
marked improvement over the \mse\ relation. These scaling relations
depend sensitively on the host galaxy morphology: early-type galaxies
follow a tighter relation than late-type galaxies.  If \vc\ is a proxy
for the dark matter halo mass, the large scatter of the \mv\ relation
then suggests that \mbh\ is more coupled to the baryonic rather than
the dark matter.
We have tested the need for a third parameter in the \mbh\ scaling
relations, through various linear correlations with bulge and galaxy
parameters, only to confirm that the fundamental plane of the SMBH is
mainly driven by \sigmae\, with a small tilt due to the effective
radius.
We provide a compendium of galaxy structural properties for most of
the SMBH hosts known to date.

\end{abstract}


\section{Introduction}
\label{sec:introduction}
The mass \mbh\ of supermassive black holes (SMBHs) is closely tied to
the properties of the spheroidal component of galaxies, such as the
bulge luminosity, \Lbu\ \citep[][ hereafter
  \citetalias{Gultekin2009}]{Dressler1989, Kormendy1995, Marconi2003a,
  Graham2007b, Gultekin2009}, the stellar velocity dispersion,
\sigmas\ (\citealt{Ferrarese2000}; \citealt{Gebhardt2000};
\citealt{Tremaine2002}; \citealt{Ferrarese2005};
\citetalias{Gultekin2009}), the mass of the bulge
\citep{Magorrian1998, Haering2004}, the central light concentration
\citep{Graham2001}, the \sersic index \citep{Graham2007}, the virial
mass of the galaxy \citep{Ferrarese2006}, the gravitational binding
energy \citep{Aller2007}, the kinetic energy of random motions of the
bulge \citep{Feoli2009}, and the stellar light and mass deficit
associated to the core ellipticals
\citep{Lauer2007b,Kormendy2009}. Most of these relations are
inter-compared in \citet{Novak2006} and in \citetalias{Gultekin2009}.
Given the \ms\ relation and the correlation between \sigmas\ and the
circular velocity, \vc, \citet{Ferrarese2002} and \citet{Pizzella2005}
suggested a link between \mbh\ and \vc\ (or equivalently, with the
mass of the dark matter halo).  However, \citet{Courteau2007} and
\citet{Ho2007} pointed out that the \vc$-$\sigmas\ relation actually
depends on galaxy morphology (or equivalently, on its light
concentration) thus precluding a simple connection between \mbh\ and
\vc. \citet{Kormendy2011b} studied a sample of bulgeless galaxies and
concluded that there is almost no correlation between the SMBH and
dark matter halo, unless the galaxy also contains a bulge.

Several authors have noted that the residuals of the \ms\ and \mlbu\
relations correlate with the galaxy effective radius
\citep[e.g.,][]{Marconi2003a}.
\citet{Hopkins2007a, Hopkins2007b} suggested the possibility of a
linear combination between different galaxy properties to reduce the
scatter of the \mbh\ scaling laws, heralding the idea of a fundamental
plane of SMBHs (BHFP).  Many SMBH scaling relations could thus be seen
as projections of the BHFP \citep{Aller2007, Barway2007}.  A
correlation of \mbh\ with more than one galaxy parameter would suggest
a SMBH growth sensitive to the overall structure of the host galaxy.

The local characterisation and cosmic evolution of the \mbh\ scaling
relations have already been examined through theoretical models for
the coevolution of galaxies and SMBHs \citep{Granato2004,
  Vittorini2005, Hopkins2006, Monaco2007}.  These studies have
revealed that the observed scaling relations could be reproduced in
models of SMBH growth with strong feedback from the active galactic
nucleus \citep[AGN,][]{Silk1998, Cox2006, Robertson2006a,
  Robertson2006b, DiMatteo2005}. In particular, these models predict
the existence of the BHFP \citep{Hopkins2007a, Hopkins2007b,
  Hopkins2009}.  However, while the observed relations can be
reproduced by the models, these still depend on the adopted slope,
zero point, and scatter \citep{Somerville2009} which remain
ill-constrained.

In this work, we make use of a large sample of galaxies with available
\mbh\ estimates to improve our understanding of the known scaling laws
over a wide range of \mbh, morphological type and nuclear activity, as
well as to test for possible correlations of \mbh\ with different
combinations of spheroid and galaxy parameters.

This paper is organised as follows.  The sample of SMBH hosts is
described in \se{sample_selection}. Their photometric, kinematic, and
dynamical properties are presented in \se{gal_properties}, while the
correlations between \mbh\ and the bulge and galaxy properties are
shown in \se{analysis}.  We discuss our results and conclude in
\se{discussion}.

\section{Sample Description}
\label{sec:sample_selection}

The \mbh\ values were retrieved from two different samples: the
compilation of \mbh\ upper limits by \citet[][hereafter
  \citetalias{Beifiori2009}]{Beifiori2009} and the compilation of
secure \mbh\ by \citetalias{Gultekin2009}.

The \mbh\ estimates of \citetalias{Beifiori2009} were obtained from
{\it Hubble Space Telescope} (\hst) archival spectra.  That sample
includes nuclear spectra for 105 nearby galaxies with $D <100$ Mpc
obtained with the Space Telescope Imaging Spectrograph (STIS) equipped
with the G750M grating.  The spectra cover the region of the \ha\ line
and \niipg\ and \siipg\ doublets.  The nebular-line widths were
modelled in terms of gas motion in a thin disc of unknown orientation
but known spatial extent following the method of \citet{Sarzi2002}.
\citetalias{Beifiori2009} adopted two different inclinations for the
gaseous disc with a nearly face-on disc ($i=33$\degree) hosting a
larger \mbh\ and a nearly edge-on disc (81\degree) harbouring a
smaller \mbh.  The two inclinations correspond to the 68\% upper and
lower confidence limits for randomly oriented discs.
We augmented the \citetalias{Beifiori2009} sample with the \mbh\ upper
limits of NGC~2892 and NGC~5921.  The STIS/G750M spectra for these
galaxies were retrieved from the HST archive and we have calculated
their \mbh\ upper limits from the nebular line widths following the
prescription of \citetalias{Beifiori2009}.
We include in Sample A the set of 105 galaxies from
\citetalias{Beifiori2009} minus 18 galaxies in common with
\citetalias{Gultekin2009} (15 of them with a secure \mbh\ and 3 with a
\mbh\ upper limit derived from the dynamical modelling of resolved
kinematics).  We are left with 87 \mbh\ upper limits from
\citetalias{Beifiori2009} based on nebular-line widths. We also
include two newly determined \mbh\ upper limits and the five upper
limits derived from the dynamical modelling of resolved kinematics by
\citetalias{Gultekin2009}. The resulting 94 galaxies constitute our
Sample A.

\citetalias{Gultekin2009} collected \mbh\ data published up to
November 2008 based on the resolved kinematics of ionised gas, stars,
and water maser for total sample of 49 galaxies with a secure \mbh\
estimate and 18 galaxies with a \mbh\ upper limit.  Our Sample B is
limited to the 49 definite values of \mbh.  The five upper limits
derived from the dynamical modelling of resolved kinematics already
belong to Sample A.  The remaining 13 upper limits by
\citetalias{Gultekin2009} are taken from the nebular line width
measurements by \citet{Sarzi2002}.  They were also measured by
\citetalias{Beifiori2009} and are therefore already included in Sample
A.

All the data relative to Samples A and B are listed in
Table~\ref{tab:allMBH_vc} and \ref{tab:MBH_gultekin}, respectively.
The upper limits by \citetalias{Beifiori2009} were rescaled assuming
$H_0=70$ \kmsmpc, $\Omega_{\rm m}=0.3$, and $\Omega_{\Lambda}=0.7$ to
conform with \citetalias{Gultekin2009}.

\Fig{comparison_b09_g09} shows the upper limits and the accurate
determinations of \mbh\ for the 18 galaxies in common between the
\citetalias{Beifiori2009} and \citetalias{Gultekin2009} samples.  The
\mbh\ estimates by \citetalias{Beifiori2009} and
\citetalias{Gultekin2009} are consistent within $1\sigma$ of each
other; no systematic offset is detected. Thus, nebular line width
measurements by \citetalias{Beifiori2009} (included in Sample A) trace
well the nuclear gravitational potential dominated by the central
SMBH, allowing a reliable estimate of \mbh\ (as those in Sample B).
Therefore, we shall use the \mbh\ upper limits from Sample A to study
the correlations of \mbh\ versus various galaxy parameters.  We adopt
for our tests the case of $i=33$\degree\ which maximises the upper
limit on \mbh.
\citetalias{Beifiori2009} already compared their \mse\ relation with
those of \citet{Ferrarese2005} and \citet{Lauer2007b} to show that
their upper limits on \mbh\ (included in Sample A) are a valuable
proxy for the more secure determinations of \mbh\ (comprised in Sample
B).  A Kolmogorov-Smirnov test on the distributions of \mbh\ and
\sigmas\ indicates that Samples A and B could be drawn from the same
parent distribution to better than the 80\% confidence level. 
For this reason, we merged Samples A and B into a joint A$+$B data
set. The combination of Samples A and B yields a total sample of 143
\mbh\ determinations.
  The union of those two samples proves most valuable for the proper
  statistical assessment of scaling relations with \mbh.

\placefigone 

\section{Galaxy Properties}
\label{sec:gal_properties}

The aim of this study is to determine the strength of the
correlations, if any, between \mbh\ and the properties of their host
galaxies. For the latter, we have compiled as large a collection as
possible of homogeneous measurements of photometric parameters
(effective radius, effective surface brightness, \sersic index, and
luminosity of the bulge, effective radius, effective surface
brightness, concentration, and total luminosity of the galaxy),
kinematic properties (stellar velocity dispersion and circular
velocity), and masses (bulge mass, galaxy stellar mass, virial and
dynamical mass of the galaxy).

The sample demographics are as follows: 29\% of the host galaxies are
ellipticals, 27\% lenticulars, and 44\% are spirals \citep[][hereafter
RC3]{RC3}. Regarding nuclear activity, 23\% of the sample galaxies are
Low-Ionisation Nuclear Emission-line Regions (LINERs), 11\% host
\hiire\ nuclei, 25\% are Seyferts, and 8\% are classified as
transition objects according to \citet{Ho1997}.  The remaining 33\% do
not show central emission.

We describe in the sub-sections below the extraction of all the galaxy
structural parameters.

\subsection{Galaxy Photometric Parameters}
\label{sec:galaxy_parameters}

We retrieved $g$ and $i$-band images from the seventh data release
\citep[DR7,][]{Abazajian2009} of the Sloan Digital Sky Survey
\citep[SDSS,][]{York2000} for as many Sample A and B galaxies as
possible.  The total SDSS sample includes 90 galaxies, 62 from Sample
A and 28 from Sample B, from which to derive structural parameters.

The SDSS images are already bias subtracted, flat-fielded and cleaned
from bright stars; however, we performed our own sky subtraction since
the SDSS pipeline sky levels may be flawed for extended galaxies
\citep{Bernardi2007b, Lauer2007b}.  This issue has been addressed in
the SDSS-III Data Release 8 \citep{Aihara2011,Blanton2011}.
For our purposes, we estimated the sky level of each galaxy image by
isolating five regions away from the galaxy, free of any contaminant,
and calculating the mode of the sky intensities per pixel within each
sky region.  The average and standard deviation of the five sky values
was then computed.
The difference between our measured sky values and those provided by
SDSS can be as large as $4\%$. The SDSS sky level is always biased
high, likely due to the inclusion of bright, extended sources while
our interactive technique ensures a contaminant-free selection of the
sky fields.
We find that the typical surface brightness error in the $g$ and $i$
bands is 0.1 mag arcsec$^{-2}$ at $\mu_g \simeq 26$ mag arcsec$^{-2}$
and $\mu_i \simeq 25$ mag arcsec$^{-2}$, respectively \citep[see][for
more details]{McDonald2011}.
The large angular extent of NGC~224 and NGC~4594 relative
to the field of view thwarted their proper sky subtraction;
these two galaxies were therefore excluded from our SDSS sample.
NGC~221 was also discarded due to improper positioning
of the image on the detector. 
The remaining images were flux-calibrated based on the SDSS
photometric zero-point, corrected for Galactic extinction
\citep{Schlegel1998} as well as for internal extinction and
$K$-correction following \citet{Shao2007}.

The galaxy surface brightness profiles were extracted using the
isophotal fitting methods outlined in \citet{Courteau1996} and
\citet{McDonald2011}. These make use of the astronomical data
reduction package {\sc XVISTA}\footnote{See {\tt
    http://astronomy.nmsu.edu/holtz/xvista/index.html}.}.  The
azimuthally-averaged surface brightness profiles, projected onto the
major axis of each galaxy, are shown in \Fig{profiles} for four sample
galaxies.  These are the elliptical galaxy NGC~5127, two high-
(NGC~4036) and low- (NGC~4477) inclination lenticular galaxies, and
the spiral galaxy NGC~3675 which boasts a remarkably high dust
content.

\placefigtwo

The $g$ and $i$-band total magnitude of each galaxy is then determined
by summing the flux at each isophote and extrapolating the light
profile to infinity.  The $g-i$ colour of each galaxy was calculated
from the difference of the fully corrected $g$ and $i$-band
magnitudes.  The remaining structural parameters were measured from
the $i$-band light profiles since, of all the SDSS band passes, the
$i$-band suffers least dust extinction. We extracted the isophotal
radius, $r_{24.5}$, corresponding to the surface brightness of 24.5
\mas, the half-light (or effective) radius of the galaxy, \reg, the
effective surface brightness of the galaxy, $\mu_{\rm e, gal}$, and
the galaxy concentration $C_{28}=5\log(r_{80}/r_{20})$, where $r_{20}$
and $r_{80}$ are the radii which enclose $20\%$ and $80\%$ of the
total luminosity, respectively.
These $i$-band structural parameters are listed in
Table~\ref{tab:MBH_photometry_i}.  The $g$-band magnitude is also
listed.  Based on simulated models of spiral galaxies
\citep{MacArthur2003}, the typical error per galaxy structural
parameter is roughly 10-20\%.

\subsection{Photometric Parameters of Bulges and Discs}
\label{sec:bulge_parameters}

The structural parameters for elliptical galaxies, modelled typically
as a single spheroid, and for spiral galaxies, modelled as the sum of
a spheroid and a disc, were derived by applying the two-dimensional
photometric decomposition algorithm {\sc GASP2D} \citep{MendezAbreu2008} to
the SDSS $i-$band images.

The surface brightness of the spheroid component (typically the entire
elliptical galaxy or the bulge component for a disc galaxy) is modelled
using a \sersic function (\citealt{Sersic1968}; see also
\citealt{Graham2005})

\begin{equation}
\label{eq:sersic}
 I_{\rm b}(r)=I_{\rm e} 10^{-b_n\{ \left(r/r_{\rm e}\right)^{1/n}-1 \}},
\end{equation}
\noindent
where $r_{\rm e}$ is the effective radius, $I_{\rm e}$ is the surface
brightness at $r_{\rm e}$ , and $n$ is a shape parameter that
describes the curvature of the radial profile. With $n=1$ or $n=4$,
the \sersic function reduces to the exponential or de~Vaucouleurs
function, respectively.  The coefficient $b_n = 0.868\,n - 0.142$
\citep{Caon1993} is a normalisation term. The spheroid model
elliptical isophotes have constant position angle PA$_{\rm b}$ and
constant axial ratio $q_{\rm b}$.

The surface brightness distribution of the disc component is assumed
to follow an exponential law \citep{Freeman1970}

\begin{equation}
\label{eq:exp}
I_{\rm d}(r)=I_0\,{\rm e}^{- r/h},
\end{equation}
\noindent
where $h$ and $I_0$ are the scale length and central surface
brightness of the disc, respectively. The disc model elliptical
isophotes have constant position angle PA$_{\rm d}$ and constant axial
ratio $q_{\rm d}$.
The fitting algorithm {\sc GASP2D} relies on a $\chi^2$ minimisation of the
intensities in counts, for which we must adopt initial trial
parameters that are as close as possible to their final values.  The
latter were estimated from our ellipse-averaged light profiles from
which basic fits to the bulge and disc were applied to estimate
structural parameters \citep[see][for details]{MendezAbreu2008}.
The $\chi^2$ minimisation is based on the robust Levenberg-Marquardt
method by \citet{More1980}.  The actual computation has been done
using the {\sc MPFIT} algorithm \citep{Markwardt2009} under the {\sc
  IDL}\footnote{Interactive Data Language is distributed by ITT Visual
  Information Solutions. It is available from http://www.ittvis.com/.}
environment.

The {\sc GASP2D} software yields structural parameters for the bulge
($I_{\rm e}$, $r_{\rm e}$, $n$, PA$_{\rm b}$, and $q_{\rm b}$) and
disc ($I_0$, $h$, PA$_{\rm d}$, and $q_{\rm d}$) and the position of
the galaxy centre $(x_0, y_0)$. In {\sc GASP2D}, each image pixel intensity
is weighted according to the variance of its total observed photon
counts due to the contribution of both galaxy and sky, and accounting
for photon and detector read-out noise.
Seeing effects were taken into account by convolving the model image
with a circular Moffat point spread function \citep[][hereafter
  PSF]{Moffat1969} with shape parameters measured from the stars in
the galaxy image.
Only the image pixels with an intensity larger than 1.5 times the sky
standard deviation were included in the fit. Foreground stars were
masked and excluded from the fit.
The initial guesses were adopted to initialise the non-linear
least-squares fit to galaxy image, where the parameters were all
allowed to vary. A model of the galaxy surface brightness distribution
was built using the fitted parameters. It was convolved with the
adopted circular two-dimensional Moffat PSF and subtracted from the
observed image to obtain a residual image.  In order to confirm the
minimum in the $\chi^2$-space found in this first pass, two more
iterations were performed. In these iterations, all the pixels and/or
regions of the residual image with values greater or less than a fixed
threshold, controlled by the user, were rejected. Those regions were
masked out and the fit was repeated assuming, as initial trials for
the free parameters, the values obtained in the previous
iteration. These masks are useful when galaxies have spiral arms and
dust lanes, which can affect the fitted parameters. We found that our
algorithm converges after three iterations.

The model decompositions for few elliptical galaxies (NGC~4473,
NGC~4636, NGC~4649) were significantly improved by including a disc
component, as found in a few other elliptical galaxies
\citep{Kormendy2009a, McDonald2009}.  For the other ellipticals, the
values of $r_{\rm e}$ obtained either directly (empirically) from the
light profile (in \se{galaxy_parameters}) or by using a \sersic
fitting function (from the two-dimensional photometric decomposition)
agree within their respective errors.

We eliminated 33 galaxies from our sample because of poor
decompositions due to either a strong central bar, a Freeman II
profile \citep{Freeman1970}, or just the overall inadequacy of our
single or double-component modelling (e.g., due to the presence of
strong dust lanes and/or spiral arms).  We successfully performed
photometric decompositions, as judged by a global $\chi^2$
figure-of-merit, for the 57 galaxies ( 38 from Sample A, 19 from
Sample B) listed in Table~\ref{tab:gasp2d_phot}.  The latter includes
the resulting bulge and disc structural parameters.  Some examples
illustrating the various fitting strategies are shown in
\Fig{profiles_gasp2d}.
These are the same galaxies, whose azimuthally-averaged surface
brightness profiles are shown in \Fig{profiles}. For all of them, the
ellipse-averaged radial profiles of surface brightness, ellipticity,
and position angle of the model image are consistent with those
measured on the galaxy image. The differences between model and data
found in the ellipticity and position angle of NGC~4477 and in the
ellipticity of NGC~3675 are due to the presence of a weak bar and
strong dust lanes, respectively. These features are clearly seen in
the galaxies' residual images. Nevertheless, the surface brightness
residuals $\Delta\mu_i$ are remarkably small; $|\Delta\mu_i|<0.2$ mag
arcsec$^{-2}$ for all the galaxies shown in \Fig{profiles_gasp2d},
except for some portions of the dust lanes of NGC~3675 where the
residuals increase to $\sim0.4$ mag arcsec$^{-2}$.  Although part of
the NGC~3675 disc is missing, this does not affect the fit
result. {\sc GASP2D} performs a reliable fit as soon as the observed galaxy
can be modelled by the sum of two axisymmetric components and the
field of view covers at least half of the galaxy
\citep{MendezAbreu2008c}.

\placefigthree

The {\sc GASP2D} formal errors obtained from the $\chi^2$ minimisation
method are not representative of the real errors in the structural
parameters \citep{MendezAbreu2008}. Instead, the estimated errors
given in Table~\ref{tab:gasp2d_phot} were obtained through a series of
Monte Carlo simulations.  To this end, we generated a set of 400
images of elliptical galaxies with a \sersic spheroid and 400 images
of disc galaxies with a \sersic bulge and an exponential disc; each
with a different PA and ellipticity.
The range of tested parameters for the simulated images was taken from
the photometric analysis of nearby elliptical galaxies by
\citet{Kormendy2009a} and face-on disc galaxies by
\citet{Gadotti2009a} (but assuming a wider range of axial ratios than
the latter).  Our tests include treatment for resolution effects
(galaxy distance, pixel scale, and seeing), colour effects (accounting
for $V$ and $i$ bands), inclination, and more.  
The two-dimensional parametric decomposition was applied to analyse
the images of the artificial galaxies as if they were real.  The
artificial and observed galaxies were divided in bins of 1 mag. The relative errors on the fitted parameters of the
artificial galaxies were estimated by comparing the input and output
values and were assumed to be normally distributed. In each magnitude
bin, the mean and standard deviation of relative errors of artificial
galaxies were adopted as the systematic and typical error on the
relevant parameter for the observed galaxies.  Overall, we find that
{\sc GASP2D} recovers the galaxy structural parameters with an uncertainty
ranging from 1\%, to 10\%, and to 20\% for brighter ($6.5\leq m_{{\rm
    tot},i} < 7.5$), intermediate ($10.5\leq m_{{\rm tot},i} < 11.5$),
and fainter ($12.5\leq m_{{\rm tot},i} < 13.5$) sample galaxies,
respectively.

The inclination of disc galaxies was calculated from the fitted disc
axial ratio in the $i$-band as

\begin{equation}
\sin^2 i = \frac{1 - q_{\rm d}^2}{1 - q_0^2},
\label{eqn:inclination}
\end{equation}

\noindent
where $q_0$ is the intrinsic disc axial ratio. We obtain the latter
from \citet{Paturel1997}:

\begin{equation}
\log q_0 = -0.43 - 0.053\;T ,
\label{eqn:intrinsic}
\end{equation}
where $T$ is the galaxy Hubble type from RC3.

The fitted $i$-band disc axial ratios agree well with the isophotal
axial ratios at a surface brightness level $\mu_B = 25$ \mas\ reported
in the RC3.  Our disc axial ratios are on average $6\%$ lower than RC3
with a standard deviation of $7\%$.  We adopted the RC3 axial ratios
for the lenticular and spiral galaxies whose {\sc GASP2D} solution for the
disc was too uncertain; this amounts to 28 galaxies in Sample A and
another 8 in Sample B.

\subsection{Stellar Velocity Dispersion}
\label{sec:sigmas}

The measured stellar velocity dispersions, \sigmas, for galaxies in
Sample A were taken from the same sources as
\citetalias{Beifiori2009}.  We applied the aperture correction of
\citet{Jorgensen1995} to transform the \sigmas\ into the equivalent of
an effective stellar dispersions, \sigmae, measured within a circular
aperture of radius \reb.  The effective radii were also taken from the
same sources as \citetalias{Beifiori2009}, except for the 35 galaxies
for which \reb\ was obtained from our own decomposition of the SDSS
images (Tables \ref{tab:MBH_photometry_i} and \ref{tab:gasp2d_phot}).
The maximum difference between our and literature values of \reb\ is
about $20\%$, though the comparison often involves different band
passes which broadens the discrepancy.

The aperture correction was also applied to the \sigmas\ measured for
NGC~2892 and NGC~5921 by \citet{Ho2009a} and \citet{Wegner2003},
respectively.

For Sample B galaxies, we adopted the values of \sigmae\ given by
\citetalias{Gultekin2009}.  These were derived as the
luminosity-weighted mean of the stellar velocity dispersion within
\reb .

\subsection{Circular Velocity}
\label{sec:vc}

The values for the galaxy circular velocity, \vc, for both elliptical
and disc galaxies were collected from different sources.  

We retrieved \vc\ from the compilation of \citet{Ho2007} for 40 disc
galaxies (31 from Sample A, 9 from Sample B).  These were derived
from the \hire\ line widths available in the HyperLeda catalogue
\citep{Paturel2003a}.  The $W_{20}$ and $W_{50}$ line widths are
measured at $20\%$ and the $50\%$ of the total \hire\ line profile
flux.
For galaxies missing in \citet{Ho2007}, line widths were found in
HyperLeda for an additional 27 galaxies (22 from the Sample A, 5 from
Sample B).
Given multiple sources in HyperLeda, we favoured the
larger survey source for each galaxy in order maximise the homogeneity
of the data base\footnote{We selected the most recent and highest
  resolution observations available in the on-line HyperLeda catalogue
  ({\tt http://leda.univ-lyon1.fr}) up to September 2009.}.
All line widths were already corrected for instrumental resolution.
We further applied a correction for cosmological stretching and
broadening by gas turbulence following \citet{Bottinelli1983} and
\citet{Verheijen2001}.
Finally, the corrected line widths $W_{\rm 20, corr}$ and $W_{\rm 50,
  corr}$ were deprojected using the prescription of
\citet{Paturel1997}
\begin{eqnarray}
V_{20} & = & 0.5 (10^{1.187\log W_{\rm 20, corr}-0.543})/\sin i , \\
V_{50} & = & 0.5 (10^{1.071\log W_{\rm 50, corr}-0.210})/\sin i,
\label{eqn:corr_turb2}
\end{eqnarray}
\noindent
where the inclination, $i$, is listed in Table~\ref{tab:allMBH_vc}
and \ref{tab:MBH_gultekin}.  We take the final circular velocity as
the average of $V_{20}$ and $V_{50}$,
\begin{equation}
V_{\rm c} = (V_{20}+V_{50})/2.
\label{eqn:corr_turb3}
\end{equation}
Following \citet{Ho2007}, we adopted a 5\% error on \vc\ for the
maximum velocity listed in Hyperleda.

For 14 early-type galaxies (7 in Sample A, 7 in Sample B), we adopted
the value of \vc\ derived from dynamical modelling (IC~1459,
\citealt{Samurovic2005}; NGC~1052, \citealt{Binney1990}; NGC~3115,
\citealt{Bender1994}; NGC~3608, \citealt{Coccato2009}; NGC~4314
\citealt{Quillen1994}; and 9 ellipticals in
\citealt{Kronawitter2000}).
For a few other galaxies (five in Sample A, one in Sample B), either
the ionised-gas (NGC~2911, \citealt{Silchenko2004}; NGC~5252,
\citealt{Morse1998}), the \hire\ (IC~342, \citealt{Pizzella2005};
NGC~3801, \citealt{Hota2009}) or the CO kinematics (NGC~4526,
\citealt{Young2008}), Milky Way (\citealt{Baes2003}) rotational
velocity at large radii are assumed to represent \vc. We adopted a
10\% error when the \vc\ uncertainty from models or observations was
not given.

Finally, we eliminated 7 galaxies (4 from Sample A, 3 from Sample B)
since their quoted circular velocities are unrealistically low
(NGC~1497, NGC~3384, and NGC~5576, \citealt{Paturel2003a}; NGC~3642,
NGC~4429, NGC~5347, and NGC~7052, \citealt{Ho2007}).

The final circular velocities of 88 galaxies from Sample A (65) and
Sample B (23) are listed in Tables~\ref{tab:allMBH_vc} and
\ref{tab:MBH_gultekin}.

\subsection{Masses}
\label{sec:masses}

We estimated the bulge mass as \Mbuvir $=\alpha\; r_{\rm e, bulge}
\sigma_{\rm e}^2/G$, where $G$ is the gravitational constant and
$\alpha=5.0\pm0.1$ \citep{Cappellari2006} is a dimensionless constant
that depends on galaxy structure. The effective radius, $r_{\rm e,
  bulge}$, for elliptical galaxies is extracted from their
azimuthally-averaged light profile (Table~\ref{tab:MBH_photometry_i}),
whereas the bulge effective radius of lenticular and spiral galaxies
is obtained from two-dimensional photometric decomposition
(Table~\ref{tab:gasp2d_phot}).
In deriving \Mbuvir, we implicitly assumed that the measured value of
\sigmae\ is dominated by the bulge component. For late-type galaxies
we expect that the disc contribution to \sigmae\ results in an
increase of the scatter of \Mbuvir\ as a function of the morphological
type. This does not affect our analysis since we could measure $r_{\rm
  e, bulge}$ (and therefore \Mbuvir ) for only a few Sb--Sbc galaxies.
Moreover, the ratio $B/T>0.1$ except for three galaxies.

\citet{Ferrarese2006} suggested a connection between \mbh\
and the mass of early-type galaxies calculated as
\begin{equation}
M_{\rm e, gal}\,=\,\alpha\,\ r_{\rm e, gal}\sigma_{\rm e}^2 /G\ {\rm with}\
\alpha = 5.0\pm0.1.
\label{eq:megal}
\end{equation}
Note that such a mass estimate is indicative of the galaxy mass
within $r_{\rm e, gal}$ and is thus an incomplete representation
of the total galaxy mass.  We calculated \megal\ for the elliptical
and lenticular galaxies in our sample by adopting $r_{\rm e, gal}$
calculated from the azimuthally-averaged light profiles
(Table~\ref{tab:MBH_photometry_i}).  For elliptical galaxies,
\Mbuvir$\,=\,$\megal .
The virial estimator by \citet{Cappellari2006} captures the entire
dynamical mass so long as total mass traces light
\citep{Thomas2011}. \citet{Wolf2010} found a different coefficient
($\alpha=4$) for their derivation of a mass estimator for stellar
systems supported by velocity dispersion.  The latter is a good
estimate of the total dynamical mass inside a radius which is just
larger than the effective radius \citep{Thomas2011}.  The appropriate
value of $\alpha$ is actually a function of the \sersic shape index
\n\ \citep{Trujillo2004, Cappellari2006}.  However the exact
application of the $\alpha(n)$ relation, which results in a zero-point
offset for \megal, does not affect our conclusions.  We thus make use
of the \citet{Cappellari2006} mass estimator.  Error estimates on
\Mbuvir\ and \megal\ account for the uncertainty on $\alpha$.

We derived the dynamical mass of the disc galaxies from
\begin{equation}
M_{\rm dyn}\,=\,R\,V_{\rm c}^2 /G,
\label{eq:mdyn}
\end{equation}
where $R = r_{24.5}$ (Table~\ref{tab:MBH_photometry_i}).
Most of the circular velocities that we derived from \hire\ data yield
no information about their radial coverage. Nevertheless, \mdyn\
corresponds to the galaxy mass within the optical radius, because
$r_{24.5}$ is roughly indicative of the galaxy optical radius and the
observations of spatially-resolved \hire\ kinematics in spirals show
that the size of \hire\ discs closely matches that of a galaxy's
optical disc \citep{Ho2008a}.

The masses computed from \Eq{megal} and \Eq{mdyn} are both available
only for the 10 lenticular galaxies in our sample, since they share
dynamical properties of both elliptical and spiral galaxies.  For
these galaxies, $\langle$\mdyn/\megal$\rangle=1.27\pm1.07$.

The galaxy stellar masses, \mstar, were derived from \li\ under the
assumption of constant mass-to-light ratio $(M/L)_i$.  $(M/L)_i$
estimates were inferred from our $g-i$ colours following
\citet{Bell2003}.  Their mass-to-light ratios were derived from the
stellar population models of \citet{Bruzual2003}, which are tuned to
reproduce the ages and metallicities of the local spiral galaxies.

\section{Analysis}
\label{sec:analysis}

\subsection{Correlations Between \mbh\ and Bulge and Galaxy
  Parameters}
\label{sec:corr_BH}

We used the data obtained in \se{gal_properties} to build scaling
relations between \mbh\ and the bulge (i.e., the velocity dispersion,
luminosity, virial mass, \sersic index, and mean effective surface
brightness) and galaxy (luminosity, circular velocity, stellar,
virial, and dynamical mass) properties.  In addition to the Sample
A$+$B, we also built a comparison sample with the 30 galaxies
for which all the desired physical parameters could be measured.  For
each of the above parameters $x$, we assume that there exists a
relation of the form
\be
\log \frac{M_{\bullet}}{{\rm M}_{\odot}}\,=\, \alpha\, +
  \,\beta\,\log{\frac{x}{x_0}}
\ee
where $x_0$ is a normalisation value chosen near the mean of the
distribution of $x-$values.

The best-fit values of $\alpha$ and $\beta$, their associated
uncertainties, the total scatter $\epsilon$ and intrinsic scatter
$\epsilon_{\rm intr}$ of the relation, the Spearman rank correlation
coefficient $\rho$, and the level of significance of the correlation
$P$ were obtained as a function of the sample at hand.  The results
are given in Table~\ref{tab:fit_single}.  The quoted errors are all
derived through a bootstrap technique.

Samples A$+$B and the comparison sample include both accurate
determinations and upper limits of \mbh.  The proper handling of these
heterogeneous data sets requires that we perform a censored regression
analysis \citep{Feigelson1985, Isobe1986} with the {\sc
  ASURV}\footnote{The {\sc FORTRAN} source code of the Astronomy
  Survival Analysis Package (v.1.3) is available at
  http://www2.astro.psu.edu/statcodes/asurv.}  package
\citep{Lavalley1992}.
Linear regressions were calculated with the EM maximum-likelihood
algorithm implementing the technique of expectation (E-step) and
maximization (M-step) by \citet{Dempster1977} which assumes normal
residuals. For each relation, we checked that the distribution of
residuals about the fitted line was normal. With uncensored data, this
analysis would be equivalent to that of a standard least-squares
linear regression.
{\sc ASURV} gives as correlation coefficient the generalised Spearman
rank correlation coefficient $\rho$ \citep{Akritas1989} and computes
the level of significance of the correlation $P$.
We derived the scatter of the relation $\epsilon$ as the root-mean
square (rms) deviation in $\log (M_{\bullet}/{\rm M}_{\odot})$ from
the fitted relation assuming no measurement errors. This assumption
does not affect our results, since we wish to compare the relative
tightness of the different scaling relations for the comparison
sample.

\citetalias{Gultekin2009} focused their analysis on \mse\ and
\mlbui. We have extended their analysis to a broader range of scaling
relations by performing least-square linear regressions with the
secure values of \mbh\ and the bulge and galaxy parameters measured
for Sample B. We used the {\sc MPFITEXY}\footnote{The {\sc IDL} source
  code of {\sc MPFITEXY} is available at
  http://purl.org/mike/mpfitexy/.} algorithm \citep{Williams2010},
which accounts for measurement errors in both variables and the
intrinsic scatter $\epsilon_{\rm intr}$ of the relation.
We used the {\sc IDL} routine {\sc R\_CORRELATE} to compute the
standard Spearman rank correlation coefficient $\rho$ and the level of
significance of the correlation.
The intrinsic scatter of each relation was assessed by varying
$\epsilon_{\rm intr}$ in the fitting process to ensure that the
reduced $\chi_\nu^2=1$. The total scatter $\epsilon$ was derived as
the rms deviation in $\log (M_{\bullet}/{\rm M}_{\odot})$ from the
fitted relation weighted by measurement errors.

In addition to fitting the \mbh--bulge and galaxy scaling relations as
outlined above, we also look in the subsections below for correlations
against morphological type or nuclear activity as listed in
Tables~\ref{tab:allMBH_vc} and \ref{tab:MBH_gultekin}.  We will see
that the latter (morphology or nuclear activity) do not play a strong
r\^ole in any of the \mbh--bulge and galaxy scaling relations.
Morphology only plays a small r\^ole in the \mmstar\ and in the \mv\
relations.
  
\subsubsection{\mbh\ versus \sigmae}
\label{sec:MBH_sigma}

We have plotted the \mse\ distribution for the Sample A$+$B in
\Fig{m_s}. The slope of our \mse\ relation is consistent within the
errors with those by \citet{Ferrarese2005} and \citet{Lauer2007b} as
discussed by \citetalias{Beifiori2009}, and with that by
\citetalias{Gultekin2009} by default.  On the other hand, the
inclusion of the upper limits measured by \citetalias{Beifiori2009}
and \citetalias{Gultekin2009} slightly changes the zero point of the
relation but does not appreciably affect its scatter.  The \mse\
relation is indeed the tightest correlation that we measure; its
scatter is consistent with \citetalias{Gultekin2009}
($\epsilon=0.44\pm0.06$ dex) and slightly larger than
\citet[][$\epsilon=0.34$ dex]{Ferrarese2005}.
According to \citetalias{Gultekin2009}, the increased fraction of
spirals may be a source of the increased scatter with respect to
previous estimates based mostly on SMBH measurements in early-type
galaxies.  Consistently, our analysis reveals that the \mse\ scatter
for late-type galaxies ($\epsilon=0.52\pm0.14$ dex) alone is slightly
larger than that of early-type galaxies ($\epsilon=0.38\pm0.04$ dex).
Nevertheless, at small \sigmae, some upper limits exceed the expected
\mbh\ as the line-widths for such low-\sigmae\ outliers are most
likely affected by the mass contribution of a conspicuous nuclear
cluster (\citetalias{Beifiori2009}). The scatter for the late-type
galaxies reduces to ($\epsilon= 0.38\pm0.05$ dex) by excluding the
nucleated galaxies from the fit. Therefore, they are the cause for the
observed difference between the scatter of early and late-type
galaxies. We conclude that the \mse\ relation is the same for both
early and late-type galaxies. S0 galaxies overlap ellipticals at high
\sigmae\ values and spirals at low \sigmae\ values.

Barred and unbarred galaxies follow the same \mse\ relation within the
errors. The same is not true for galaxies with classical and
pseudo-bulge. We have classified our bulges according to their
measured \sersic index \citep[$n>2$ for classical and $n\leq2$ for
pseudo-bulges;][]{Kormendy2004}. We find slope differences in the
\mse\ relations of our 46 classical ($\beta=3.86\pm 0.71$) and 11
pseudo-bulges ($\beta=5.63\pm 0.86$), whereas their zero points
($\alpha=8.09\pm 0.09$ and $\alpha=7.78\pm 0.33$, respectively) are in
closer agreement within the errors.  Conversely, \citet{Hu2008} found
similar slopes and different zero points. At face values, our and Hu
et al.'s correlation coefficients are in agreement. This is mostly due
to the large error bars on the \mse\ coefficients for pseudo-bulges,
which translate into a lower level of significance of the \mse\
relation for pseudo-bulges ($P<2\%$) with respect to that for
classical bulges ($P<0.1\%$).  This results supports the recent
findings by \citet{Kormendy2011a} that there is little or no
correlation between \mbh\ and pseudo-bulges.

\placefigfour

At high \sigmae\ values, the \mbh\ of the Sample A$+$B shows a weak
steepening with respect to \mse\ confirming previous results by
\citet{Wyithe2006}, \citet{Lauer2007b}, and \citet{DallaBonta2009}.

Finally, \Fig{m_s} shows the absence of trends of the \mse\
relation with nuclear activity. The different types of nuclear
activity cover all the \sigmae\ range, with most LINERs at high
\sigmae\ and most Seyferts at low \sigmae.  This is expected
considering the strong dependence of nuclear spectral class on Hubble
type (and therefore on \sigmae), with LINERs and Seyfert nuclei being
more frequent in ellipticals and spirals, respectively \citep[see][for
a review]{Ho2008}.

\subsubsection{\mbh\ versus \Lbui}
\label{sec:MBH_Lbulge}

The correlation between \mbh\ and the luminosity of the spheroidal
component of a galaxy is also fairly tight (see \citealt{Graham2007b},
and references therein).
\citet{Graham2007b} compared the different versions of the \mlbu\ by
\citet{Kormendy2001a}, \citet{McLure2002}, \citet{Marconi2003a}, and
\citet{Erwin2004}.  {Recently, \citet{Sani2011} derived the
  \mlbu\ relation by measuring the 3.6 $\mu$m bulge luminosity.} Since
the differing relations do not predict the same SMBH mass,
\citet{Graham2007b} investigated the effects of possible biases on the
\mlbu\ relation including any dependency on the Hubble constant, the
correction for dust attenuation in the bulges of disc galaxies, the
mis-classification of lenticular galaxies as elliptical galaxies and
the adopted regression analysis. These adjustments resulted in
relations which are consistent with each other and suitable for
predicting similar \mbh. In particular, a detailed photometric
decomposition of the galaxy light profile is crucial to obtaining a
representative bulge luminosity and therefore a reliable \mlbu.

\citetalias{Gultekin2009} derived an updated value of the intrinsic
scatter of the \mlbu\ relation, which is comparable to that of the
\mse\ relation in early-type galaxies, confirming the results of
\citet{Marconi2003a}. According to them, the scatter of the \mlbu\
relation is significantly reduced when the bulge effective radius is
extracted from a careful two-dimensional image decomposition.

\Fig{ml_bulge} shows our \mlbui\ relation. The best-fit coefficients
are in agreement within the errors with \citet{Graham2007b} and
\citet{Sani2011}. The scatter is slightly higher than
\citetalias{Gultekin2009} who fit only early-type galaxies, but larger
than that of the \mse\ relation.  All galaxies show a similar \mlbui\
distribution with no dependence on morphological type or nuclear
activity.
We cannot test the claim that barred and unbarred galaxies follow
different \mlbui\ relations \citep{Graham2008, Graham2009,
  Gadotti2009b, Hu2009}, since the sample galaxies with a strong bar
were excluded for photometric decomposition and thus lack a \Lbui\
measurement.  For all the remaining galaxies, we did not model any
other components than the bulge and disc (see
Table~\ref{tab:gasp2d_phot}).

\placefigfive

\subsubsection{\mbh\ versus \Mbuvir}\label{sec:MBH_Massbulge}

The connection between the \mbh\ and \Lbu\ suggests a linear
correlation with mass of the spheroidal component of the galaxy
\citep{Magorrian1998, McLure2002, Marconi2003a, Haering2004}.

The slope of our \mmbuvir\ relation (Fig.~\ref{fig:mmass_bulge}) is
consistent within the errors with \citet{McLure2002},
\citet{Marconi2003a}, \citet{Aller2007} and \citet{Sani2011} and
slightly shallower than \citet{Haering2004}.
This regression is not a marked improvement over the \mse\ relation,
as one might expect considering the additional fitting parameter in
the mass measurement (namely the radius).  Indeed, the \mmbuvir\ is
worse than the \mse\ relation.  However, \Mbuvir\ is still a better
proxy for \mbh\ than \Lbui .
Different Hubble types follow the same \mmbuvir\ relation, with the
lenticular galaxies covering the whole range of masses. There is also
no dependence of this relation on nuclear activity, in agreement with
\citet{McLure2002}.

\placefigsix

\subsubsection{\mbh\ versus \sersic {\it n} and \muemean}
\label{sec:MBH_phot}

The two-dimensional photometric decompositions yield a panoply of
galaxy structural parameters, including the \sersic shape index \n\
and mean effective surface brightness \muemean, both used as measure
of the concentration of the bulge light.  The \sersic index \n\ has
indeed been adopted by some as a good tracer for \mbh\
\citep{Graham2001, Graham2003a, Graham2007}.

Our \mbh$-$\n\ relation (\Fig{mbh_n}) differs significantly from the
linear and quadratic relations with $\log$~\mbh\ and $\log$~$n$ by
\citet{Graham2001, Graham2003a} and \citet{Graham2007},
respectively. In particular, the values of $n$ for galaxies with high
\mbh\ are not as large as those in \citet{Graham2007}.  The relation
is characterised by a large scatter, small Spearman correlation
coefficient, and low level of significance. Therefore, the correlation
between \mbh\ and $n$ is poor and the \mbh$-$\n\ relation is not
reliable for predicting \mbh . Our findings are in agreement with
\citet{Hopkins2007b} who found no correlation between \mbh\ and
\sersic index with both observations and hydrodynamical simulations.

The correlation between \mbh\ and \muemean\ (\Fig{mbh_mue}) is also
poor, lending further support to the idea that \mbh\ is unrelated to
the light concentration of the bulge, regardless of galaxy morphology
and nuclear activity.

\placefigseven
\placefigeight

\subsubsection{\mbh\ versus \li}
\label{sec:MBH_Litot}

\citet{Kormendy2001b} compared the $B$-band total magnitude with
dynamically secure \mbh\ estimates in spheroids and few bulgeless disc
galaxies. The poor correlation between \mbh\ and \li\ led him to
conclude that the evolution of SMBHs is linked to bulges rather than
discs.

We compared our own \li\ with \mbh\ as a function of the morphological
type in \Fig{mltot}. The Spearman rank correlation coefficient
suggests a correlation at 17\% significance level, which is tighter
than \citet{Kormendy2001b} but not as tight as that for the \mlbu\
relation.
Our sample is short of bulgeless galaxies which may explain the better
agreement for our \mbh\ and \li\ with respect to
\citet{Kormendy2001b}.
On the other hand, total and bulge luminosity differ for disc
galaxies.  The later the morphological type, the larger the
discrepancy resulting in both a larger slope and scatter of the \mli\
relation with respect to \mlbu . The scatter is larger and the
correlation weaker when the $g-$band total luminosity is considered.
Our findings are consistent with \citet[][see their Fig.~6]{Hu2009}
and confirm that bulge luminosity is a better tracer of the mass of
SMBHs than total light of the galaxy.

\placefignine

\subsubsection{\mbh\ versus \mstar , \megal , and \mdyn}
\label{sec:MBH_Mgalaxy}

We have tested whether the scatter in predicting \mbh\ could be
reduced by adopting \mstar\ rather than \li, since the galaxy
luminosity is a proxy for its stellar mass.  The \mmstar\ relation
shown in \Fig{m_stellar_mass} is a slight improvement over the \mli\
relation.  The slopes are comparable but the Spearman rank correlation
coefficient of the former is higher with a 3\% significance level.
Ellipticals and lenticulars follow a tighter \mmstar\ relation than
late-type galaxies; the disc light clearly plays an anti-correlating
r\^ole, indicating once more that the bulge parameters drive SMBH
correlations.

\placefigten

We adopted the mass estimator from \citet{Cappellari2006} (i.e., the
mass within the effective radius) for our ellipticals and lenticulars.
The distribution of \mbh\ as a function of \megal\ for the early-type
galaxies is plotted in \Fig{mbh_vir}. They follow the same trend as
the \mmegal\ relation by \citet{Ferrarese2006}.  This is especially
true for \megal$>10^{11}$ \msun\ where the fit slope and normalisation
are consistent within the errors with
\citet{Ferrarese2006}. Less-massive galaxies seem to follow a steeper
\mmegal\ relation with a smaller normalisation.  More data are however
needed in this mass range to carefully address any trend differences.
Introducing a new parameter \reg\ and a specific exponent for \sigmae\
ought to considerably reduce the scatter of the \mse\ relation but, in
fact, it does not.  In particular, the scatter of the \mse\ relation
is smaller than that of \mmegal\ if the same sample of early-type
galaxies is considered.

The dynamical mass \mdyn\ (i.e., the mass within the optical radius)
was taken as the galaxy mass estimator for lenticulars and spirals. It
does not correlate with \mbh\ (\Fig{MBH_Mdyn}), irrespective of Hubble
type and nuclear activity.
Our analysis reveals that the coarse \mbh$-$\mdyn\ relation found by
\citet[][$\epsilon=0.61$ dex]{Ho2008b} does not hold when galaxies
with \mdyn$\leq10^{10}$ \msun\ (i.e., at the lower mass end of the
\mdyn\ range) are taken into account.  The scatter of the \mbh--galaxy
mass relation is larger and the correlation weaker when \mstar\ and
\mdyn\ are considered.  Since the stellar and dynamical masses include
the mass contribution of the disc component, we conclude that the
bulge mass is a better tracer of \mbh . Nevertheless, the \mse\ is
tighter and yet again more fundamental than the \mmegal\ relation.

\placefigeleven
\placefigtwelve

\subsubsection{\mv\ and \vs\ relations}
\label{sec:MBH_vs}

Assuming that \vc\ at large radii trace the dark matter halo, we study
the possible link between SMBHs and dark matter halos by comparing
\mbh\ with \vc.  A possible relation between \mbh\ and the dark matter
depends clearly on the radius at which \vc\ is measured. Theoretical
models that reproduce the observed luminosity function of quasars
\citep{Cattaneo2001, Adams2003, Volonteri2003, Hopkins2005a,
  Springel2005} suggest that \mbh\ scales as a power law of the virial
velocity (i.e., the circular velocity of the galactic halo at the
virial radius) of the galactic halo of the SMBH host.  However, the
conversion between \vc\ measured at large radii and virial velocity,
depends on the assumed model.

We show \mbh\ against \vc\ in \Fig{m_v}. There is a weak correlation
between these quantities, which is mostly driven by the late-type
galaxies.

\placefigthirteen

This is in agreement with previous studies by \citet{Zasov2005} and
\citet{Ho2008b}. \citet{Zasov2005} used a collection of 41 galaxies
with \mbh , \sigmae , and \vc\ from the literature to conclude that
there is a coarse \mbh$-$\vc\ relation and that, for a given \vc,
early-type galaxies have larger \mbh\ than late-type galaxies.
\citet{Ho2008b} studied the \mv\ relation for a sample of 154 nearby
galaxies comprising both early and late-type systems for which \mbh\
was estimated from the mass-luminosity-line width relation
\citep{Kaspi2000, Greene2005, Peterson2006} and \vc\ from \hire\ and
optical data \citep{Ho2008a}.  They found that the correlation between
\mbh\ and \vc\ improves if only the galaxies with the most reliable
\vc\ measurements are considered. The distribution of our spiral
galaxies is consistent with the \mbh$-$\vc\ relation by
\citet{Ho2008b}, whereas the value of \mbh\ for elliptical and
lenticular galaxies is almost constant over the full observed \vc\
range (\Fig{m_v}).

\citet{Ho2008b} suggested that the main source of scatter in the
\mbh$-$\vc\ relation is related to the dependence of \vc/\sigmae\ on
the bulge-to-disc ratio \citep{Courteau2007, Ho2007}.
Since the bulge-to-disc ratio scales with the light concentration
\citep{Doi1993}, which is another way to estimate the degree of bulge
dominance, we derived \vc/\sigmae\ as a function of $C_{28}$.  We
found a general agreement with the relation found by
\citet{Courteau2007}.
The Jeans equation evokes a relation between the circular velocity of
galaxy and its velocity dispersion for a pressure-supported system.
The \vc/\sigmae\ relation was first reported by \citet{Whitmore1979}
and more recently by \citet{Ferrarese2002}, \citet{Buyle2004}, and
\citet{Pizzella2005}.  \citet{Courteau2007} and \citet{Ho2007} further
demonstrated that the \vc$-$\sigmae\ relation must depend on a third
parameter which depends on the galaxy structure.

The existence of the \mse\ relation and the absence of a single
universal \vs\ for all the morphological types is in contrast with the
hypothesis that \mbh\ is more fundamentally connected to halo than to
bulge, as suggested by \citet{Ferrarese2002} and \citet{Pizzella2005}.
We conclude that the SMBH mass is associated with the bulge and not
the halo \citep[see also][]{Peng2010} by analysing the \vs\ relation
as a function of \mbh\ for our sample galaxies (\Fig{v_sigma_MBH}).
For a given \mbh, the range of \vc\ is on average 1.6 times
  larger than the range of \sigmae\ demonstrating that \mbh\ is driven
  by \sigmae\ and not by \vc.
The few bulgeless galaxies known to host a SMBH (NGC~4395,
\citealt{Filippenko2003}; POX 52, \citealt{Barth2004}; NGC~1042,
\citealt{Shields2008}) are an exception to this scenario, which is
supported by cases like M33. The latter is a pure disc galaxy, which
does not show any evidence of a SMBH \citep{Gebhardt2001} but has a
massive dark matter halo \citep{Corbelli2003}.

More recently, \citet{Kormendy2011b} confirmed our early findings
\citep{Beifiori2010} by extending the analysis of the \vc$-$\sigmae\
relation to bulgeless galaxies.  They reported an absence of
correlation between \vc\ and \sigmae\ unless the galaxy also contains
a bulge, suggesting that the fundamental driver for \mbh\ is \sigmae\
for all Hubble types and that \vc\ plays only a small r\^ole in
galaxies with a bulge.
Although at this later cosmic time the \mbh$-$\vc\ is coarser than the
\mse\ relation, \citet{Volonteri2011} remind us that a tighter
connection in the past is not precluded.  At earlier epochs, the SMBH
assembly was more likely coupled to the dark matter halo properties
based on a merger-driven hierarchical formation scenario.  Still, the
scatter we measure for the \mbh$-$\vc\ relation is a factor $\sim2$
larger than that of the \mse\ relation, which is significantly more
than argued by \citet{Volonteri2011} in their analysis of
\citet{Kormendy2011b}'s data set.

\placefigfourteen

\subsection{Correlations Between \mbh\ and Linear Combinations of
Bulge and Galaxy Parameters}
\label{sec:linearcombination}

We wish to understand whether the relations between \mbh\ and bulge
and galaxy properties studied in \se{corr_BH} can be improved by the
addition of a third parameter.
For the bulge or the galaxy parameters $x$ and $y$ in
Table~\ref{tab:fit_combination}, a correlation of the form
\be
\log \frac{M_{\bullet}}{{\rm M}_{\odot}}\,=\, \gamma\, +
  \,\alpha\,\log{\frac{x}{x_0}} +
  \,\beta\,\log{\frac{y}{y_0}}
\ee
is assumed, where $x_0$ and $y_0$ are normalisation values chosen near
the mean of the distribution of $x$ and $y$ values respectively.  The
fit parameters are the offset $\gamma$, and the logarithmic slopes
$\alpha$ and $\beta$.
 The best-fit values of $\alpha$, $\beta$, and $\gamma$, their
  associated uncertainties, the total scatter $\epsilon$ and intrinsic
  scatter $\epsilon_{\rm intr}$ of the relation are given in
  Table~\ref{tab:fit_combination}.
  The Spearman rank test cannot be evaluated in the same way as
  described in \se{corr_BH} as it applies to two variables.  To
  estimate the degree of correlations between the parameters in use,
  we applied the Spearman rank statistic using the projections of the
  planes once the combination of the two independent variables has
  been fixed to its best-fit values.  This provides an indication of
  the degree of correlation between several linear combinations.  The
  Spearman rank correlation coefficient $\rho$, and the significance
  of the correlation $P$ derived for the projections are also given in
   Table~\ref{tab:fit_combination}.
  The quoted errors are all computed through a bootstrap technique.

 For both Samples A$+$B and the comparison sample, we performed a
 censored regression analysis with two independent variables assuming
 normal residuals.  The best-fit parameters were computed with the EM
 algorithm implemented in the {\sc ASURV} package which yields maximum
 likelihood estimates from a censored data set allowing, in the
 general form, N observables distributed via a multivariate Gaussian
 distribution.  We applied this technique to the special case of two
 independent variables and another dependent variable containing the
 censored data. 
The regression problem solves like a least-square
fit but accounting for censored data.  It uses initial guesses
from an ordinary linear regression of the non-censored data
and then converges on the new censored values.

 The standard deviation is estimated in the
 standard fashion, weighted for a factor which accounts for the
 censored data \citep[see][for details]{Isobe1986}.
  The generalised Spearman rank correlation coefficient, and the
  significance of the correlation for the projections of planes, are
  based on the technique described in \citet{Akritas1989}.
The total scatter was computed as the rms deviation in $\log
(M_{\bullet}/{\rm M}_{\odot})$ from the fitted relation assuming no
measurement errors.

We used our own modified version of the {\sc MPFITEXY} algorithm to
perform a least-squares linear regression of Sample B's data with two
independent variables. Measurement errors were invoked in the
derivation of best-fit parameters, total, and intrinsic scatter. The
Spearman rank correlation coefficient and level of significance of the
correlation were also derived by using the projections of the best-fit
plane given all possible parameter combinations. The actual
computation was done using the {\sc IDL} routine {\sc R\_CORRELATE}.

\subsubsection{\mbh\ versus Bulge Parameters}
\label{sec:local}

So far we have found that \mbh\ is more strongly related with
\sigmae\ than \Lbui\ and \Mbuvir. Nevertheless, the correlation
between the \mbh\ and \Mbuvir\ suggests that a possible combination
of the bulge properties can result in a tighter correlation with \mbh.
Note that the BHFP derives from the FP which connects \reb, \sigmae,
and \muemean\ together \citep[][and reference therein]{Dressler1987,
 Djorgovski1987,Jorgensen2007}.
Therefore, we have analysed the residuals of \mse, \mlbu, and
\mmbuvir\ relations to test whether there is the evidence
for an additional dependence with an additional parameter. The
residuals were obtained as the difference between the observed
data and their expected values from the fitting relations in
Table~\ref{tab:fit_single}.
We compared the residuals in \msres, \mlibures, and \mmassbures\ with
the bulge photometric and kinematic properties (i.e., \muemean, \reb,
\n, and \sigmae; \Fig{residual_mlbule}).  There is no trend between
\msres\ and other bulge properties (\Fig{residual_mlbule}, left
column), whereas \mlibures\ and \mmassbures\ show a very weak
dependence on \reb\ according to the the Spearman rank correlation
coefficient and their significance intervals $\sim 6$\% and $\sim2$\%,
respectively (\Fig{residual_mlbule}, middle and left columns,
respectively). This was confirmed by \citet{Sani2011}, who found no
correlation of the residual with \re.

\placefigfifteen

We have tested for the more fundamental driver of \mbh.
To this aim we considered different linear combinations of the bulge
parameters with \mbh\ (\Fig{FP}).
The correlation between \mbh, \sigmae, and \reb\ is as tight as the
\mse\ (\Fig{FP}a). Taking into account \reb\ does not significantly
improve the \mse\ fit.  The relation between \mbh, \Lbui, and
\reb\ (\Fig{FP}b) is not as strong as that between \mbh, \sigmae, and
\reb, but it is a slight improvement over \mlbui.
The relation between \mbh, \sigmae, and \Lbui\ is as tight as that
between \mbh, \sigmae, and \reb\ (\Fig{FP}c).
The relation between \mbh, \sigmae, and \Mbuvir\ (\Fig{FP}d) is
expected since \Mbuvir\ is known to correlate with \mbh\ and it
depends on both \sigmae\ and \reb. For this reason the two
correlations with \mbh\ and either \sigmae\ and \Mbuvir\ or
\reb\ and \Mbuvir\ are different (but not independent) expressions
of the same relationship.
Thus, larger SMBHs are associated with more massive and larger bulges,
as understood in the framework of coevolution of spheroids and SMBHs.
\citet{Hopkins2007a, Hopkins2007b} first studied the residuals of the
\mmbuvir\ and \mse\ relations with bulge properties, finding tighter
correlations between the combination of the bulge parameters and
\mbh. This bolstered the notion of a BHFP.  The bulge parameters
\Mbuvir\ and \Lbui\ are indeed physically related to each other
through the fundamental plane relation (FP) and virial theorem.
The existence of such BHFP has important implications for the largest
\mbh\ and resolves the apparent conflict between expected and measured
values of \mbh\ for the outliers in both \mse\ and
\mmbuvir\ relations \citep[e.g.][]{Bernardi2007, Lauer2007b}.
A similar correlation, but between \mbh, \reb, and \muemean, was
reported by \citet{Barway2007} for nearby ellipticals with measured
\mbh. This correlation has a smaller scatter than the \mse\ and
\mlbug\ relations and gives further support to the existence
of a FP-type relation for SMBHs.

\placefigsixteen

Furthermore, by comparing the tightness of the correlations between
\mbh\ and linear combination of the bulge parameters, we argue that
\sigmae\ is the parameter that drives the connection with \mbh\ in the
BHFP. A small contribution is due to \reb\ or \Lbui.
Similar results were obtained by \citet{Aller2007} who compared
\mbh\ with \sigmae, $I_{\rm e, bulge}$ and \reb.  Thus, including an
additional parameter to the relation does not improve the quality of
the fit, with \sigmae\ always being the dominant parameter.

\subsubsection{\mbh\ versus Galaxy Parameters}
\label{sec:global}

We also wish to test if the shallow \mli\ and \mbh$-$\mstar\ relations
can be improved by the addition of another galaxy parameter.
Therefore, we calculated the residuals of \mse, \mli, and
\mbh$-$\mstar\ relations to look for the signature of an additional
galaxy parameter. The residuals were obtained as the difference
between the data points and the relation fits listed in
Table~\ref{tab:fit_single}.
We compared the residuals in \msres, \mlires, and \mstarres\ with the
galaxy photometric and kinematic properties (i.e., \co, $I_{\rm e,
  gal}$, \reg, and \sigmae ; \Fig{mbh_residual_ms}).
For \msres, no trends with other galaxy properties are found whereas
the \mlires\ and \mstarres\ relations show a weak dependence on \re\
according to the Spearman rank correlation coefficient with a
significance levels of 0.3\% and 0.9\%, respectively.
We also examined the residuals of the \mbh$- (g-i)$ relation to check
that our results are not affected by the propagated errors in the
determination of \mstar . \Fig{FP_li} shows the different linear
combinations of two galaxy parameters with \mbh .  A few tight
correlations are found when considering the total galaxy rather than
bulge parameters.  The tightness always depends on having \sigmae\ as
a fitting parameter (Table~\ref{tab:fit_combination}), i.e., the
Spearman rank correlation coefficient is higher when \sigmae\ is taken
into account.  The constancy of the Spearman rank correlation
coefficient despite the addition of other fitting parameters, argues
that the main acting parameter is \sigmae.

\placefigseventeen

We conclude that the addition of bulge (e.g., \reb, \Lbui, \Mbuvir) or
galaxy (e.g., \reg, \li, \mstar) structural parameters does not
appreciably modify and improve the \mse\ relation.
We exclude that parameter covariances would produce the same
scatter in both the \mse\ and \mse$-$\re\ relations.  Such a 
conspiracy would require an anti-correlation between \sigmae\
and \reb\ (or \reg) which is not observed. 

\placefigeighteen

\section{Discussion}
\label{sec:discussion}

The existence of the \mbh\ scaling relations implies that SMBH and
galaxy formation processes are closely linked.  Some of the challenges
of the current models of SMBH formation and evolution include
reproducing and maintaining the relations regardless of the sequence
of galaxy evolution during the hierarchical mass assembly
\citep{Robertson2006a, Schawinski2006, Croton2009}.  A more
comprehensive assessment the \mbh\ scaling relations scatter thus
enables a better characterisation of the different SMBH/galaxy
formation models.

Our large sample of galaxies with secure determination or upper limits
of \mbh\ has enabled a thorough investigation of the SMBH demography
over a wide range of \mbh , morphological type, and nuclear activity.
After establishing the unbiased mapping of \mbh\ upper limits against
that of secure \mbh, we tested whether \mbh\ is more fundamentally
driven by one of the several bulge (i.e., the velocity dispersion,
luminosity, virial mass, \sersic index, and mean effective surface
brightness) and galaxy (luminosity, circular velocity, stellar,
virial, and dynamical mass) parameters known to correlate with the
SMBH mass, and if the known scaling relations can be improved with the
addition of a third parameter.

\begin{itemize}

\item We argue that \mbh\ is fundamentally driven by \sigmae,
  considering that the \mse\ relation has the tightest scatter with
  respect to all other scaling relations.  The scatter of our \mse\
  relation is comparable to \citetalias{Gultekin2009} but slightly
  larger than that of \citet{Ferrarese2005}. At small \sigmae, some
  \mbh\ upper limits exceed the expectation value as they likely
  account for the mass of a conspicuous nuclear cluster
  (\citetalias{Beifiori2009}).  Excluding these galaxies from the fit
  reduces the scatter to \citet{Ferrarese2005}'s value. Barred and
  unbarred galaxies as in \citetalias{Beifiori2009} and
  \citetalias{Gultekin2009} follow the same \mse\ relation within the
  errors, contrary to \citet{Graham2008}'s findings. Classical and
  pseudo-bulges were identified according to their \sersic index
  \citep{Kormendy2004}. The \mse\ relations of classical and
  pseudo-bulges have a different slope. The low significance of the
  \mse\ relation for pseudo-bulges is in agreement with more recent
  findings by \citet{Kormendy2011a} and suggests that the formation
  and growth histories of SMBHs depend on the host bulge type.

\item The \mlbui\ relation is clearly not as tight as the \mse\ one.
  However, the fact that \citetalias{Gultekin2009} found a \mlbui\
  relation as tight as the \mse\ one is not in contradiction with our
  results, since \citetalias{Gultekin2009} accounted only for
  early-type galaxies.  The same is true for the \mmbuvir\
  correlation, although the bulge mass proved to be a better proxy of
  \mbh\ than \Lbu\ and it includes \reb\ as an additional fitting
  parameter.

\item Contrary to previous findings \citep{Graham2001, Graham2003a,
    Graham2007}, we find little or no correlation between \mbh\ and
  \sersic $n$ in agreement with \citet{Hopkins2007b}.  The latter
  stated that \mbh\ is unrelated with the light concentration of the
  bulge based on observations and simulations. Consistently, we
  confirm that \mbh\ and \muemean\ are poorly correlated.

\item The correlations between \mbh\ and galaxy luminosity or mass are
  not a marked improvement over the \mse\ relation. These scaling
  relations are strongly sensitive to the morphology of the host
  galaxies, with the presence of a disc playing an anti-correlating
  r\^ole, as first pointed out by \citep{Kormendy2001b}. This is a
  further indication that bulges are driving the SMBH correlations and
  SMBH evolution.

\item We found that the \mbh$-$\vc\ relation is significantly coarser
  than that of the \mse\ relation, with a scatter about twice larger,
  suggesting that \mbh\ is more strongly controlled by the baryons
  than the dark matter.

\item To assess the need for an additional parameter in the
  \mbh-bulge/galaxy scaling relations, we performed both a residual
  analysis and third-parameter fits as in \citet{Hopkins2007b},
  \citet{Barway2007}, and \citet{Aller2007}. To this aim we considered
  different linear combinations of bulge or galaxy parameters with
  \mbh.
  The strongest correlations always include \sigmae\ as a fundamental
  structural parameter. The tightest relation is found between \mbh,
  \sigmae, and \reb\ (also known as the BHFP, \citealt{Hopkins2007a,
    Hopkins2007b}). Since its scatter does not change appreciably
  compared to that of the \mse\ relation, the addition of \reb\ is
  barely an improvement.  This is a further confirmation that \sigmae\
  is the fundamental parameter which drives also the BHFP.

\end{itemize}

Our findings about the tightness of scaling relations such as the \mse\
may be interpreted in the framework of self-regulating feedback in
galaxies \citet{Hopkins2009}. These authors argued that the energy
released from the accretion of gas onto a SMBH is enough to stop
further gas accretion, drive away the gas, and quench star formation.
\citet{Hopkins2009} studied the relation between \mbh\ and the mass of
the gas accreting onto the SMBH as a function of the distance from the
central engine.
The scatter of the \mbh-stellar mass relation increases at smaller
radii (i.e., approaching the radius of influence of the SMBH) and
decreases at larger radii (i.e., where the SMBH scaling relations are
defined).  This has been interpreted in terms of self-regulated growth
of SMBHs, where the SMBH accretes and regulates itself without
accounting for the gas supply but just being controlled by the bulge
properties.

In this scenario, \sigmae\ is the property which gives the tightest
connection with MBH since it determines the depth of the local
potential well from which the gas must be expelled.
\citet{Younger2008} studied self-regulated models of SMBHs growth in
different scenarios of major mergers, minor mergers, and disc
instabilities, to find that SMBHs depend on the scale at which
self-regulation occurs. They compared the bulge binding energy and
total binding energy with \mbh, finding that the total binding energy
is not a good indicator of \mbh\ mass in disc-dominated systems. This
agrees with our findings that the late-type systems deviate most
significantly from the \mbh-galaxy and BHFP scaling relations.

Several SMBH formation models predict a connection between \mbh\ and
the total mass of the galaxy \citep{Haehnelt1998, Silk1998,
  Adams2001,Croton2006,Croton2009} such that if the dark and baryonic
matter act to form the bulge and SMBH, the dark halo determines the
bulge and SMBH properties.  Therefore, the mass of the SMBH and dark
matter halo should be connected \citep{Cattaneo2001, Hopkins2005a,
  Hopkins2005b}. 
Numerous recent studies have erred along those lines.  For instance,
\citet{Bandara2009} studied the correlation between \mbh\ and total
(luminous+dark) mass of the galaxy estimated from numerical
simulations of gravitational lensing.  Their relation suggests that
the more massive halos are more efficient at forming SMBHs than the
less massive ones; the slope of their relation suggests merger-driven,
feedback-regulated processes of SMBH growth.  Likewise,
\citet{Booth2010} used self-consistent simulations of the coevolution
of SMBHs and galaxies to confirm the relation by \citet{Bandara2009}
and to prove that self-regulation of the SMBH growth occurs on dark
matter halo scales.  \citet{Volonteri2009} investigated through
numerical simulations the observational signature of the
self-regulated SMBH growth by analysing the mass assembly history of
black hole seeds.  They found that the \mse\ relation stems from the
merging history of massive dark halos, and that its slope and scatter
depend on the halo seed and specific SMBH self-regulation process.

Our observational results are not supportive of such results.  We
found only a weak correlation of \mbh\ with \vc.  This agrees with our
expectations based on the observed tightness of the \mse\ relation and
given the fact that the scatter of the \vs\ relation is large and
morphologically-dependent \citep{Courteau2007, Ho2007}.

The \mbh-\vc\ relation may be improved with homogeneous measurements
of the \vc\ data base.  Indeed, while our collection of observed
velocities is as current and reliable as possible, especially for
spiral galaxies, we still lack a fully homogeneous data base.  Many of
us are working on improvements of this nefarious situation.  However
it is clear, as this work demonstrates, that the \mbh-\sigmas\ is the
definitive fundamental relation of the SMBH-bulge connection.

\bigskip
\bigskip

\section*{Acknowledgements}
We are indebted to Ralf Bender, Michele Cappellari, Lodovico Coccato,
Elena Dalla Bont\`a, Victor Debattista, John Kormendy, Tod Lauer,
Lorenzo Morelli, Alessandro Pizzella, Marc Sarzi, and Roberto Saglia
for many useful discussions and suggestions.  We also thank Jairo
M\'endez Abreu for his {\sc GASP2D} package which we used to measure the
photometric parameters of our sample galaxies.
We acknowledge the anonymous referee for valuable comments that led to
an improved presentation.  
AB is grateful to Queen's University for its hospitality while part of
this paper was being written.  AB was also supported by a grant from
Accademia dei Lincei and Royal Society as well as STFC rolling grant
ST/I001204/1 ``Survey Cosmology and Astrophysics''.
EMC was supported by Padua University through grants CPDA089220/08
and CPDR095001/09 and by Italian Space Agency through grant ASI-INAF
I/009/10/0.
SC and YZ acknowledge support from the Natural Science and Engineering
Science Council of Canada.
This research has made use of the HyperLeda database, NASA/IPAC
Extragalactic Database (NED), and Sloan Digital Sky Survey (SDSS).

\newpage
\onecolumn

\clearpage


\defcitealias{Tully1988}{1}
\defcitealias{RC3}{2}
\defcitealias{Tonry2001}{3}
\defcitealias{Freedman2001}{4}
\defcitealias{Pizzella2005}{6}
\defcitealias{Paturel2003a}{7}
\defcitealias{Kronawitter2000}{8}
\defcitealias{Ho2007}{9}
\defcitealias{Binney1990}{10}
\defcitealias{Silchenko2004}{11} 
\defcitealias{Hota2009}{12}	         
\defcitealias{Quillen1994}{13}      
\defcitealias{Young2008}{14}       
\defcitealias{Morse1998}{15}       

\begin{landscape}
\begin{center}
\begin{tiny}
\begin{longtable}{l l l r c c c c c c c c c c}
\caption{Properties of the galaxies of the Sample A}\\
\hline
\hline
\noalign{\smallskip}
\multicolumn{1}{c}{Galaxy} &
\multicolumn{1}{c}{Morph. T.} &
\multicolumn{1}{c}{Spec. Cl.} &
\multicolumn{1}{c}{$D$} &
\multicolumn{1}{c}{Ref.} &
\multicolumn{1}{c}{$M_{B}^0$} &
\multicolumn{1}{c}{$\sigma_{\rm e}$} &
\multicolumn{1}{c}{$i$} &
\multicolumn{1}{c}{Ref} &
\multicolumn{1}{c}{$V_{\rm c}$} &
\multicolumn{1}{c}{Ref} &
\multicolumn{1}{c}{\mbh\ ($33$\degree)} &
\multicolumn{1}{c}{\mbh\ ($81$\degree)} &
\multicolumn{1}{c}{\mbh} \\
\multicolumn{1}{c}{} &
\multicolumn{1}{c}{} &
\multicolumn{1}{c}{} &
\multicolumn{1}{c}{(Mpc)} &
\multicolumn{1}{c}{} &
\multicolumn{1}{c}{(mag)} &
\multicolumn{1}{c}{(km s$^{-1}$)} &
\multicolumn{1}{c}{(\degree)}  &
\multicolumn{1}{c}{} &
\multicolumn{1}{c}{(km s$^{-1}$)} &
\multicolumn{1}{c}{} &
\multicolumn{1}{c}{(\msun)}  &
\multicolumn{1}{c}{(\msun)}  &
\multicolumn{1}{c}{(\msun)}  \\
\multicolumn{1}{c}{(1)} &
\multicolumn{1}{c}{(2)} &
\multicolumn{1}{c}{(3)} &
\multicolumn{1}{c}{(4)} &
\multicolumn{1}{c}{(5)} &
\multicolumn{1}{c}{(6)} &
\multicolumn{1}{c}{(7)} &
\multicolumn{1}{c}{(8)} &
\multicolumn{1}{c}{(9)} &
\multicolumn{1}{c}{(10)} &
\multicolumn{1}{c}{(11)} &
\multicolumn{1}{c}{(12)} &
\multicolumn{1}{c}{(13)} &
\multicolumn{1}{c}{(14)} \\
\noalign{\smallskip}
\hline
\noalign{\smallskip}
\endfirsthead
\multicolumn{14}{c}%
{\tablename\ \thetable{} -- Continued} \\
\hline
\hline
\noalign{\smallskip}
\multicolumn{1}{c}{Galaxy} &
\multicolumn{1}{c}{Morph. T.} &
\multicolumn{1}{c}{Spec. Cl.} &
\multicolumn{1}{c}{$D$} &
\multicolumn{1}{c}{Ref.} &
\multicolumn{1}{c}{$M_{B}^0$} &
\multicolumn{1}{c}{$\sigma_{\rm e}$} &
\multicolumn{1}{c}{$i$} &
\multicolumn{1}{c}{Ref.} &
\multicolumn{1}{c}{$V_{\rm c}$} &
\multicolumn{1}{c}{Ref.} &
\multicolumn{1}{c}{\mbh\ ($33$\degree)} &
\multicolumn{1}{c}{\mbh\ ($81$\degree)} &
\multicolumn{1}{c}{\mbh} \\
\multicolumn{1}{c}{} &
\multicolumn{1}{c}{} &
\multicolumn{1}{c}{} &
\multicolumn{1}{c}{(Mpc)} &
\multicolumn{1}{c}{} &
\multicolumn{1}{c}{(mag)} &
\multicolumn{1}{c}{(km s$^{-1}$)} &
\multicolumn{1}{c}{(\degree)}  &
\multicolumn{1}{c}{} &
\multicolumn{1}{c}{(km s$^{-1}$)} &
\multicolumn{1}{c}{} &
\multicolumn{1}{c}{(\msun)}  &
\multicolumn{1}{c}{(\msun)}  &
\multicolumn{1}{c}{(\msun)}  \\
\multicolumn{1}{c}{(1)} &
\multicolumn{1}{c}{(2)} &
\multicolumn{1}{c}{(3)} &
\multicolumn{1}{c}{(4)} &
\multicolumn{1}{c}{(5)} &
\multicolumn{1}{c}{(6)} &
\multicolumn{1}{c}{(7)} &
\multicolumn{1}{c}{(8)} &
\multicolumn{1}{c}{(9)} &
\multicolumn{1}{c}{(10)} &
\multicolumn{1}{c}{(11)} &
\multicolumn{1}{c}{(12)} &
\multicolumn{1}{c}{(13)} &
\multicolumn{1}{c}{(14)} \\
\noalign{\smallskip}
\hline
\noalign{\smallskip}
\endhead
\noalign{\smallskip}
\hline
\endfoot
%
%
\multicolumn{13}{c}{Upper limits on \mbh\ from \citetalias{Beifiori2009} } \\
\noalign{\smallskip}
\hline
\noalign{\smallskip}
 IC~342         & SABcd(rs)     & H      &  3.73 & \citetalias{Tully1988}    & $-21.97$ & $ 65.1\pm 23.8$ & 12 & \citetalias{RC3} & $ 185.0\pm  8.9$ & \citetalias{Pizzella2005}  & 5.0E6 & 1.3E6 & ... \\
 IC~3639        & SBbc(rs):     & S2*    & 44.80 & \citetalias{RC3}          & $-20.70$ & $ 91.4\pm  4.8$ & 18 & \citetalias{RC3} & $ 248.0\pm 11.6$ & \citetalias{Paturel2003a}  & 2.3E7 & 4.5E6 & ... \\
 NGC~193        & SAB0$^-$(s):  & ...    & 49.65 & \citetalias{RC3}          & $-20.22$ & $185.0\pm 17.1$ & 40 & \citetalias{RC3} & ...             & ...                         & 5.2E8 & 1.2E8 & ... \\
 NGC~289        & SBbc(rs)      & ...    & 17.08 & \citetalias{RC3}          & $-19.91$ & $108.9\pm 11.6$ & 47 & \citetalias{RC3} & $240.1\pm  1.8$ & \citetalias{Paturel2003a}   & 4.7E7 & 1.2E7 & ... \\
 NGC~315        & E$^+$:        & L1.9   & 57.68 & \citetalias{RC3}          & $-22.09$ & $313.4\pm 27.2$ &... & ...              & $  569\pm   59$*& \citetalias{Kronawitter2000}& 1.7E9 & 4.0E8 & ... \\
 NGC~383        & SA0$^-$:      & ...    & 59.17 & \citetalias{RC3}          & $-21.33$ & $238.8\pm 16.7$ & 32 & \citetalias{RC3} & ...             & ...                         & 1.2E9 & 2.7E8 & ... \\
 NGC~541        & S0$^-$:       & ...    & 63.65 & \citetalias{RC3}          & $-21.19$ & $191.4\pm  4.0$ & 15 & \citetalias{RC3} & ...             & ...                         & 8.6E8 & 1.8E8 & ... \\
 NGC~613        & SBbc(rs)      & H*     & 15.40 & \citetalias{RC3}          & $-20.56$ & $125.3\pm 18.9$ & 42 & \citetalias{RC3} & $240.1\pm  3.9$ & \citetalias{Ho2007}         & 9.0E7 & 1.8E7 & ... \\
 NGC~741        & E0:           & ...    & 65.71 & \citetalias{RC3}          & $-22.17$ & $232.2\pm 11.2$ & 54 & 5                & ...             & ...                         & 1.1E9 & 2.0E8 & ... \\
 NGC~788        & SA0/a(s)      & S1/S2* & 47.88 & \citetalias{RC3}          & $-20.75$ & $127.3\pm 18.2$ & 45 & \citetalias{RC3} & ...             & ...                         & 1.6E8 & 4.4E7 & ... \\
 NGC~1052       & E4            & L1.9   & 18.11 & \citetalias{Tonry2001}    & $-20.09$ & $191.4\pm  6.2$ & 82 & \citetalias{RC3} & $189  \pm 18.9$*& \citetalias{Binney1990}     & 3.4E8 & 8.8E7 & ... \\
 NGC~1358       & SAB0/a(r)     & S2     & 48.16 & \citetalias{RC3}          & $-20.86$ & $171.5\pm 18.5$ & 41 & \citetalias{RC3} & $180.4\pm 14.3$ & \citetalias{Paturel2003a}   & 4.9E8 & 1.1E8 & ... \\
 NGC~1497       & S0            & ...    & 75.32 & \citetalias{RC3}          & $-21.24$ & $245.6\pm 22.5$ & 57 & \citetalias{RC3} & ...             &  ...                        & 6.6E8 & 2.8E8 & ... \\
 NGC~1667       & SAB(r)c       & S2     & 56.09 & \citetalias{RC3}          & $-21.49$ & $178.0\pm 28.9$ & 40 & \citetalias{RC3} & $236.6\pm 13.1$ & \citetalias{Ho2007}         & 2.7E8 & 9.3E7 & ... \\
 NGC~1961       & SABc(rs)      & L2     & 48.63 & \citetalias{RC3}          & $-22.58$ & $222.2\pm 38.0$ & 51 & \citetalias{RC3} & $415.4\pm 13.0$ & \citetalias{Ho2007}         & 4.3E8 & 8.8E7 & ... \\
 NGC~2110       & SAB0$^-$      & S2*    & 29.12 & \citetalias{RC3}          & $-20.62$ & $201.4\pm 22.9$ & 53 & \citetalias{RC3} & ...             & ...                         & 5.8E8 & 3.0E7 & ... \\
 NGC~2179       & SA0/a(s)      & ...    & 35.84 & \citetalias{RC3}          & $-20.09$ & $154.3\pm 11.2$ & 51 & \citetalias{RC3} & $251.7\pm 17.4$ & \citetalias{Ho2007}         & 3.5E8 & 1.2E8 & ... \\
 NGC~2273       & SBa(r):       & S2     & 23.33 & \citetalias{RC3}          & $-19.97$ & $116.7\pm 10.6$ & 44 & \citetalias{RC3} & $223.3\pm  6.5$ & \citetalias{Ho2007}         & 6.6E6 & 2.0E6 & ... \\
 NGC~2329       & S0$^-$:       & ...    & 72.33 & \citetalias{RC3}          & $-21.36$ & $217.6\pm 13.3$ & 38 & \citetalias{RC3} & ...             & ...                         & 2.3E8 & 1.0E8 & ... \\
 NGC~2685       & (R)SB0$^+$ pec& S2/T2: & 12.51 & \citetalias{RC3}          & $-18.81$ & $ 82.3\pm  7.2$ & 70 & \citetalias{RC3} & $149.0\pm  4.3$ & \citetalias{Paturel2003a}   & 1.0E7 & 1.5E6 & ... \\
 NGC~2892       & E+ pec:       & ..     & 86.24 & \citetalias{RC3}          & $-20.78$ & $297.1\pm 18.8$ & 22 & 5                & ...             & ...                         & 3.5E8 & 2.1E8 & ... \\
 NGC~2903       & SABbc(rs)     & H      & 10.45 & \citetalias{RC3}          & $-21.14$ & $ 94.0\pm 12.1$ & 64 & \citetalias{RC3} & $185.8\pm  4.1$ & \citetalias{Ho2007}         & 2.2E7 & 6.0E6 & ... \\
 NGC~2911       & SA0(s): pec   & L2     & 43.49 & \citetalias{RC3}          & $-21.09$ & $211.6\pm 15.2$ & 62 & 5                & $253.0\pm 25.3$ & \citetalias{Silchenko2004}  & 2.4E9 & 6.2E8 & ... \\
 NGC~2964       & SABbc(r):     & H      & 19.69 & \citetalias{RC3}          & $-20.03$ & $ 95.4\pm 18.7$ & 43 & 5                & $182.4\pm  5.0$ & \citetalias{Paturel2003a}   & 2.2E7 & 1.3E6 & ... \\
 NGC~3021       & SAbc(rs)      & ...    & 22.40 & \citetalias{RC3}          & $-19.37$ & $ 56.4\pm 24.8$ & 58 & \citetalias{RC3} & $127.4\pm  4.1$ & \citetalias{Ho2007}         & 3.7E7 & 9.1E6 & ... \\
 NGC~3078       & E2-3          & ...    & 32.85 & \citetalias{Tonry2001}    & $-20.79$ & $207.8\pm 12.2$ & 50 & \citetalias{RC3} & ...             & ...                         & 2.1E8 & 3.1E7 & ... \\
 NGC~3081       & (R)SAB0/a(r)  & S2*    & 33.51 & \citetalias{RC3}          & $-20.19$ & $123.0\pm  7.6$ & 43 & \citetalias{RC3} & $136.9\pm  5.1$ & \citetalias{Paturel2003a}   & 3.2E7 & 7.9E6 & ... \\
 NGC~3351       & SBb(r)        & H      &  9.33 & \citetalias{Freedman2001} & $-19.74$ & $ 95.4\pm 15.1$ & 50 & \citetalias{RC3} & $149.7\pm  5.2$ & \citetalias{Ho2007}         & 6.0E6 & 1.8E6 & ... \\
 NGC~3368       & SABab(rs)     & L2     &  9.71 & \citetalias{Tonry2001}    & $-20.28$ & $105.3\pm  4.1$ & 49 & \citetalias{RC3} & $205.2\pm  5.8$ & \citetalias{Ho2007}         & 4.5E7 & 1.4E7 & ... \\
 NGC~3393       & (R$'$)SBa(rs):& S2*    & 50.59 & \citetalias{RC3}          & $-21.03$ & $167.5\pm 25.6$ & 26 & \citetalias{RC3} & $191.4\pm 14.4$ & \citetalias{Paturel2003a}   & 2.3E8 & 8.8E7 & ... \\
 NGC~3627       & SABb(s)       & T2/S2  &  9.43 & \citetalias{Freedman2001} & $-20.88$ & $ 97.2\pm  7.5$ & 68 & 5                & $168.4\pm  4.1$ & \citetalias{Ho2007}         & 1.4E7 & 6.3E6 & ... \\
 NGC~3642       & SAbc(r):      & L1.9   & 21.65 & \citetalias{RC3}          & $-20.37$ & $ 96.1\pm 24.8$ & 35 & \citetalias{RC3} &  ...            & ...                         & 2.9E7 & 2.4E7 & ... \\
 NGC~3675       & SAb(s)        & T2     & 12.41 & \citetalias{RC3}          & $-20.10$ & $105.3\pm  4.4$ & 66 & 5                & $201.8\pm  5.1$ & \citetalias{Ho2007}         & 3.6E7 & 9.3E6 & ... \\
 NGC~3801       & S0?           & ...    & 46.29 & \citetalias{RC3}          & $-20.69$ & $210.1\pm 17.7$ & 65 & \citetalias{RC3} & $280.0\pm 28  $ & \citetalias{Hota2009}       & 3.9E8 & 9.3E7 & ... \\
 NGC~3862       & E             & ...    & 84.56 & \citetalias{RC3}          & $-21.27$ & $210.4\pm 13.0$ &  2 & 5                & ...             & ...                         & 6.0E8 & 1.1E8 & ... \\
 NGC~3953       & SBbc(r)       & T2     & 15.40 & \citetalias{RC3}          & $-20.71$ & $128.5\pm  0.9$ & 63 & \citetalias{RC3} & $207.5\pm  4.3$ & \citetalias{Ho2007}         & 4.2E7 & 1.1E7 & ... \\
 NGC~3982       & SABb(r):      & S1.9   & 15.87 & \citetalias{RC3}          & $-19.47$ & $ 78.0\pm  2.0$ & 31 & 5                & $161.5\pm  6.8$ & \citetalias{Ho2007}         & 1.6E7 & 4.9E6 & ... \\
 NGC~3992       & SBbc(rs)      & T2:    & 15.31 & \citetalias{RC3}          & $-20.81$ & $124.4\pm 17.8$ & 60 & 5                & $245.9\pm  4.2$ & \citetalias{Paturel2003a}   & 6.3E7 & 1.7E7 & ... \\
 NGC~4036       & S0$^-$        & L1.9   & 19.04 & \citetalias{RC3}          & $-20.06$ & $163.8\pm  5.4$ & 87 & 5                & ...             & ...                         & 1.8E8 & 3.4E7 & ... \\
 NGC~4088       & SABbc(rs)     & H      & 11.85 & \citetalias{RC3}          & $-20.00$ & $ 85.7\pm  3.8$ & 71 & \citetalias{RC3} & $162.0\pm  3.0$ & \citetalias{Ho2007}         & 1.2E7 & 3.2E6 & ... \\
 NGC~4143       & SAB0$^0$(s)   & L1.9   & 14.84 & \citetalias{Tonry2001}    & $-19.11$ & $201.2\pm  5.7$ & 67 & 5                & ...             & ...                         & 1.9E8 & 3.7E7 & ... \\
 NGC~4150       & SA0$^0$(r)?   & T2     & 12.79 & \citetalias{Tonry2001}    & $-18.29$ & $ 85.7\pm  3.0$ & 60 & 5                & ...             & ...                         & 2.4E6 & 3.2E5 & ... \\
 NGC~4203       & SAB0$^-$:     & L1.9   & 14.09 & \citetalias{Tonry2001}    & $-19.29$ & $156.5\pm  2.8$ & 30 & 5                & $204.4\pm  6.8$ & \citetalias{Paturel2003a}   & 1.2E8 & 3.6E7 & ... \\
 NGC~4212       & SAc:          & H      &  3.17 & \citetalias{RC3}          & $-16.28$ & $ 67.8\pm  2.8$ & 54 & \citetalias{RC3} & $133.6\pm  5.0$ & \citetalias{Paturel2003a}   & 2.6E6 & 3.7E5 & ... \\
 NGC~4245       & SB0/a(r):     & H      & 14.56 & \citetalias{RC3}          & $-18.96$ & $ 83.1\pm  2.9$ & 55 & 5                & $112.1\pm  5.4$ & \citetalias{Paturel2003a}   & 4.7E7 & 5.2E6 & ... \\
 NGC~4278       & E1-2          & L1.9   & 15.03 & \citetalias{Tonry2001}    & $-20.06$ & $232.5\pm  7.1$ &... & ...              & $  416\pm   13$*& \citetalias{Kronawitter2000}& 1.7E8 & 4.9E7 & ... \\
 NGC~4314       & SBa(rs)       & L2     & 15.49 & \citetalias{RC3}          & $-19.93$ & $106.6\pm  3.6$ & 43 & 5                & $ 175 \pm 17.5$*& \citetalias{Quillen1994}    & 1.6E7 & 4.1E6 & ... \\
 NGC~4321       & SABbc(s)      & T2     & 14.19 & \citetalias{Freedman2001} & $-20.93$ & $ 83.0\pm  3.6$ & 33 & \citetalias{RC3} & $211.8\pm  4.0$ & \citetalias{Ho2007}         & 6.9E6 & 3.2E6 & ... \\
 NGC~4335       & E             & ...    & 59.08 & \citetalias{RC3}          & $-20.67$ & $259.2\pm  5.0$ & 61 & 5                & ...             & ...                         & 5.1E8 & 1.2E8 & ... \\
 NGC~4429       & SA0$^+$(r)    & T2     & 18.20 & \citetalias{RC3}          & $-20.48$ & $169.8\pm  7.0$ & 82 & 5                & ...             & ...                         & 1.6E8 & 3.2E7 & ... \\
 NGC~4450       & SAab(s)       & L1.9   & 28.28 & \citetalias{RC3}          & $-21.66$ & $108.2\pm 14.9$ & 62 & 5                & $153.3\pm  4.9$ & \citetalias{Ho2007}         & 2.2E8 & 6.1E7 & ... \\
 NGC~4477       & SB0(s):?      & S2     & 20.81 & \citetalias{RC3}          & $-20.44$ & $144.8\pm  8.9$ & 37 & 5                & ...             & ...                         & 7.1E7 & 1.8E7 & ... \\
 NGC~4501       & SAb(rs)       & S2     & 32.29 & \citetalias{RC3}          & $-22.84$ & $140.2\pm 15.8$ & 61 & \citetalias{RC3} & $277.5\pm  5.1$ & \citetalias{Ho2007}         & 7.5E7 & 1.4E7 & ... \\
 NGC~4507       & (R$'$)SABb(rs)& S2*    & 47.04 & \citetalias{RC3}          & $-21.23$ & $143.9\pm  6.9$ & 39 & \citetalias{RC3} &  ...            & ...                         & 3.4E7 & 6.8E6 & ... \\
 NGC~4526       & SAB0(s):      & H      & 15.77 & \citetalias{Tonry2001}    & $-20.61$ & $195.4\pm  2.7$ & 83 & 5                & $357\pm  17.85$ & \citetalias{Young2008}      & 3.0E8 & 5.7E7 & ... \\
 NGC~4548       & SBb(rs)       & L2     & 17.92 & \citetalias{Tonry2001}    & $-20.63$ & $143.7\pm 13.0$ & 38 & 5                & $176.9\pm  5.8$ & \citetalias{Ho2007}         & 3.5E7 & 9.0E6 & ... \\
 NGC~4552       & E0-1          & T2:    & 14.37 & \citetalias{Tonry2001}    & $-20.36$ & $233.5\pm 10.5$ & 28 & 5                & ...             & ...                         & 1.8E9 & 6.4E8 & ... \\
 NGC~4579       & SABb(rs)      & S1.9/L & 22.96 & \citetalias{RC3}          & $-21.66$ & $109.4\pm 14.6$ & 48 & 5                & $211.5\pm  8.5$ & \citetalias{Ho2007}         & 2.1E8 & 4.0E7 & ... \\
 NGC~4636       & E0-1          & L1.9   & 13.72 & \citetalias{Tonry2001}    & $-20.40$ & $174.3\pm  8.7$ & ...& ...              & $  341\pm   13$*& \citetalias{Kronawitter2000}& 6.3E8 & 2.3E8 & ... \\
 NGC~4698       & SAab(s)       & S2     & 16.43 & \citetalias{RC3}          & $-19.99$ & $123.9\pm  8.3$ & 62 & 5                & $256.1\pm  3.9$ & \citetalias{Ho2007}         & 8.3E7 & 4.0E7 & ... \\
 NGC~4736       & (R)SAab(r)    & L2     &  4.85 & \citetalias{Tonry2001}    & $-19.83$ & $ 97.9\pm  2.6$ & 38 & \citetalias{RC3} & $150.7\pm  4.7$ & \citetalias{Ho2007}         & 1.3E7 & 3.1E6 & ... \\
 NGC~4800       & SAb(rs)       & H      & 12.51 & \citetalias{RC3}          & $-18.51$ & $102.2\pm  1.8$ & 44 & \citetalias{RC3} & $238.5\pm 22.9$ & \citetalias{Ho2007}         & 3.6E7 & 3.1E6 & ... \\
 NGC~4826       & (R)SAab(rs)   & T2     &  7.00 & \citetalias{Tonry2001}    & $-20.55$ & $105.7\pm 12.2$ & 62 & \citetalias{RC3} & $145.9\pm  4.5$ & \citetalias{Ho2007}         & 3.9E7 & 1.7E7 & ... \\
 NGC~5005       & SABbc(rs)     & L1.9   & 14.65 & \citetalias{RC3}          & $-20.79$ & $198.2\pm  7.0$ & 72 & 5                & $257.9\pm  7.3$ & \citetalias{Paturel2003a}   & 3.2E8 & 1.1E8 & ... \\
 NGC~5127       & E pec         & ...    & 62.53 & \citetalias{RC3}          & $-21.32$ & $199.4\pm 48.6$ & 61 & 5                & ...             & ...                         & 4.8E8 & 7.2E7 & ... \\
 NGC~5194       & SAbc(s) pec   & S2     &  7.93 & \citetalias{RC3}          & $-20.99$ & $ 70.3\pm  9.4$ & 54 & \citetalias{RC3} & $ 83.5\pm  2.2$ & \citetalias{Ho2007}         & 2.1E6 & 4.0E5 & ... \\
 NGC~5248       & SABbc(rs)     & H      & 17.92 & \citetalias{RC3}          & $-20.78$ & $126.5\pm 11.2$ & 57 & 5                & $137.5\pm  2.2$ & \citetalias{Paturel2003a}   & 4.8E6 & 8.4E5 & ... \\
 NGC~5252       & S0            & S1.9*  & 89.32 & \citetalias{RC3}          & $-20.97$ & $168.8\pm 24.0$ & 87 & 5                & $120  \pm  12$  & \citetalias{Morse1998}      & 6.5E8 & 1.2E8 & ... \\
 NGC~5283       & S0?           & S2*    & 34.53 & \citetalias{RC3}          & $-18.72$ & $136.1\pm 12.9$ & 28 & \citetalias{RC3} & ...             & ...                         & 5.5E7 & 1.2E7 & ... \\
 NGC~5347       & (R$'$)SBab(rs)& S2*    & 32.29 & \citetalias{RC3}          & $-19.59$ & $ 64.9\pm 12.4$ & 39 & \citetalias{RC3} & ...             & ...                         & 4.3E7 & 6.1E6 & ... \\
 NGC~5427       & SAc(s)pec     & S2*    & 36.31 & \citetalias{RC3}          & $-21.22$ & $ 64.8\pm 11.4$ & 32 & \citetalias{RC3} & $230.1\pm 30.7$ & \citetalias{Paturel2003a}   & 7.6E7 & 1.9E7 & ... \\
 NGC~5490       & E             & ...    & 65.24 & \citetalias{RC3}          & $-21.34$ & $260.6\pm 24.9$ & 59 & 5                & ...             & ...                         & 1.2E9 & 2.4E8 & ... \\
 NGC~5643       & SABc(rs)      & S2*    & 17.36 & \citetalias{RC3}          & $-21.11$ & $ 89.5\pm  1.0$ & 30 & \citetalias{RC3} & $155.1\pm  3.9$ & \citetalias{Paturel2003a}   & 4.4E7 & 2.9E6 & ... \\
 NGC~5695       & S?            & S2*    & 54.60 & \citetalias{RC3}          & $-20.46$ & $143.1\pm  1.9$ & 55 & 5                & $203.0\pm 18.0$ & \citetalias{Ho2007}         & 2.1E8 & 4.8E7 & ... \\
 NGC~5728       & SABa(r):      & S2*    & 37.61 & \citetalias{RC3}          & $-21.37$ & $193.4\pm 40.4$ & 60 & \citetalias{RC3} & $205.3\pm  7.5$ & \citetalias{Ho2007}         & 2.2E8 & 5.8E7 & ... \\
 NGC~5879       & SAbc(rs):?    & T2/L2  & 10.55 & \citetalias{RC3}          & $-18.88$ & $ 54.3\pm  8.0$ & 76 & \citetalias{RC3} & $117.8\pm  2.8$ & \citetalias{Ho2007}         & 7.9E6 & 2.2E6 & ... \\
 NGC~5921       & SB(r)bc       & L      & 20.49 & \citetalias{RC3}          & $-20.22$ & $ 84.9\pm  9.3$ & 37 & \citetalias{RC3} & $112.3\pm  2.9$ & \citetalias{Paturel2003a}   & 3.1E7 & 4.4E6 & ... \\
 NGC~6300       & SBb(rs)       & S2*    & 14.19 & \citetalias{RC3}          & $-20.71$ & $ 86.5\pm  4.7$ & 51 & \citetalias{RC3} & $165.9\pm  2.0$ & \citetalias{Paturel2003a}   & 2.2E7 & 8.8E6 & ... \\
 NGC~6500       & SAab:         & L2     & 36.49 & \citetalias{RC3}          & $-20.50$ & $211.5\pm  5.9$ & 45 & \citetalias{RC3} & $341.7\pm 16.0$ & \citetalias{Ho2007}         & 4.0E8 & 1.2E8 & ... \\
 NGC~6861       & SA0$^-$(s):   & ...    & 26.13 & \citetalias{Tonry2001}    & $-20.32$ & $404.2\pm 36.2$ & 65 & \citetalias{RC3} & ...             & ...                         & 1.4E9 & 3.4E8 & ... \\
 NGC~6951       & SABbc(rs)     & S2     & 15.96 & \citetalias{RC3}          & $-20.45$ & $ 95.5\pm  9.8$ & 35 & \citetalias{RC3} & $241.3\pm  5.4$ & \citetalias{Ho2007}         & 1.3E7 & 5.5E6 & ... \\
 NGC~7331       & SAb(s)        & T2     & 12.23 & \citetalias{Tonry2001}    & $-21.21$ & $115.8\pm  4.1$ & 75 & \citetalias{RC3} & $244.6\pm  4.1$ & \citetalias{Ho2007}         & 1.6E8 & 6.9E7 & ... \\
 NGC~7626       & E pec:        & L2::   & 38.08 & \citetalias{RC3}          & $-20.99$ & $232.5\pm 10.6$ & 37 & 5                & $401  \pm   32$*& \citetalias{Kronawitter2000}& 8.1E8 & 1.8E8 & ... \\
 NGC~7682       & SBab(r)       & S2*    & 59.27 & \citetalias{RC3}          & $-20.34$ & $112.8\pm 15.6$ & 28 & \citetalias{RC3} & $169.3\pm  5.5$ & \citetalias{Paturel2003a}   & 7.7E7 & 1.7E7 & ... \\
 UGC~1214       &(R)SAB0$^+$(rs):& S2*   & 59.92 & \citetalias{RC3}          & $-20.47$ & $106.8\pm 13.9$ & 14 & \citetalias{RC3} & $225.0\pm 23.4$ & \citetalias{Paturel2003a}   & 8.3E7 & 3.6E7 & ... \\
 UGC~1395       & SAb(rs)       & S1.9*  & 60.85 & \citetalias{RC3}          & $-20.21$ & $ 64.2\pm  5.8$ & 39 & \citetalias{RC3} & $147.4\pm  9.1$ & \citetalias{Paturel2003a}   & 1.3E7 & 3.5E6 & ... \\
 UGC~1841       & E             & ...    & 74.95 & \citetalias{RC3}          & $-21.28$ & $295.1\pm 24.6$ &  5 & \citetalias{RC3} & ...             & ...                         & 4.6E8 & 1.9E8 & ... \\
 UGC~7115       & E             & ...    & 88.20 & \citetalias{RC3}          & $-20.70$ & $183.0\pm 34.3$ & 32 & \citetalias{RC3} & ...             & ...                         & 2.8E9 & 3.6E8 & ... \\
 UGC~12064      & S0$^-$:       & ...    & 60.11 & \citetalias{RC3}          & $-20.19$ & $258.7\pm 17.6$ &  5 & \citetalias{RC3} & ...             & ...                         & 2.3E9 & 2.1E8 & ... \\
\noalign{\smallskip}
\hline
\noalign{\smallskip}
\multicolumn{13}{c}{Upper limits on \mbh\ from \citetalias{Gultekin2009} } \\
\noalign{\smallskip}
\hline
\noalign{\smallskip}
 Abell~2052   & E     & ...     & 151.1 & \citetalias{Gultekin2009}&$-22.14$ & $233\pm 11$ & 71 &2&...              & ...                                        & ... & ...& 4.9E9 \\
 NGC~3310     & SB(r)bc & H     &  17.4 & \citetalias{Gultekin2009}&$-20.25$ & $ 83\pm  4$ & 40 &\citetalias{RC3}  &$158.2\pm   3.8$ & \citetalias{Paturel2003a} & ... & ...& 4.2E7 \\
 NGC~4041     & S(rs)bc & H     &  20.9 & \citetalias{Gultekin2009}&$-19.84$ & $ 88\pm  4$ & 28 & 2 &$255.3\pm   3.8$ & \citetalias{Paturel2003a}                & ... & ...& 6.4E6 \\
 NGC~4435     & SB0   & T2/H:   &  17.0 & \citetalias{Gultekin2009}&$-19.54$ & $150\pm  7$ & 51 &\citetalias{RC3}  &$207.7\pm  17.9$ & \citetalias{Ho2007}       & ... & ...& 8.0E6 \\
 NGC~4486B    & E1    & ...     &  17.0 & \citetalias{Gultekin2009}&$-16.91$ & $185\pm  9$ & 56 & 2 &$248.7\pm    17$*& \citetalias{Kronawitter2000}             & ... & ...& 1.1E9 \\

\label{tab:allMBH_vc}

\end{longtable}
\end{tiny}

\newpage

\begin{scriptsize}
\begin{minipage}{21cm}
{\sc Notes.} ---
Col.(1): Galaxy name.
Col.(2): Morphological type from RC3.
Col.(3): Nuclear spectral class from \citet{Ho1997}, where H =
\hiire\ nucleus, L = LINER, S = Seyfert , T = transition object
(LINER/\hiire), 1 = Seyfert 1, 2 = Seyfert 2, and a fractional number
between 1 and 2 denotes various intermediate types; uncertain and
highly uncertain classifications are followed by a single and double
colon, respectively. The nuclear spectral class of galaxies marked
with * is from the NASA/IPAC Extragalactic Database (NED).
Col.(4): distance.  All the distances were taken from the literature
except those we obtained from $ V_{\rm 3K}$ the weighted mean
recessional velocity corrected to the reference frame of the microwave
background radiation given in RC3. These were derived as $D=V_{\rm
  3K}/H_0$ with and $H_0 = 70$ km s$^{-1}$ Mpc$^{-1}$. For \mbh\ upper
limits of \citetalias{Gultekin2009} we adopted the distances from their paper.
Col.(5): References for Col.(4).
Col.(6): Absolute corrected $B$ magnitude derived from $B^0_T$ (RC3)
with the adopted distance. 
Col.(7): Central \sigmas\ of the stellar component within $r_{\rm e}$
derived from the measured values taken from the same sources as
\citetalias{Beifiori2009} and corrected following
\citet{Jorgensen1995}. The value of UGC~1841 reported by
\citetalias{Beifiori2009} has been corrected adopting the effective
radius by \citet{Donzelli2007}.
Col.(8): Inclination of the disc galaxies derived from either the
fitted disc axial ratio (from this paper) or the isophotal axial ratio
at a surface brightness of $\mu_B = 25$ \mas\ (from RC3) and the
intrinsic disc axial ratio (derived following \citet{Paturel1997}).
Col.(9): References for the apparent axial ratio adopted to calculate
the inclination given in Col.(8).
Col.(10): Circular velocity. The circular velocities were derived from
either \hire\ line-width measurements or gas resolved kinematics
available in literature corrected according to the adopted
inclination. Only the values marked with * are from dynamical models
of the stellar component.
Col.(11): References for Col.(10).
Col.(12): \mbh\ upper limit of \citetalias{Beifiori2009} 
for a Keplerian disc model assuming $i=33$\degree\ and adopted distance
Col.(13): \mbh\ upper limit of \citetalias{Beifiori2009} 
for $i=81$\degree\ and adopted distance.
Col.(14): \mbh\ upper limit of \citetalias{Gultekin2009} and 
adopted distance.

{\sc References.} ---
(\citetalias{Tully1988})	   \citet{Tully1988},
(\citetalias{RC3})		   RC3,
(\citetalias{Tonry2001})           \citet{Tonry2001},
(\citetalias{Freedman2001})	   \citet{Freedman2001},
(5)                                this paper,
(\citetalias{Pizzella2005})	   \citet{Pizzella2005},
(\citetalias{Paturel2003a})        HyperLeda,
(\citetalias{Kronawitter2000})     \citet{Kronawitter2000},
(\citetalias{Ho2007})	           \citet{Ho2007},
(\citetalias{Binney1990})	          \citet{Binney1990},
(\citetalias{Silchenko2004})	   \citet{Silchenko2004},
(\citetalias{Hota2009})	           \citet{Hota2009},
(\citetalias{Quillen1994})         \citet{Quillen1994},
(\citetalias{Young2008})           \citet{Young2008},
(\citetalias{Morse1998})           \citet{Morse1998}.
\end{minipage}
\end{scriptsize}
\end{center}
\end{landscape}

\clearpage


\defcitealias{RC3}{1}
\defcitealias{Ho2007}{3}
\defcitealias{Samurovic2005}{4} 
\defcitealias{Baes2003}{5}
\defcitealias{Debattista2002}{6}
\defcitealias{Kronawitter2000}{7}
\defcitealias{Paturel2003a}{8}
\defcitealias{Bender1994}{9}       
\defcitealias{Coccato2009}{10}     

\begin{landscape}
\begin{center}
\begin{tiny}
\begin{longtable}{l l l r c c c c c c c c}
\caption{Properties of the galaxies of Sample B}\\
\hline
\hline
\noalign{\smallskip}
\multicolumn{1}{c}{Galaxy} &
\multicolumn{1}{c}{Morph. T.} &
\multicolumn{1}{c}{Spec. Cl.} &
\multicolumn{1}{c}{$D$} &
\multicolumn{1}{c}{$M_{V, T}^0$} &
\multicolumn{1}{c}{$M_{V, \rm bulge}^0$} &
\multicolumn{1}{c}{\sigmae} &
\multicolumn{1}{c}{$i$} &
\multicolumn{1}{c}{Ref.} &
\multicolumn{1}{c}{$V_{\rm c}$} &
\multicolumn{1}{c}{Ref.} &
\multicolumn{1}{c}{\mbh(low,high)} \\
\multicolumn{1}{c}{} &
\multicolumn{1}{c}{} &
\multicolumn{1}{c}{} &
\multicolumn{1}{c}{(Mpc)} &
\multicolumn{1}{c}{(mag)} &
\multicolumn{1}{c}{(mag)} &
\multicolumn{1}{c}{(km s$^{-1}$)} &
\multicolumn{1}{c}{(\degree)}  &
\multicolumn{1}{c}{} &
\multicolumn{1}{c}{(km s$^{-1}$)} &
\multicolumn{1}{c}{} &
\multicolumn{1}{c}{(\msun)}\\
\multicolumn{1}{c}{(1)} &
\multicolumn{1}{c}{(2)} &
\multicolumn{1}{c}{(3)} &
\multicolumn{1}{c}{(4)} &
\multicolumn{1}{c}{(5)} &
\multicolumn{1}{c}{(6)} &
\multicolumn{1}{c}{(7)} &
\multicolumn{1}{c}{(8)} &
\multicolumn{1}{c}{(9)} &
\multicolumn{1}{c}{(10)} &
\multicolumn{1}{c}{(11)} &
\multicolumn{1}{c}{(12)} \\
\noalign{\smallskip}
\hline
\noalign{\smallskip}
\endfirsthead
\multicolumn{12}{c}%
{\tablename\ \thetable{} -- Continued} \\
\hline
\hline
\noalign{\smallskip}
\multicolumn{1}{c}{Galaxy} &
\multicolumn{1}{c}{Morph. T.} &
\multicolumn{1}{c}{Spec. Cl.} &
\multicolumn{1}{c}{$D$} &
\multicolumn{1}{c}{$M_{V, T}^0$} &
\multicolumn{1}{c}{$M_{V, \rm bulge}^0$} &
\multicolumn{1}{c}{\sigmae} &
\multicolumn{1}{c}{$i$} &
\multicolumn{1}{c}{Ref.} &
\multicolumn{1}{c}{$V_{\rm c}$} &
\multicolumn{1}{c}{Ref.} &
\multicolumn{1}{c}{\mbh(low,high)} \\
\multicolumn{1}{c}{} &
\multicolumn{1}{c}{} &
\multicolumn{1}{c}{} &
\multicolumn{1}{c}{(Mpc)} &
\multicolumn{1}{c}{(mag)} &
\multicolumn{1}{c}{(mag)} &
\multicolumn{1}{c}{(km s$^{-1}$)} &
\multicolumn{1}{c}{(\degree)}  &
\multicolumn{1}{c}{} &
\multicolumn{1}{c}{(km s$^{-1}$)} &
\multicolumn{1}{c}{} &
\multicolumn{1}{c}{(\msun)}  \\
\multicolumn{1}{c}{(1)} &
\multicolumn{1}{c}{(2)} &
\multicolumn{1}{c}{(3)} &
\multicolumn{1}{c}{(4)} &
\multicolumn{1}{c}{(5)} &
\multicolumn{1}{c}{(6)} &
\multicolumn{1}{c}{(7)} &
\multicolumn{1}{c}{(8)} &
\multicolumn{1}{c}{(9)} &
\multicolumn{1}{c}{(10)} &
\multicolumn{1}{c}{(11)} &
\multicolumn{1}{c}{(12)} \\
\noalign{\smallskip}
\hline
\noalign{\smallskip}
\endhead
\noalign{\smallskip}
\hline
\endfoot
%
%
 Abell~1836-BCG & E   & ...     & 157.5 & $-23.31$ &$-23.31\pm0.15$ & $288\pm 14$ & ...& ...              &...              & ...                           & 3.9E9(3.3,4.3)\\
 Abell~3565-BCG & E   & ...     &  54.4 & $-23.27$ &$-23.27\pm0.15$ & $322\pm 16$ & ...& ...              &...              & ...                           & 5.2E8(4.4,6.0)\\
 Circinus     & Sb    & S2*     &   4.0 & $-17.36$ &...             & $158\pm 18$ & 69 & \citetalias{RC3} &$115.0\pm  5.9$  & \citetalias{Ho2007}           & 1.7E6(1.4,2.1)\\
 IC~1459      & E4    & L*      &  30.9 & $-22.57$ &$-22.57\pm0.15$ & $340\pm 17$ & 71 & \citetalias{RC3} &$278  \pm  27.8$*& \citetalias{Samurovic2005}    & 2.8E9(1.6,3.9)\\
 MW           & Sbc   &...      & 0.008 & ...      &...             & $105\pm 20$ & ...& ...              &$180.0\pm 20.0$  & \citetalias{Baes2003}         & 4.1E6(3.5,4.7)\\
 NGC~221      & E2    &...      &  0.86 & $-16.83$ &$-16.83\pm0.05$ & $ 75\pm  3$ & 57 & 2                &...              & ...                           & 3.1E6(2.5,3.7)\\
 NGC~224      & Sb    & L       &  0.80 & $-21.84$ &...             & $160\pm  8$ & 78 & \citetalias{RC3} &$244.2\pm  6.0$  & \citetalias{Ho2007}           & 1.5E8(1.2,2.4)\\
 NGC~821      & E4    & ...     &  25.5 & $-21.24$ &$-21.24\pm0.13$ & $209\pm 10$ & 73 & 2                &...              & ...                           & 4.2E7(3.4,7.0)\\
 NGC~1023     & SB0   & ...     &  12.1 & $-21.26$ &$-20.61\pm0.28$ & $205\pm 10$ & 88 & \citetalias{RC3} &$  270\pm   31$* & \citetalias{Debattista2002}   & 4.6E7(4.1,5.1)\\
 NGC~1068     & Sb    & S1.8    &  15.4 & $-22.17$ &...             & $151\pm  7$ & 10 & 2                &$219.2\pm  5.0$  & \citetalias{Ho2007}           & 8.6E6(8.3,8.9)\\
 NGC~1300     & SBbc  & ...     &  20.1 & $-21.34$ &...             & $218\pm 10$ & 50 & \citetalias{RC3} &$153.7\pm  1.8$  & \citetalias{Ho2007}           & 7.1E7(3.6,14)\\
 NGC~1399     & E1    & ...     &  21.1 & $-22.13$ &$-22.13\pm0.10$ & $337\pm 16$ & ...& ...              &$  424\pm   46$* & \citetalias{Kronawitter2000}  & 1.3E9(0.6,1.8)\\
 NGC~2748     & Sc    & H       &  24.9 & $-20.97$ &...             & $115\pm  5$ & 72 & \citetalias{RC3} &$128.4\pm  3.8$  & \citetalias{Paturel2003a}     & 4.7E7(0.9,8.5)\\
 NGC~2778     & E2    & ...     &  24.2 & $-19.62$ &$-19.62\pm0.13$ & $175\pm  8$ & ...& ...              &...              & ...                           & 1.6E7(0.6,2.5)\\
 NGC~2787     & SB0   & L1.9    &   7.9 & $-18.90$ &...             & $189\pm  9$ & 57 & \citetalias{RC3} &$191.2\pm 20.5$  & \citetalias{Ho2007}           & 4.3E7(3.8,4.7)\\
 NGC~3031     & Sb    & S1.5    &   4.1 & $-21.51$ &...             & $143\pm  7$ & 65 & 2                &$215.9\pm  7.0$  & \citetalias{Ho2007}           & 8.0E7(6.9,10.0)\\
 NGC~3115     & S0    & ...     &  10.2 & $-21.25$ &$-21.18\pm0.05$ & $230\pm 11$ & 88 & \citetalias{RC3} &$315  \pm  10$ * & \citetalias{Bender1994}       & 9.6E8(6.7,15)\\
 NGC~3227     & SBa   & S1.5    &  17.0 & $-20.73$ &...             & $133\pm 12$ & 51 & \citetalias{RC3} &$125.9\pm  4.8$  & \citetalias{Paturel2003a}     & 1.5E7(0.7,2)\\
 NGC~3245     & S0    & T2:     &  22.1 & $-20.96$ &...             & $205\pm 10$ & ...& ...              &...              & ...                           & 2.2E8(1.7,2.7)\\
 NGC~3377     & E6    & ...     &  11.7 & $-20.11$ &$-20.11\pm0.10$ & $145\pm  7$ & ...& ...              &...              & ...                           & 1.1E8(1.0,2.2)\\
 NGC~3379     & E0    & L2/T2:: &  11.7 & $-21.10$ &$-21.10\pm0.03$ & $206\pm 10$ & ...& ...              &$  259\pm   23$* & \citetalias{Kronawitter2000}  & 1.2E8(0.6,2)\\
 NGC~3384     & SB0   & ...     &  11.7 & $-20.50$ &$-19.93\pm0.22$ & $143\pm  7$ & 77 & 2                & ...  & ...     & 1.8E7(1.5,1.9)\\
 NGC~3585     & S0    & ...     &  21.2 & $-21.88$ &$-21.80\pm0.20$ & $213\pm 10$ & ...& ...              &...              & ...                           & 3.4E8(2.8,4.9)\\
 NGC~3607     & E1    & L2      &  19.9 & $-21.62$ &$-21.62\pm0.10$ & $229\pm 11$ & ...& ...              &...              & ...                           & 1.2E8(0.8,1.6)\\
 NGC~3608     & E1    & L2/S2:  &  23.0 & $-21.05$ &$-21.05\pm0.10$ &$182\pm  9$ & 57 & 2                &$191 \pm  43$ *    & \citetalias{Coccato2009}      & 2.1E8(1.4,3.2)\\
 NGC~3998     & S0    & L1.9    &  14.9 & $-20.32$ &...             & $305\pm 15$ & 46 & \citetalias{RC3} &$462.3\pm 26.9$  & \citetalias{Ho2007}           & 2.4E8(0.6,4.5)\\
 NGC~4026     & S0    & ...     &  15.6 & $-20.28$ &$-19.83\pm0.20$ & $180\pm  9$ & ...& ...              &...              & ...                           & 2.1E8(1.7,2.8)\\
 NGC~4258     & SABbc & S1.9    &   7.2 & $-21.31$ &...             & $115\pm 10$ & 71 & \citetalias{RC3} &$202.4\pm  5.2$  & \citetalias{Ho2007}           &3.78E7(3.77,3.79)\\
 NGC~4261     & E2    & L2      &  33.4 & $-22.72$ &$-22.72\pm0.06$ & $315\pm 15$ & ...& ...              &...              & ...                           & 5.5E8(4.3,6.6)\\
 NGC~4291     & E2    & ...     &  25.0 & $-20.67$ &$-20.67\pm0.13$ & $242\pm 12$ & ...& ...              &...              & ...                           & 3.2E8(0.8,4.1)\\
 NGC~4342     & S0    & ...     &  18.0 & $-18.84$ &...             & $225\pm 11$ & ...& ...              &...              & ...                           & 3.6E8(2.4,5.6)\\
 NGC~4374     & E1    & L2      &  17.0 & $-22.45$ &$-22.45\pm0.05$ & $296\pm 14$ & ...& ...              &$  410\pm   31$* & \citetalias{Kronawitter2000}  & 1.5E9(0.9,2.6)\\
 NGC~4459     & E2    & H/L     &  17.0 & $-21.06$ &$-21.06\pm0.04$ & $167\pm  8$ & ...& ...              &...              & ...                           & 7.4E7(6.0,8.8)\\
 NGC~4473     & E4    & ...     &  17.0 & $-21.14$ &$-21.14\pm0.04$ & $190\pm  9$ & ...& ...              &...              & ...                           & 1.3E8(0.4,1.8)\\
 NGC~4486     & E1    & L2      &  17.0 & $-22.92$ &$-22.92\pm0.04$ & $375\pm 18$ & ...& ...              &$  507\pm   38$* & \citetalias{Kronawitter2000}  & 3.6E9(2.6,4.6)\\
 NGC~4486A    & E2    & ...     &  17.0 & $-18.70$ &$-18.70\pm0.05$ & $111\pm  5$ & ...& ...              &...              & ...                           & 1.3E7(0.9,1.8)\\
 NGC~4564     & S0    & ...     &  17.0 & $-20.10$ &$-19.60\pm0.32$ & $162\pm  8$ & ...& ...              &...              & ...                           & 6.9E7(5.9,7.3)\\
 NGC~4594     & Sa    & L2      &  10.3 & $-22.52$ &$-22.44\pm0.15$ & $240\pm 12$ & 75 & \citetalias{RC3} &$389.7\pm  9.4$  & \citetalias{Ho2007}           & 5.7E8(1.7,11)\\
 NGC~4596     & SB0   & L2::    &  18.0 & $-20.70$ &...             & $136\pm  6$ & 54 & 2                &$154.9\pm  5.4$  & \citetalias{Paturel2003a}     & 8.4E7(5.9,12)\\
 NGC~4649     & E2    & ...     &  16.5 & $-22.65$ &$-22.65\pm0.05$ & $385\pm 19$ & ...& ...              &...              & ...                           & 2.1E9(1.5,2.6)\\
 NGC~4697     & E6    & ...     &  12.4 & $-21.29$ &$-21.29\pm0.11$ & $177\pm  8$ & ...& ...              &...              & ...                           & 2.0E8(1.8,2.2)\\
 NGC~5077     & E3    & L1.9    &  44.9 & $-22.04$ &$-22.04\pm0.13$ & $222\pm 11$ & ...& ...              &...              & ...                           & 8.0E8(4.7,13)\\
 NGC~5128     & S0/E  & S2      &   4.4 & $-21.82$ &$-21.82\pm0.08$ & $150\pm  7$ & 46 & \citetalias{RC3} &$325.3\pm 16.7$  & \citetalias{Paturel2003a}     & 7.0E7(3.2,8.3)\\
 NGC~5576     & E3    &  ...    &  27.1 & $-21.26$ &$-21.26\pm0.13$ & $183\pm  9$ & 81 & 2                &...              & ...                              & 1.8E8(1.4,2.1)\\
 NGC~5845     & E3    & ...     &  28.7 & $-19.77$ &$-19.77\pm0.13$ & $234\pm 11$ & ...& ...              &...              & ...                           & 2.9E8(1.2,3.4)\\
 NGC~6251     & E1    & S2      & 106.0 & ...      &...             & $290\pm 14$ & ...& ...              &...              & ...                           & 6.0E8(4.0,8.0)\\
 NGC~7052     & E3    & ...     &  70.9 & ...      &...             & $266\pm 13$ & 71 & \citetalias{RC3} &...              & ...                              & 4.0E8(2.4,6.8)\\
 NGC~7457     & S0    & ...     &  14.0 & $-19.80$ &$-18.72\pm0.11$ & $ 67\pm  3$ & ...& ...              &...              & ...                           & 4.1E6(2.4,5.3)\\
 NGC~7582     & SBab  & S2      &  22.3 & $-21.51$ &...             & $156\pm 19$ & 72 & \citetalias{RC3} &$121.4\pm  3.2$  & \citetalias{Paturel2003a}     & 5.5E7(4.4,7.1)\\

\label{tab:MBH_gultekin}

\end{longtable}
\end{tiny}

\begin{scriptsize}
\begin{minipage}{21cm}

{\sc Notes.} ---
Col.(1): Galaxy name.
Col.(2): Morphological type from \citetalias{Gultekin2009}.
Col.(3): Nuclear spectral class from \citet{Ho1997}, where H = \hiire\
nucleus, L = LINER, S = Seyfert , T = transition object
(LINER/\hiire), 1 = Seyfert 1, 2 = Seyfert 2, and a fractional number
between 1 and 2 denotes various intermediate types; uncertain and
highly uncertain classifications are followed by a single and double
colon, respectively. The nuclear spectral class of galaxies marked
with * is from NED.
Col.(4): Distance from \citetalias{Gultekin2009}.
Col.(5): Absolute corrected total $V$ magnitude from \citetalias{Gultekin2009}.
Col.(6): Absolute corrected $V$ magnitude of the bulge from \citetalias{Gultekin2009}.
Col.(7): Central velocity dispersion of the stellar component within
$r_{\rm e}$ from \citetalias{Gultekin2009}.
Col.(8): Inclination of the disc galaxies derived from either the
fitted disc axial ratio (from this paper) or the isophotal axial ratio
at a surface brightness of $\mu_B = 25$ \mas\ (from RC3) and the
intrinsic disc axial ratio (derived following \citet{Paturel1997}).
Col.(9): References for the apparent axial ratio adopted to calculate
the inclination given in Col.(8).
Col.(10): Circular velocity. The circular velocities were derived from
either \hire\ line-width measurements or gas resolved kinematics
available in literature corrected according to the adopted
inclination. Only the values marked with * are from dynamical models
of the stellar component.
Col.(11): References for Col.(10).
Col.(12): Mass (and confidence interval) of the SMBH derived from
modelling based on the resolved kinematics from
\citetalias{Gultekin2009}.
{\sc References.} ---
(\citetalias{RC3})                        RC3,
(2)                                              this paper,
(\citetalias{Ho2007})	          \citet{Ho2007},
(\citetalias{Samurovic2005})      \citet{Samurovic2005},
(\citetalias{Baes2003})               \citet{Baes2003},
(\citetalias{Debattista2002})      \citet{Debattista2002},
(\citetalias{Kronawitter2000})    \citet{Kronawitter2000},
(\citetalias{Paturel2003a})          \citet{Paturel2003a},
(\citetalias{Bender1994})            \citet{Bender1994},
(\citetalias{Coccato2009})         \citet{Coccato2009}.
\end{minipage}
\end{scriptsize}
\end{center}
\end{landscape}

\clearpage


\begin{landscape}
\begin{center}
\begin{tiny}
\begin{longtable}{l r r r c r r c r c}
\caption{Structural parameters of the galaxies from the
  azimuthally-averaged light profiles measured in $g$ and
  $i$-band SDSS images.}\\
\hline
\hline
\noalign{\smallskip}
\multicolumn{1}{c}{Galaxy} &
\multicolumn{1}{c}{$m_{\rm tot, g}$}&
\multicolumn{1}{c}{$m_{\rm tot, i}$}&
\multicolumn{1}{c}{PA} &
\multicolumn{1}{c}{$\epsilon$} &
\multicolumn{1}{c}{$m_{24.5}$} &
\multicolumn{1}{c}{$R_{24.5}$}&
\multicolumn{1}{c}{$\mu_{\rm e,gal}$}&
\multicolumn{1}{c}{\reg} &
\multicolumn{1}{c}{$C_{28}$}\\
\multicolumn{1}{c}{} &
\multicolumn{1}{c}{(mag)}&
\multicolumn{1}{c}{(mag)}&
\multicolumn{1}{c}{(\degree)} &
\multicolumn{1}{c}{} &
\multicolumn{1}{c}{(mag)} &
\multicolumn{1}{c}{(\farcsn)}&
\multicolumn{1}{c}{(\mas)}&
\multicolumn{1}{c}{(\farcsn) } &
\multicolumn{1}{c}{}\\
\multicolumn{1}{c}{(1)} &
\multicolumn{1}{c}{(2)} &
\multicolumn{1}{c}{(3)} &
\multicolumn{1}{c}{(4)} &
\multicolumn{1}{c}{(5)} &
\multicolumn{1}{c}{(6)} &
\multicolumn{1}{c}{(7)} &
\multicolumn{1}{c}{(8)} &
\multicolumn{1}{c}{(9)} &
\multicolumn{1}{c}{(10)} \\
\noalign{\smallskip}
\noalign{\smallskip}
\noalign{\smallskip}
\hline
\noalign{\smallskip}
\endfirsthead
\multicolumn{10}{c}%
{\tablename\ \thetable{} -- Continued} \\
\hline
\hline
\noalign{\smallskip}
\multicolumn{1}{c}{Galaxy} &
\multicolumn{1}{c}{$m_{\rm tot, g}$}&
\multicolumn{1}{c}{$m_{\rm tot, i}$}&
\multicolumn{1}{c}{PA} &
\multicolumn{1}{c}{$\epsilon$} &
\multicolumn{1}{c}{$m_{24.5}$} &
\multicolumn{1}{c}{$R_{24.5}$}&
\multicolumn{1}{c}{$\mu_{\rm e,gal}$}&
\multicolumn{1}{c}{\reg} &
\multicolumn{1}{c}{$C_{28}$}\\
\multicolumn{1}{c}{} &
\multicolumn{1}{c}{(mag)}&
\multicolumn{1}{c}{(mag)}&
\multicolumn{1}{c}{(\degree)} &
\multicolumn{1}{c}{} &
\multicolumn{1}{c}{(mag)} &
\multicolumn{1}{c}{(\farcsn)}&
\multicolumn{1}{c}{(\mas)}&
\multicolumn{1}{c}{(\farcsn) } &
\multicolumn{1}{c}{} \\
\multicolumn{1}{c}{(1)} &
\multicolumn{1}{c}{(2)} &
\multicolumn{1}{c}{(3)} &
\multicolumn{1}{c}{(4)} &
\multicolumn{1}{c}{(5)} &
\multicolumn{1}{c}{(6)} &
\multicolumn{1}{c}{(7)} &
\multicolumn{1}{c}{(8)} &
\multicolumn{1}{c}{(9)} &
\multicolumn{1}{c}{(10)} \\
\noalign{\smallskip}
\hline
\noalign{\smallskip}
\endhead
\hline
\endfoot
\multicolumn{10}{c}{Galaxies from Sample A: upper limit on \mbh\ from \citetalias{Beifiori2009}} \\
\noalign{\smallskip}
\hline
\noalign{\smallskip}
 NGC~ 741 & $ 11.35 \pm   0.12$ &$  10.55 \pm   0.11$&    8.11 &   0.25 &  10.74  &  142.60 &  22.84 &  64.76  &  5.33 \\
 NGC~1052 & $ 10.80 \pm   0.11$ &$   9.65 \pm   0.10$&  141.62 &   0.19 &   9.71  &  151.82 &  19.87 &  28.98  &  4.77 \\
 NGC~1667 & $ 12.23 \pm   0.12$ &$  11.69 \pm   0.12$&   67.00 &   0.26 &  11.71  &   59.57 &  20.49 &  18.22  &  3.15 \\
 NGC~2685 & $ 11.15 \pm   0.11$ &$  10.22 \pm   0.10$&    3.71 &   0.52 &  10.30  &  141.32 &  20.21 &  31.26  &  5.10 \\
 NGC~2892 & $ 12.77 \pm   0.13$ &$  11.71 \pm   0.12$&  164.31 &   0.18 &  11.82  &   72.27 &  21.79 &  21.44  &  5.28 \\
 NGC~2903 & $  8.92 \pm   0.09$ &$   8.08 \pm   0.08$&   54.50 &   0.46 &   8.10  &  308.12 &  20.17 &  94.85  &  1.89 \\
 NGC~2911 & $ 11.73 \pm   0.12$ &$  10.63 \pm   0.11$&  129.46 &   0.35 &  10.75  &  138.18 &  21.96 &  42.13  &  4.90 \\
 NGC~2964 & $ 11.30 \pm   0.11$ &$  10.50 \pm   0.11$&  162.05 &   0.48 &  10.53  &  102.10 &  20.01 &  31.16  &  2.94 \\
 NGC~3021 & $ 12.18 \pm   0.12$ &$  11.40 \pm   0.11$&  142.43 &   0.37 &  11.42  &   59.09 &  19.80 &  19.69  &  3.20 \\
 NGC~3351 & $ 10.06 \pm   0.10$ &$   9.03 \pm   0.09$&   64.60 &   0.10 &   9.06  &  199.27 &  20.79 &  62.43  &  4.16 \\
 NGC~3368 & $  9.54 \pm   0.10$ &$   8.47 \pm   0.09$&  104.57 &   0.19 &   8.51  &  239.62 &  19.94 &  57.12  &  4.30 \\
 NGC~3627 & $  8.76 \pm   0.09$ &$   7.98 \pm   0.08$&   93.08 &   0.52 &   8.01  &  359.49 &  20.21 &  79.79  &  3.56 \\
 NGC~3642 & $ 11.73 \pm   0.12$ &$  10.96 \pm   0.11$&  107.72 &   0.13 &  11.14  &   95.44 &  21.44 &  28.22  &  4.54 \\
 NGC~3675 & $ 10.06 \pm   0.10$ &$   8.97 \pm   0.09$&   79.91 &   0.39 &   9.04  &  239.51 &  20.45 &  73.02  &  3.54 \\
 NGC~3801 & $ 12.05 \pm   0.12$ &$  10.97 \pm   0.11$&  144.11 &   0.39 &  11.07  &  112.49 &  21.17 &  30.81  &  4.49 \\
 NGC~3862 & $ 12.17 \pm   0.13$ &$  11.14 \pm   0.11$&  121.80 &   0.34 &  11.40  &  116.84 &  22.73 &  42.87  &  5.94 \\
 NGC~3953 & $ 10.15 \pm   0.10$ &$   9.19 \pm   0.09$&   69.23 &   0.52 &   9.22  &  204.72 &  20.88 &  62.31  &  3.72 \\
 NGC~3982 & $ 11.75 \pm   0.12$ &$  11.05 \pm   0.11$&   89.38 &   0.13 &  11.07  &   65.24 &  19.93 &  18.09  &  2.33 \\
 NGC~3992 & $  9.83 \pm   0.10$ &$   8.98 \pm   0.09$&   17.60 &   0.51 &   9.02  &  239.99 &  21.29 &  90.88  &  2.80 \\
 NGC~4036 & $ 10.74 \pm   0.11$ &$   9.69 \pm   0.10$&    4.46 &   0.51 &   9.72  &  140.08 &  19.33 &  29.33  &  4.43 \\
 NGC~4088 & $ 10.15 \pm   0.10$ &$   9.47 \pm   0.09$&   34.26 &   0.64 &   9.49  &  182.01 &  20.54 &  68.32  &  2.52 \\
 NGC~4143 & $ 11.33 \pm   0.11$ &$  10.21 \pm   0.10$&  121.10 &   0.32 &  10.23  &   88.69 &  19.17 &  17.19  &  4.64 \\
 NGC~4150 & $ 11.85 \pm   0.12$ &$  10.86 \pm   0.11$&  123.60 &   0.33 &  10.89  &   78.23 &  20.00 &  17.19  &  4.78 \\
 NGC~4203 & $ 10.93 \pm   0.11$ &$   9.77 \pm   0.10$&   82.54 &   0.09 &   9.80  &  122.53 &  20.43 &  29.20  &  4.89 \\
 NGC~4212 & $ 11.19 \pm   0.11$ &$  10.38 \pm   0.10$&   13.46 &   0.34 &  10.40  &   99.19 &  20.38 &  39.46  &  2.68 \\
 NGC~4245 & $ 11.55 \pm   0.12$ &$  10.47 \pm   0.11$&   97.00 &   0.20 &  10.52  &  112.39 &  20.71 &  33.72  &  4.38 \\
 NGC~4278 & $ 10.52 \pm   0.11$ &$   9.40 \pm   0.09$&   53.47 &   0.09 &   9.45  &  156.47 &  19.90 &  26.21  &  4.89 \\
 NGC~4314 & $ 10.81 \pm   0.11$ &$   9.70 \pm   0.10$&  178.77 &   0.12 &   9.75  &  152.44 &  20.55 &  59.58  &  4.32 \\
 NGC~4321 & $  9.53 \pm   0.10$ &$   8.67 \pm   0.09$&   79.18 &   0.16 &   8.71  &  261.30 &  21.08 & 114.94  &  2.58 \\
 NGC~4335 & $ 12.32 \pm   0.13$ &$  11.27 \pm   0.11$&  120.07 &   0.22 &  11.37  &   83.81 &  20.75 &  17.63  &  5.32 \\
 NGC~4429 & $  9.95 \pm   0.10$ &$   8.93 \pm   0.09$&  178.12 &   0.56 &   8.96  &  231.68 &  20.19 &  57.68  &  4.24 \\
 NGC~4450 & $ 10.13 \pm   0.10$ &$   9.13 \pm   0.09$&   93.86 &   0.43 &   9.17  &  218.24 &  20.74 &  58.04  &  4.08 \\
 NGC~4477 & $ 10.73 \pm   0.11$ &$   9.58 \pm   0.10$&   19.31 &   0.14 &   9.61  &  142.18 &  20.51 &  34.81  &  4.39 \\
 NGC~4501 & $  9.81 \pm   0.10$ &$   9.05 \pm   0.09$&  131.35 &   0.48 &   9.07  &  236.04 &  20.60 &  73.04  &  3.23 \\
 NGC~4526 & $  9.46 \pm   0.10$ &$   8.50 \pm   0.09$&  159.75 &   0.70 &   8.54  &  337.08 &  19.75 &  45.57  &  5.23 \\
 NGC~4548 & $ 10.32 \pm   0.10$ &$   9.20 \pm   0.09$&  123.32 &   0.20 &   9.28  &  217.25 &  21.52 &  80.08  &  3.82 \\
 NGC~4552 & $ 10.15 \pm   0.10$ &$   9.02 \pm   0.09$&  142.26 &   0.13 &   9.07  &  202.90 &  20.07 &  34.11  &  5.37 \\
 NGC~4579 & $  9.64 \pm   0.10$ &$   8.61 \pm   0.09$&  177.58 &   0.25 &   8.68  &  251.94 &  20.76 &  77.27  &  4.23 \\
 NGC~4636 & $  9.73 \pm   0.10$ &$   8.63 \pm   0.09$&  144.03 &   0.20 &   8.68  &  264.19 &  20.94 &  72.16  &  4.43 \\
 NGC~4698 & $ 10.94 \pm   0.11$ &$   9.84 \pm   0.10$&  111.10 &   0.11 &   9.89  &  134.68 &  20.31 &  35.10  &  4.47 \\
 NGC~4736 & $  8.24 \pm   0.09$ &$   7.55 \pm   0.08$&    2.81 &   0.47 &   7.79  &  383.92 &  20.04 &  63.14  &  6.64 \\
 NGC~4800 & $ 11.68 \pm   0.12$ &$  10.70 \pm   0.11$&   71.94 &   0.07 &  10.73  &   75.43 &  19.81 &  20.53  &  3.63 \\
 NGC~4826 & $  8.84 \pm   0.09$ &$   8.12 \pm   0.08$&  159.61 &   0.49 &   8.13  &  299.77 &  19.93 &  77.97  &  3.23 \\
 NGC~5005 & $  9.74 \pm   0.10$ &$   8.78 \pm   0.09$&   32.04 &   0.53 &   8.81  &  207.82 &  19.73 &  51.96  &  3.97 \\
 NGC~5127 & $ 12.56 \pm   0.13$ &$  11.46 \pm   0.12$&   25.55 &   0.29 &  11.58  &   82.72 &  21.41 &  23.31  &  4.63 \\
 NGC~5194 & $  8.67 \pm   0.09$ &$   8.11 \pm   0.08$&   84.80 &   0.36 &   8.12  &  328.36 &  20.91 & 115.51  &  3.18 \\
 NGC~5248 & $ 10.51 \pm   0.11$ &$   9.56 \pm   0.10$&  138.61 &   0.16 &   9.60  &  165.17 &  20.72 &  66.22  &  3.58 \\
 NGC~5252 & $ 12.83 \pm   0.13$ &$  11.93 \pm   0.12$&   83.50 &   0.45 &  12.04  &   68.31 &  20.83 &  17.92  &  4.64 \\
 NGC~5347 & $ 12.87 \pm   0.13$ &$  11.91 \pm   0.12$&  165.10 &   0.07 &  11.98  &   59.64 &  21.34 &  30.66  &  3.46 \\
 NGC~5490 & $ 12.27 \pm   0.12$ &$  11.15 \pm   0.11$&   92.15 &   0.22 &  11.23  &   84.70 &  20.67 &  16.69  &  5.50 \\
 NGC~5695 & $ 12.99 \pm   0.13$ &$  11.95 \pm   0.12$&  151.27 &   0.26 &  12.00  &   57.14 &  20.48 &  14.98  &  4.07 \\
 NGC~5879 & $ 11.30 \pm   0.11$ &$  10.62 \pm   0.11$&  122.79 &   0.61 &  10.67  &  125.47 &  20.36 &  26.91  &  4.09 \\
 NGC~5921 & $ 11.03 \pm   0.11$ &$  10.27 \pm   0.10$&  147.44 &   0.25 &  10.44  &  160.99 &  22.27 &  61.04  &  3.86 \\
 NGC~7331 & $  9.61 \pm   0.10$ &$   8.75 \pm   0.09$&  161.29 &   0.58 &   8.80  &  285.70 &  20.30 &  67.00  &  4.47 \\
 NGC~7626 & $ 11.56 \pm   0.12$ &$  10.69 \pm   0.11$&    5.53 &   0.14 &  10.72  &  104.39 &  21.05 &  27.18  &  4.52 \\
 UGC~1395 & $ 13.66 \pm   0.14$ &$  13.01 \pm   0.13$&  112.82 &   0.08 &  13.05  &   38.71 &  21.56 &  18.01  &  3.15 \\
 UGC~7115 & $ 13.15 \pm   0.14$ &$  12.04 \pm   0.12$&   96.41 &   0.28 &  12.28  &   66.21 &  21.95 &  17.77  &  6.33 \\
\noalign{\smallskip}
\hline
\noalign{\smallskip}
\multicolumn{10}{c}{Galaxies from Sample A: upper limit on \mbh\ from \citetalias{Gultekin2009}} \\
\noalign{\smallskip}
\hline
\noalign{\smallskip}
Abell~2052  & $ 12.47 \pm 0.13$ & $ 11.58 \pm 0.12$&   53.05 &   0.36 &  11.74& 104.37&   23.17&   52.44 &   4.37 \\
NGC~3310    & $ 10.83 \pm 0.11$ & $ 10.48 \pm 0.11$&   78.70 &   0.04 &  10.56& 107.72&   19.66&   16.30 &   4.16 \\
NGC~4041    & $ 11.42 \pm 0.12$ & $ 10.57 \pm 0.11$&   12.03 &   0.13 &  10.63& 100.88&   20.47&   21.22 &   3.90 \\
NGC~4435    & $ 11.01 \pm 0.11$ & $  9.93 \pm 0.10$&   76.36 &   0.24 &   9.95& 111.64&   19.48&   23.96 &   4.73 \\
NGC~4486B   & $ 13.74 \pm 0.14$ & $ 12.75 \pm 0.13$&  175.94 &   0.29 &  12.79&  28.21&   18.17&    2.99 &   4.47 \\

\noalign{\smallskip}
\hline
\noalign{\smallskip}
\multicolumn{10}{c}{Galaxies from Sample B} \\
\noalign{\smallskip}
\hline
\noalign{\smallskip}
 Abell~1836& $ 13.24\pm 0.14$  & $  12.03\pm  0.12$  & 107.86 &   0.22&   12.17 &  63.47 &  21.57 &  17.61 &   4.80   \\
 NGC~0821  & $ 11.10\pm 0.11$  & $  10.33\pm  0.10$  & 169.09 &   0.28&   10.39 & 139.90 &  21.15 &  35.01 &   5.03   \\
 NGC~1068  & $  8.97\pm 0.09$  & $   8.12\pm  0.08$  &   5.51 &   0.30&    8.17 & 329.45 &  19.67 &  44.48 &   5.63   \\
 NGC~2778  & $ 12.85\pm 0.13$  & $  11.74\pm  0.12$  &  13.21 &   0.19&   11.77 &  53.45 &  20.19 &  12.02 &   4.57   \\
 NGC~3031  & $  7.76\pm 0.08$  & $   6.76\pm  0.07$  &  86.85 &   0.42&    6.79 & 396.00 &  19.40 &  94.61 &   4.09   \\
 NGC~3227  & $ 10.78\pm 0.11$  & $   9.84\pm  0.10$  & 106.11 &   0.47&    9.91 & 162.41 &  20.77 &  59.08 &   3.30   \\
 NGC~3245  & $ 11.03\pm 0.11$  & $   9.97\pm  0.10$  &  78.83 &   0.38&   10.05 & 127.01 &  19.95 &  26.68 &   5.15   \\
 NGC~3377  & $ 10.24\pm 0.11$  & $   9.34\pm  0.10$  &   2.80 &   0.11&    9.51 & 201.63 &  21.23 &  59.52 &   5.00   \\
 NGC~3379  & $  9.67\pm 0.10$  & $   8.49\pm  0.09$  &   4.39 &   0.11&    8.54 & 246.37 &  20.00 &  56.99 &   5.26   \\
 NGC~3384  & $  9.80\pm 0.10$  & $   8.85\pm  0.09$  &  28.09 &   0.56&    8.88 & 227.66 &  20.14 &  37.69 &   5.74   \\
 NGC~3607  & $ 10.32\pm 0.10$  & $   9.18\pm  0.09$  & 136.66 &   0.10&    9.24 & 183.12 &  20.22 &  36.42 &   4.79   \\
 NGC~3608  & $ 11.23\pm 0.11$  & $  10.13\pm  0.10$  &  15.55 &   0.14&   10.19 & 123.25 &  20.48 &  28.42 &   4.87   \\
 NGC~3998  & $ 10.81\pm 0.11$  & $   9.68\pm  0.10$  & 129.80 &   0.19&    9.73 & 128.72 &  19.67 &  20.40 &   5.47   \\
 NGC~4026  & $ 10.65\pm 0.11$  & $   9.69\pm  0.10$  &  90.02 &   0.63&    9.70 & 138.74 &  18.88 &  21.11 &   5.50   \\
 NGC~4258  & $  8.63\pm 0.09$  & $   7.94\pm  0.08$  & 118.60 &   0.45&    7.97 & 517.97 &  20.97 & 135.85 &   4.13   \\
 NGC~4261  & $ 10.18\pm 0.11$  & $   9.15\pm  0.09$  &  99.42 &   0.32&    9.36 & 236.04 &  21.38 &  56.63 &   5.78   \\
 NGC~4342  & $ 12.51\pm 0.13$  & $  11.26\pm  0.11$  & 102.67 &   0.26&   11.34 &  57.41 &  18.06 &   7.59 &   5.91   \\
 NGC~4374  & $  9.56\pm 0.10$  & $   8.37\pm  0.08$  & 175.20 &   0.06&    8.42 & 258.44 &  20.38 &  53.79 &   4.89   \\
 NGC~4459  & $ 10.45\pm 0.11$  & $   9.23\pm  0.09$  & 176.05 &   0.25&    9.34 & 200.03 &  20.66 &  40.99 &   5.24   \\
 NGC~4473  & $ 10.34\pm 0.10$  & $   9.27\pm  0.09$  & 178.13 &   0.35&    9.33 & 192.68 &  20.00 &  34.79 &   5.49   \\
 NGC~4486  & $  9.21\pm 0.09$  & $   8.03\pm  0.08$  & 139.27 &   0.16&    8.06 & 301.70 &  20.20 &  68.38 &   4.46   \\
NGC~4486A  & $ 11.11\pm 0.12$  & $  11.20\pm  0.11$  & 170.01 &   0.34&   11.04 & 111.52 &  18.32 &   5.82 &   4.44   \\
 NGC~4564  & $ 11.21\pm 0.11$  & $  10.15\pm  0.10$  &  41.86 &   0.41&   10.18 & 113.23 &  19.45 &  22.41 &   5.14   \\
 NGC~4596  & $ 10.63\pm 0.11$  & $   9.56\pm  0.10$  & 161.17 &   0.24&    9.61 & 164.73 &  20.73 &  56.20 &   4.50   \\
 NGC~4649  & $  9.18\pm 0.09$  & $   7.96\pm  0.08$  & 167.95 &   0.21&    8.01 & 322.54 &  20.14 &  67.47 &   4.80   \\
 NGC~4697  & $  9.64\pm 0.10$  & $   8.53\pm  0.09$  & 119.54 &   0.25&    8.57 & 253.13 &  20.24 &  59.72 &   4.36   \\
 NGC~5576  & $ 10.61\pm 0.11$  & $   9.52\pm  0.10$  &   6.18 &   0.46&    9.76 & 202.00 &  21.43 &  49.36 &   6.82   \\
 NGC~5845  & $ 12.64\pm 0.13$  & $  11.45\pm  0.11$  & 140.39 &   0.22&   11.48 &  50.59 &  17.68 &   4.34 &   4.91   \\
\label{tab:MBH_photometry_i}
\end{longtable}
\end{tiny}

\begin{scriptsize}
\begin{minipage}{22cm}
{\sc Notes.} ---
Col.(1): Galaxy name.
Col.(2): Total extrapolated $g$-band magnitude of the galaxy.
Col.(3): Total extrapolated $i$-band magnitude of the galaxy.
Col.(4): Position angle of the isophotes.
Col.(5): Ellipticity of the isophotes.
Col.(6): Integrated $i$-band magnitude within the isophote at a
surface brightness level of 24.5 \mas\ corrected for Galactic
extinction, K-correction, and internal extinction following
\citep{Shao2007}.
Col.(7): Radius of isophote at a surface brightness level of 
$\mu_i = 24.5$ \mas .
Col.(8): Effective surface brightness of the galaxy.
Col.(9): Effective radius of the galaxy.
Col.(10): Concentration index $C_{28}=5 \log(r_{80}/r_{20})$ where
$r_{80}$ and $r_{20}$ are the radii that enclose 80\% and 20\% of
total light, respectively.

\end{minipage}
\end{scriptsize}
\end{center}
\end{landscape}

\clearpage


\begin{landscape}
\begin{center}
\begin{tiny}
\begin{longtable}{l l l c r c c r c r c r c r}
\caption{Structural properties of the sample galaxies from
  two-dimensional decomposition of the $i$-band SDSS images}\\
\hline
\hline
\noalign{\smallskip}
\multicolumn{1}{c}{Galaxy}    &
\multicolumn{1}{c}{2D Fit}    &
\multicolumn{1}{c}{Comments}  &
\multicolumn{1}{c}{$\mu_{\rm e}$}  &
\multicolumn{1}{c}{$r_{\rm e}$}    &
\multicolumn{1}{c}{$n$}           &
\multicolumn{1}{c}{$q_{\rm b}$}    &
\multicolumn{1}{c}{PA$_{\rm b}$}   &
\multicolumn{1}{c}{$\mu_0$}        &
\multicolumn{1}{c}{$h$}           &
\multicolumn{1}{c}{$q_{\rm d}$}    &
\multicolumn{1}{c}{PA$_{\rm d}$}   &
\multicolumn{1}{c}{$B/T$}         &
\multicolumn{1}{c}{$m_{i, {\rm bulge}}$}\\
\multicolumn{1}{c}{}         &
\multicolumn{1}{c}{}         &
\multicolumn{1}{c}{}         &
\multicolumn{1}{c}{(\mas) }  &
\multicolumn{1}{c}{(\farcsn)} &
\multicolumn{1}{c}{}         &
\multicolumn{1}{c}{}         &
\multicolumn{1}{c}{(\degree)}  &
\multicolumn{1}{c}{(\mas) }  &
\multicolumn{1}{c}{(\farcsn)} &
\multicolumn{1}{c}{}         &
\multicolumn{1}{c}{(\degree)}  &
\multicolumn{1}{c}{}         &
\multicolumn{1}{c}{(mag)}        \\
\multicolumn{1}{c}{(1)} &
\multicolumn{1}{c}{(2)} &
\multicolumn{1}{c}{(3)} &
\multicolumn{1}{c}{(4)} &
\multicolumn{1}{c}{(5)} &
\multicolumn{1}{c}{(6)} &
\multicolumn{1}{c}{(7)} &
\multicolumn{1}{c}{(8)} &
\multicolumn{1}{c}{(9)} &
\multicolumn{1}{c}{(10)} &
\multicolumn{1}{c}{(11)} &
\multicolumn{1}{c}{(12)} &
\multicolumn{1}{c}{(13)} &
\multicolumn{1}{c}{(14)} \\

\noalign{\smallskip}
\hline
\noalign{\smallskip}
\endfirsthead
\multicolumn{14}{c}%
{\tablename\ \thetable{} -- Continued} \\
\hline
\hline
\noalign{\smallskip}
\multicolumn{1}{c}{Galaxy}    &
\multicolumn{1}{c}{2D Fit}    &
\multicolumn{1}{c}{Comments}  &
\multicolumn{1}{c}{$\mu_{\rm e}$}  &
\multicolumn{1}{c}{$r_{\rm e}$}    &
\multicolumn{1}{c}{$n$}           &
\multicolumn{1}{c}{$q_{\rm b}$}    &
\multicolumn{1}{c}{PA$_{\rm b}$}   &
\multicolumn{1}{c}{$\mu_0$}        &
\multicolumn{1}{c}{$h$}           &
\multicolumn{1}{c}{$q_{\rm d}$}    &
\multicolumn{1}{c}{PA$_{\rm d}$}   &
\multicolumn{1}{c}{$B/T$}         &
\multicolumn{1}{c}{$m_{i, {\rm bulge}}$}\\
\multicolumn{1}{c}{}         &
\multicolumn{1}{c}{}         &
\multicolumn{1}{c}{}         &
\multicolumn{1}{c}{(\mas) }  &
\multicolumn{1}{c}{(\farcsn)} &
\multicolumn{1}{c}{}         &
\multicolumn{1}{c}{}         &
\multicolumn{1}{c}{(\degree)}  &
\multicolumn{1}{c}{(\mas) }  &
\multicolumn{1}{c}{(\farcsn)} &
\multicolumn{1}{c}{}         &
\multicolumn{1}{c}{(\degree)}  &
\multicolumn{1}{c}{}         &
\multicolumn{1}{c}{(mag)}        \\
\multicolumn{1}{c}{(1)} &
\multicolumn{1}{c}{(2)} &
\multicolumn{1}{c}{(3)} &
\multicolumn{1}{c}{(4)} &
\multicolumn{1}{c}{(5)} &
\multicolumn{1}{c}{(6)} &
\multicolumn{1}{c}{(7)} &
\multicolumn{1}{c}{(8)} &
\multicolumn{1}{c}{(9)} &
\multicolumn{1}{c}{(10)} &
\multicolumn{1}{c}{(11)} &
\multicolumn{1}{c}{(12)} &
\multicolumn{1}{c}{(13)} &
\multicolumn{1}{c}{(14)} \\
\noalign{\smallskip}
\hline
\noalign{\smallskip}
\endhead
\hline
\endfoot
\multicolumn{12}{c}{Galaxies from Sample A: upper limits on \mbh\ from
\citetalias{Beifiori2009}} \\
\noalign{\smallskip}
\hline
\noalign{\smallskip}
NGC~741  & yes & $r^{1/n}$         & $22.56\pm 0.07$ & $61.75\pm 2.15$ & $4.50\pm 0.09$ & $0.81\pm 0.01$ & $177.00\pm  0.02$ & $  ...        $ & $   ...        $ &  $ ...         $ & $   ...       $ & $    1.00    $ & $10.38\pm 0.03$ \\
NGC~1052 & no  & bad              & $  ...         $ & $  ...       $  & $  ...       $ & $  ...       $ & $  ...        $   & $  ...        $ & $  ...        $ &  $  ...        $ & $  ...        $ &  $  ...      $ &  $  ...      $  \\
NGC~1667 & no  & bad              & $  ...         $ & $  ...       $  & $  ...       $ & $  ...       $ & $  ...        $   & $  ...        $ & $  ...        $ &  $  ...        $ & $  ...        $ &  $  ...      $ &  $  ...      $  \\
NGC~2685 & no  & bad              & $  ...         $ & $  ...       $  & $  ...       $ & $  ...       $ & $  ...        $   & $  ...        $ & $  ...        $ &  $  ...        $ & $  ...        $ &  $  ...      $ &  $  ...      $  \\
NGC~2892 & yes & $r^{1/n}+$exp.   &  $20.07\pm 0.34$ & $ 7.04\pm 1.86$ & $3.46\pm 0.41$ & $0.97\pm 0.01$ & $101.84\pm 25.04$ & $20.81\pm 0.42$ & $20.79\pm 9.29 $ & $0.95\pm 0.09$  & $ 51.86\pm 0.55$ & $0.44\pm 0.18$ & $12.56\pm 0.28$ \\
NGC~2903 & no  & bad              & $  ...         $ & $  ...       $  & $  ...       $ & $  ...       $ & $  ...        $   & $  ...        $ & $  ...        $ &  $  ...        $ & $  ...        $ &  $  ...      $ &  $  ...      $  \\
NGC~2911 & yes & $r^{1/n}+$exp.   &  $20.82\pm 0.13$ & $16.73\pm 1.44$ & $3.66\pm 0.20$ & $0.74\pm 0.01$ & $ 56.01\pm  4.87$ & $21.53\pm 0.21$ & $60.99\pm 12.04$ & $0.62\pm 0.02$  & $ 48.66\pm 0.14$ & $0.37\pm 0.05$ & $11.69\pm 0.07$ \\
NGC~2964 & yes & $r^{1/n}+$exp.   &  $17.94\pm 0.39$ & $ 1.44\pm 0.47$ & $2.68\pm 0.34$ & $0.85\pm 0.02$ & $  1.34\pm  0.14$ & $18.86\pm 0.60$ & $18.73\pm 14.78$ & $0.75\pm 0.22$  & $ 30.87\pm 1.42$ & $0.04\pm 0.01$ & $14.16\pm 0.61$ \\
NGC~3021 & no  & bar              & $  ...         $ & $  ...       $  & $  ...       $ & $  ...       $ & $  ...        $   & $  ...        $ & $  ...        $ &  $  ...        $ & $  ...        $ &  $  ...      $ &  $  ...      $  \\
NGC~3351 & no  & bar              & $  ...         $ & $  ...       $  & $  ...       $ & $  ...       $ & $  ...        $   & $  ...        $ & $  ...        $ &  $  ...        $ & $  ...        $ &  $  ...      $ &  $  ...      $  \\
NGC~3368 & no  & bar              & $  ...         $ & $  ...       $  & $  ...       $ & $  ...       $ & $  ...        $   & $  ...        $ & $  ...        $ &  $  ...        $ & $  ...        $ &  $  ...      $ &  $  ...      $  \\
NGC~3627 & yes & $r^{1/n}+$exp.   &  $21.30\pm 0.02$ & $98.00\pm 0.96$ & $4.56\pm 0.01$ & $0.45\pm 0.01$ & $ 69.30\pm  0.45$ & $19.61\pm 0.02$ & $78.64\pm  0.32$ & $0.44\pm 0.01$  & $100.77\pm 0.02$ & $0.56\pm 0.01$ & $ 7.40\pm 0.01$ \\
NGC~3642 & no  & embedded disc    & $  ...         $ & $  ...       $  & $  ...       $ & $  ...       $ & $  ...        $   & $  ...        $ & $  ...        $ &  $  ...        $ & $  ...        $ &  $  ...      $ &  $  ...      $  \\
NGC~3675 & yes & $r^{1/n}+$exp.   &  $20.96\pm 0.15$ & $30.13\pm 3.90$ & $4.73\pm 0.23$ & $0.49\pm 0.01$ & $100.94\pm 10.52$ & $19.36\pm 0.29$ & $48.77\pm 12.75$ & $0.47\pm 0.02$  & $ 98.86\pm 0.23$ & $0.26\pm 0.03$ & $10.85\pm 0.14$ \\
NGC~3801 & no  & bad             &  $  ...         $ & $  ...       $  & $  ...       $ & $  ...       $ & $  ...        $   & $  ...        $ & $  ...        $ &  $  ...        $ & $  ...        $ &  $  ...      $ &  $  ...      $  \\
NGC~3862 & yes & $r^{1/n}+$exp.   &  $24.13\pm 0.16$ & $75.40\pm 4.88$ & $7.90\pm 0.30$ & $1.00\pm 0.01$ & $ 78.66\pm  0.02$ & $ ...         $ & $ ...          $ & $...         $  & $ ...          $ & $1.00        $ & $10.99\pm 0.04$ \\
NGC~3953 & no  & bar             &  $  ...         $ & $  ...       $  & $  ...       $ & $  ...       $ & $  ...        $   & $  ...        $ & $  ...        $ &  $  ...        $ & $  ...        $ &  $  ...      $ &  $  ...      $  \\
NGC~3982 & yes & $r^{1/n}+$exp.   &  $17.87\pm 0.39$ & $ 1.58\pm 0.51$ & $0.80\pm 0.10$ & $0.79\pm 0.01$ & $106.83\pm 11.36$ & $18.38\pm 0.58$ & $12.00\pm  9.47$ & $0.87\pm 0.26$  & $108.84\pm 5.00$ & $0.04\pm 0.01$ & $14.55\pm 0.62$ \\
NGC~3992 & yes & $r^{1/n}+$exp.   &  $21.44\pm 0.16$ & $31.64\pm 4.10$ & $5.33\pm 0.26$ & $0.67\pm 0.01$ & $149.86\pm 15.61$ & $20.15\pm 0.30$ & $80.39\pm 21.02$ & $0.54\pm 0.02$  & $155.01\pm 0.35$ & $0.20\pm 0.02$ & $10.83\pm 0.14$ \\
NGC~4036 & yes & $r^{1/n}+$exp.   &  $17.97\pm 0.12$ & $ 4.05\pm 0.35$ & $1.68\pm 0.09$ & $0.88\pm 0.01$ & $153.72\pm 13.35$ & $17.83\pm 0.17$ & $19.91\pm  3.93$ & $0.48\pm 0.01$  & $175.07\pm 0.50$ & $0.17\pm 0.02$ & $12.12\pm 0.08$ \\
NGC~4088 & no  & Freeman II disc &  $  ...         $ & $  ...       $  & $  ...       $ & $  ...       $ & $  ...        $   & $  ...        $ & $  ...        $ &  $  ...        $ & $  ...        $ &  $  ...      $ &  $  ...      $  \\
NGC~4143 & yes & $r^{1/n}+$exp.  &   $16.79\pm 0.11$ & $ 3.07\pm 0.26$ & $1.05\pm 0.06$ & $0.78\pm 0.01$ & $ 59.33\pm  5.15$ & $17.84\pm 0.17$ & $14.27\pm  2.82$ & $0.58\pm 0.02$  & $ 58.73\pm 0.17$ & $0.24\pm 0.03$ & $11.91\pm 0.08$ \\
NGC~4150 & yes & $r^{1/n}+$exp.  &   $17.27\pm 0.30$ & $ 2.32\pm 0.61$ & $1.68\pm 0.20$ & $0.76\pm 0.01$ & $ 53.89\pm 13.25$ & $18.83\pm 0.38$ & $14.58\pm  6.52$ & $0.65\pm 0.06$  & $ 57.08\pm 0.60$ & $0.23\pm 0.09$ & $12.80\pm 0.28$ \\
NGC~4203 & yes & $r^{1/n}+$exp.  &   $18.49\pm 0.13$ & $ 7.99\pm 1.03$ & $2.62\pm 0.13$ & $0.87\pm 0.01$ & $ 99.39\pm 10.35$ & $19.92\pm 0.30$ & $34.02\pm  8.89$ & $0.91\pm 0.04$  & $102.13\pm 0.23$ & $0.37\pm 0.04$ & $10.95\pm 0.14$ \\
NGC~4212 & no  & Freeman II disc &  $  ...         $ & $  ...       $  & $  ...       $ & $  ...       $ & $  ...        $   & $  ...        $ & $  ...        $ &  $  ...        $ & $  ...        $ &  $  ...      $ &  $  ...      $  \\
NGC~4245 & yes & $r^{1/n}+$exp.  &   $18.92\pm 0.12$ & $ 5.79\pm 0.50$ & $2.00\pm 0.11$ & $0.83\pm 0.01$ & $ 60.60\pm  5.26$ & $19.49\pm 0.19$ & $24.55\pm  4.84$ & $0.64\pm 0.02$  & $ 52.31\pm 0.15$ & $0.24\pm 0.03$ & $12.27\pm 0.08$ \\
NGC~4278 & yes & $r^{1/n}$       &   $20.11\pm 0.07$ & $28.34\pm 0.99$ & $3.70\pm 0.07$ & $0.87\pm 0.01$ & $107.34\pm  0.01$ & $ ...         $ & $  ...         $ & $...         $  & $...           $ & $1.00        $ & $ 9.65\pm 0.03$ \\
NGC~4314 & yes & $r^{1/n}+$exp.  &   $19.10\pm 0.10$ & $16.24\pm 1.09$ & $2.90\pm 0.10$ & $0.86\pm 0.01$ & $105.17\pm  4.12$ & $20.88\pm 0.12$ & $48.11\pm  2.07$ & $0.76\pm 0.01$  & $170.28\pm 0.11$ & $0.67\pm 0.03$ & $ 9.99\pm 0.09$ \\
NGC~4321 & no  & bar             &  $  ...         $ & $  ...       $  & $  ...       $ & $  ...       $ & $  ...        $   & $  ...        $ & $  ...        $ &  $  ...        $ & $  ...        $ &  $  ...      $ &  $  ...      $  \\
NGC~4335 & yes & $r^{1/n}$        &  $21.06\pm 0.14$ & $19.34\pm 1.25$ & $5.10\pm 0.19$ & $0.77\pm 0.01$ & $ 64.27\pm  0.02$ & $ ...         $ & $ ...          $ & $...         $  & $ ...          $ & $1.00        $ & $11.40\pm 0.04$ \\
NGC~4429 & yes & $r^{1/n}+$exp.  &   $20.37\pm 0.10$ & $29.74\pm 2.00$ & $3.31\pm 0.11$ & $0.65\pm 0.01$ & $  2.18\pm  0.09$ & $19.31\pm 0.11$ & $52.21\pm  2.25$ & $0.44\pm 0.01$  & $  8.23\pm 0.01$ & $0.44\pm 0.02$ & $10.17\pm 0.09$ \\
NGC~4450 & yes & $r^{1/n}+$exp.  &   $20.27\pm 0.15$ & $18.05\pm 2.34$ & $3.98\pm 0.20$ & $0.77\pm 0.01$ & $ 92.95\pm  9.68$ & $19.35\pm 0.29$ & $41.51\pm 10.85$ & $0.53\pm 0.02$  & $ 90.80\pm 0.21$ & $0.30\pm 0.03$ & $10.89\pm 0.14$ \\
NGC~4477 & yes & $r^{1/n}+$exp.  &   $19.22\pm 0.14$ & $14.52\pm 1.88$ & $2.62\pm 0.13$ & $0.75\pm 0.01$ & $105.63\pm 11.00$ & $20.22\pm 0.30$ & $42.62\pm 11.14$ & $0.85\pm 0.03$  & $164.34\pm 0.37$ & $0.43\pm 0.04$ & $10.56\pm 0.14$ \\
NGC~4501 & no  & Freeman II disc &   $  ...        $ & $  ...       $  & $  ...       $ & $  ...       $ & $  ...        $   & $  ...        $ & $  ...        $ &  $  ...        $ & $  ...        $ &  $  ...      $ &  $  ...      $  \\
NGC~4526 & yes & $r^{1/n}+$exp.  &   $18.14\pm 0.09$ & $11.92\pm 0.80$ & $1.90\pm 0.06$ & $0.68\pm 0.01$ & $ 23.35\pm  0.91$ & $19.00\pm 0.11$ & $53.19\pm  2.29$ & $0.43\pm 0.01$  & $ 22.59\pm 0.01$ & $0.37\pm 0.02$ & $10.17\pm 0.09$ \\
NGC~4548 & yes & $r^{1/n}+$exp.  &   $21.08\pm 0.15$ & $24.98\pm 3.23$ & $4.47\pm 0.22$ & $0.85\pm 0.01$ & $174.33\pm 18.16$ & $20.12\pm 0.30$ & $51.94\pm 13.58$ & $0.81\pm 0.03$  & $167.78\pm 0.38$ & $0.28\pm 0.03$ & $10.83\pm 0.14$ \\
NGC~4552 & yes & $r^{1/n}$       &   $20.61\pm 0.07$ & $45.24\pm 1.57$ & $4.30\pm 0.08$ & $0.94\pm 0.01$ & $ 33.69\pm  0.01$ & $ ...         $ & $ ...          $ & $...         $  & $ ...          $ & $1.00        $ & $ 8.98\pm 0.03$ \\
NGC~4579 & yes & $r^{1/n}+$exp.  &   $20.77\pm 0.10$ & $39.13\pm 2.63$ & $5.24\pm 0.18$ & $0.70\pm 0.01$ & $157.41\pm  6.16$ & $19.76\pm 0.11$ & $53.30\pm  2.30$ & $0.69\pm 0.01$  & $153.18\pm 0.10$ & $0.47\pm 0.02$ & $ 9.67\pm 0.08$ \\
NGC~4636 & yes & $r^{1/n}+$exp.  &   $19.49\pm 0.10$ & $19.42\pm 1.30$ & $2.18\pm 0.07$ & $0.95\pm 0.01$ & $149.80\pm  5.87$ & $20.21\pm 0.11$ & $83.59\pm  3.60$ & $0.69\pm 0.01$  & $142.29\pm 0.09$ & $0.28\pm 0.01$ & $10.03\pm 0.09$ \\
NGC~4698 & yes & $r^{1/n}+$exp.  &   $19.47\pm 0.14$ & $10.47\pm 1.36$ & $3.00\pm 0.15$ & $0.90\pm 0.01$ & $162.53\pm 16.93$ & $19.39\pm 0.29$ & $34.24\pm  8.95$ & $0.53\pm 0.02$  & $ 75.80\pm 0.17$ & $0.32\pm 0.03$ & $11.25\pm 0.15$ \\
NGC~4736 & no  & bar             &  $  ...         $ & $  ...       $  & $  ...       $ & $  ...       $ & $  ...        $   & $  ...        $ & $  ...        $ &  $  ...        $ & $  ...        $ &  $  ...      $ &  $  ...      $  \\
NGC~4800 & no  & Freeman II disc &  $  ...         $ & $  ...       $  & $  ...       $ & $  ...       $ & $  ...        $   & $  ...        $ & $  ...        $ &  $  ...        $ & $  ...        $ &  $  ...      $ &  $  ...      $  \\
NGC~4826 & yes & $r^{1/n}+$exp.   &  $16.94\pm 0.12$ & $5.01\pm 0.65$  & $1.49\pm 0.07$ & $0.65\pm 0.01$ &    $18.62\pm 1.94$ & $17.97\pm 0.27$ & $50.79\pm 13.28$& $0.60\pm 0.02$  & $16.50\pm 0.04$ &  $0.06\pm 0.01$ & $11.04\pm 0.15$  \\
NGC~5005 & yes & $r^{1/n}+$exp.   &  $18.95\pm 0.09$ & $23.21\pm 1.56$ & $2.04\pm 0.07$ & $0.43\pm 0.01$ & $153.25\pm  6.00$ & $19.31\pm 0.11$ & $58.58\pm  2.52$ & $0.38\pm 0.01$  & $147.97\pm 0.09$ & $0.40\pm 0.02$ & $10.00\pm 0.09$ \\
NGC~5127 & yes & $r^{1/n}$        &  $22.56\pm 0.15$ & $39.95\pm 2.59$ & $4.90\pm 0.19$ & $0.77\pm 0.01$ & $151.27\pm  0.04$ & $ ...         $ & $ ...          $ & $...         $  & $...           $ & $1.00        $ & $11.34\pm 0.04$ \\
NGC~5194 & no  & bad             &  $  ...         $ & $  ...       $  & $  ...       $ & $  ...       $ & $  ...        $   & $  ...        $ & $  ...        $ &  $  ...        $ & $  ...        $ &  $  ...      $ &  $  ...      $  \\
NGC~5248 & yes & $r^{1/n}+$exp.  &   $18.62\pm 0.12$ & $ 7.05\pm 0.61$ & $0.99\pm 0.05$ & $0.69\pm 0.01$ & $ 19.26\pm  1.67$ & $19.53\pm 0.19$ & $45.68\pm  9.02$ & $0.58\pm 0.02$  & $ 33.52\pm 0.10$ & $0.11\pm 0.01$ & $12.09\pm 0.08$ \\
NGC~5252 & yes & $r^{1/n}+$exp.  &   $20.56\pm 0.35$ & $ 9.92\pm 2.62$ & $4.82\pm 0.57$ & $0.41\pm 0.01$ & $102.66\pm 25.24$ & $19.61\pm 0.39$ & $11.57\pm  5.17$ & $0.48\pm 0.05$  & $100.86\pm 1.07$ & $0.51\pm 0.21$ & $13.06\pm 0.29$ \\
NGC~5347 & no  & Freeman II disc &  $  ...         $ & $  ...       $  & $  ...       $ & $  ...       $ & $  ...        $   & $  ...        $ & $  ...        $ &  $  ...        $ & $  ...        $ &  $  ...      $ &  $  ...      $  \\
NGC~5490 & yes & $r^{1/n}$       &   $22.13\pm 0.15$ & $35.09\pm 2.27$ & $6.90\pm 0.26$ & $0.78\pm 0.01$ & $ 82.69\pm  0.02$ & $ ...         $ & $ ...          $ & $...         $  & $ ...          $ & $1.00        $ & $11.00\pm 0.04$ \\
NGC~5695 & yes & $r^{1/n}+$exp.  &   $18.50\pm 0.40$ & $ 2.00\pm 0.65$ & $3.75\pm 0.47$ & $0.69\pm 0.01$ & $ 14.84\pm  1.58$ & $19.11\pm 0.61$ & $10.79\pm  8.51$ & $0.63\pm 0.19$  & $ 27.16\pm 1.25$ & $0.19\pm 0.05$ & $14.04\pm 0.60$ \\
NGC~5879 & no  & bad            &   $  ...         $ & $  ...       $  & $  ...       $ & $  ...       $ & $  ...        $   & $  ...        $ & $  ...        $ &  $  ...        $ & $  ...        $ &  $  ...      $ &  $  ...      $  \\
NGC~5921 & no  & bar            &   $  ...         $ & $  ...       $  & $  ...       $ & $  ...       $ & $  ...        $   & $  ...        $ & $  ...        $ &  $  ...        $ & $  ...        $ &  $  ...      $ &  $  ...      $  \\
NGC~7331 & no  & bad            &   $  ...         $ & $  ...       $  & $  ...       $ & $  ...       $ & $  ...        $   & $  ...        $ & $  ...        $ &  $  ...        $ & $  ...        $ &  $  ...      $ &  $  ...      $  \\
NGC~7626 & yes & $r^{1/n}$      &    $22.69\pm 0.08$ & $67.08\pm 2.33$ & $5.54\pm 0.11$ & $0.90\pm 0.01$ & $101.47\pm  0.01$ & $  ...        $ & $   ...        $ & $ ...        $  & $  ...         $ & $1.00        $ & $10.11\pm 0.03$ \\
UGC~1395 & no  & bad            &   $  ...         $ & $  ...       $  & $  ...       $ & $  ...       $ & $  ...        $   & $  ...        $ & $  ...        $ &  $  ...        $ & $  ...        $ &  $  ...      $ &  $  ...      $  \\
UGC~7115 & yes & $r^{1/n}$       &   $22.43\pm 0.34$ & $22.01\pm 3.91$ & $7.05\pm 0.52$ & $0.92\pm 0.01$ & $ 89.26\pm  0.06$ & $ ...         $ & $ ...          $ & $...         $  & $ ...          $ & $1.00        $ & $12.12\pm 0.19$ \\

\noalign{\smallskip}
\hline
\noalign{\smallskip}
\multicolumn{12}{c}{Galaxies from Sample A: upper limit on \mbh\ from \citetalias{Gultekin2009}} \\
\noalign{\smallskip}
\hline
\noalign{\smallskip}
Abell~2052 & no  & $r^{1/n}$      & $24.58\pm 0.17$  & $ 127.48\pm 8.26$ & $4.57\pm 0.17$ & $0.72\pm 0.01 $& $116.52\pm 0.03$ & $  ...         $ & $  ...         $ & $  ...        $ &$  ...           $ & $1.00         $ &$10.94\pm 0.04$   \\
NGC~3310   & no  & bar            &  $  ...        $ & $  ...         $  & $  ...       $ & $  ...       $ & $  ...         $ & $  ...        $ & $  ...         $ & $  ...        $ &  $  ...         $ & $  ...        $  &$  ...       $  \\
NGC~4041   & yes & $r^{1/n}+$exp. & $19.09\pm 0.14$  & $  10.46\pm 1.35$ & $1.15\pm 0.06$ & $0.79\pm 0.01 $& $  5.85\pm 0.61$ & $ 20.29\pm 0.31$ & $ 29.80\pm 7.79$ & $ 0.89\pm 0.04$ &$ 122.05  \pm  0.28 $ & $0.40\pm 0.04 $ &$11.49\pm 0.15$   \\
NGC~4435   & no  & bar            &  $  ...        $ & $  ...          $ & $  ...       $ & $  ...       $ & $  ...         $ & $  ...        $ & $  ...         $ & $  ...        $ &  $  ...         $ & $  ...         $ &$  ...       $  \\
NGC~4486B  & yes & $r^{1/n}$      & $17.91\pm 0.40$   & $  2.73\pm 1.37$ & $3.58\pm 0.35$ & $0.85\pm 0.01 $& $  4.90\pm 0.01$ & $  ...         $ & $  ...         $ & $  ...        $ &$  ...           $ & $1.00         $ &$12.58\pm 0.24$   \\
\noalign{\smallskip}
\hline
\noalign{\smallskip}
\multicolumn{12}{c}{Galaxies from Sample B} \\
\noalign{\smallskip}
\hline
\noalign{\smallskip}

Abell~1836& yes & $r^{1/n}+$exp.  & $21.01\pm 0.36$ & $  9.49\pm 2.51$ & $2.73\pm 0.33$ & $0.89\pm 0.01$ & $ 79.32\pm 19.50$ & $21.92\pm 0.44$ & $ 27.88\pm 12.46$ & $0.60\pm 0.06$ & $  68.58\pm  0.73$ & $ 0.54\pm 0.22$ & $ 13.06\pm 0.29 $\\
NGC~821   & yes & $r^{1/n}$       & $23.15\pm 0.08$ & $111.25\pm 3.87$ & $7.70\pm 0.15$ & $0.59\pm 0.01$ & $  6.97\pm  0.01$ & $  ...        $ & $   ...         $ & $  ...       $ & $ ...            $ & $   1.00      $ & $  9.75\pm 0.03 $\\
NGC~1068  & yes & $r^{1/n}+$exp.  & $17.14\pm 0.09$ & $ 10.27\pm 0.69$ & $1.27\pm 0.04$ & $0.65\pm 0.01$ & $140.97\pm  5.52$ & $18.30\pm 0.10$ & $ 29.35\pm  1.26$ & $0.99\pm 0.01$ & $ 180.00\pm  0.11$ & $ 0.33\pm 0.01$ & $  9.74\pm 0.09 $\\
NGC~2778  & no  & embedded disc   &  $  ...      $ &  $  ...        $  & $  ...       $ & $  ...       $ & $  ...         $  & $  ...        $ & $  ...         $ &  $  ...       $ & $  ...          $ &  $  ...       $ &  $  ...          $\\
NGC~3031  & yes & $r^{1/n}+$exp.  & $18.75\pm 0.02$ & $ 50.02\pm 0.49$ & $2.57\pm 0.02$ & $0.72\pm 0.01$ & $ 91.24\pm  0.60$ & $18.79\pm 0.01$ & $150.21\pm  0.60$ & $0.50\pm 0.01$ & $  95.77\pm  0.02$ & $ 0.33\pm 0.01$ & $  7.45\pm 0.01 $\\
NGC~3227  & no  & bar             &  $  ...      $ &  $  ...        $  & $  ...       $ & $  ...       $ & $  ...         $  & $  ...        $ & $  ...         $ &  $  ...       $ & $  ...          $ &  $  ...       $ &  $  ...          $\\
NGC~3245  & yes & $r^{1/n}+$exp.  & $17.45\pm 0.11$ & $  4.00\pm 0.35$ & $1.60\pm 0.09$ & $0.75\pm 0.01$ & $100.76\pm  8.75$ & $18.59\pm 0.18$ & $ 20.66\pm  4.08$ & $0.50\pm 0.02$ & $ 101.64\pm  0.29$ & $ 0.27\pm 0.04$ & $ 11.83\pm 0.08 $\\
NGC~3377  & yes & $r^{1/n}$       & $20.41\pm 0.07$ & $ 43.52\pm 1.51$ & $3.47\pm 0.07$ & $0.70\pm 0.01$ & $137.05\pm  0.02$ & $ ...         $ & $  ...          $ & $...         $ & $ ...            $ & $ 1.00        $ & $  9.63\pm 0.03 $\\
NGC~3379  & no  & bad             &  $  ...      $ &  $  ...        $  & $  ...       $ & $  ...       $ & $  ...         $  & $  ...        $ & $  ...          $ &  $  ...      $ & $  ...           $ & $  ...       $ &  $  ...         $\\   
NGC~3384  & yes & $r^{1/n}+$exp.  & $17.68\pm 0.09$ & $  8.31\pm 0.56$ & $2.33\pm 0.08$ & $0.84\pm 0.01$ & $144.36\pm  5.65$ & $19.57\pm 0.11$ & $ 53.79\pm  2.32$ & $0.47\pm 0.01$ & $ 148.30\pm  0.09$ & $ 0.40\pm 0.02$ & $ 10.17\pm 0.09 $\\
NGC~3607  & yes & $r^{1/n}$       & $21.14\pm 0.01$ & $ 56.34\pm 0.22$ & $4.70\pm 0.01$ & $0.80\pm 0.01$ & $ 38.34\pm  0.01$ & $ ...         $ & $  ...          $ & $...         $ & $ ...            $ & $ 1.00        $ & $  9.16\pm 0.01 $\\
NGC~3608  & yes & $r^{1/n}$       & $24.20\pm 0.01$ & $182.19\pm 0.72$ & $9.03\pm 0.03$ & $0.79\pm 0.01$ & $175.16\pm  0.02$ & $ ...         $ & $  ...          $ & $...         $ & $ ...            $ & $ 1.00        $ & $  9.33\pm 0.01 $\\
NGC~3998  & yes & $r^{1/n}+$exp.  & $17.47\pm 0.13$ & $  5.65\pm 0.73$ & $2.29\pm 0.11$ & $0.84\pm 0.01$ & $ 45.48\pm  4.74$ & $19.33\pm 0.29$ & $ 25.51\pm  6.67$ & $0.78\pm 0.03$ & $  54.96\pm  0.13$ & $ 0.45\pm 0.05$ & $ 10.80\pm 0.14 $\\
NGC~4026  & no  & bad             & $  ...       $ &  $  ...        $  & $  ...       $ & $  ...       $ & $  ...         $  & $  ...        $ & $  ...         $ &  $  ...       $ & $  ...          $ &  $  ...        $ & $  ...         $\\ 
NGC~4258  & no  & bad             & $  ...       $ &  $  ...        $  & $  ...       $ & $  ...       $ & $  ...         $  & $  ...        $ & $  ...         $ &  $  ...       $ & $  ...          $ &  $  ...        $ & $  ...         $\\
NGC~4261  & yes & $r^{1/n}$       & $21.06\pm 0.01$ & $ 48.82\pm 0.19$ & $4.31\pm 0.01$ & $0.77\pm 0.01$ & $ 67.96\pm  0.01$ & $ ...         $ & $  ...          $ & $...         $ & $ ...           $  & $ 1.00        $ & $  9.47\pm 0.01 $\\
NGC~4342  & no  & embedded disc   & $  ...       $ & $  ...         $  & $  ...       $ & $  ...       $ & $  ...         $  & $  ...        $ & $  ...         $ &  $  ...       $ & $  ...          $ &  $  ...        $ & $  ...          $\\
NGC~4374  & yes & $r^{1/n}$       & $20.63\pm 0.01$ & $ 63.61\pm 0.06$ & $4.10\pm 0.01$ & $0.86\pm 0.01$ & $ 37.32\pm  0.01$ & $ ...         $ & $  ...          $ & $...         $ & $ ...            $ & $ 1.00        $ & $  8.38\pm 0.01 $\\
NGC~4459  & yes & $r^{1/n}$       & $23.23\pm 0.01$ & $155.21\pm 0.61$ & $7.44\pm 0.02$ & $0.82\pm 0.01$ & $ 13.91\pm  0.01$ & $ ...         $ & $  ...          $ & $...         $ & $ ...            $ & $ 1.00        $ & $  8.78\pm 0.01 $\\
NGC~4473  & yes & $r^{1/n}+$exp.  & $18.05\pm 0.09$ & $ 10.60\pm 0.71$ & $2.23\pm 0.07$ & $0.57\pm 0.01$ & $  3.26\pm  0.13$ & $19.64\pm 0.11$ & $ 38.14\pm  1.64$ & $0.56\pm 0.01$ & $   1.46\pm  0.01$ & $ 0.48\pm 0.02$ & $ 10.44\pm 0.09 $\\
NGC~4486  & no  & light excess    &  $  ...      $ &  $  ...        $  & $  ...       $ & $  ...       $ & $  ...         $  & $  ...        $ & $  ...         $ &  $  ...       $ & $  ...          $ &  $  ...        $ & $  ...          $\\
NGC~4486A & no  & bad             &  $  ...      $ &  $  ...        $  & $  ...       $ & $  ...       $ & $  ...         $  & $  ...        $ & $  ...         $ &  $  ...       $ & $  ...          $ &  $  ...        $ & $  ...          $\\
NGC~4564  & no  & embedded disc   &  $  ...      $ &  $  ...        $  & $  ...       $ & $  ...       $ & $  ...         $  & $  ...        $ & $  ...         $ &  $  ...       $ & $  ...          $ &  $  ...        $ & $  ...          $\\
NGC~4596  & yes & $r^{1/n}+$exp.  & $21.49\pm 0.11$ & $ 44.91\pm 3.01$ & $4.43\pm 0.15$ & $0.87\pm 0.01$ & $ 14.90\pm  0.58$ & $19.48\pm 0.11$ & $ 32.71\pm  1.41$ & $0.67\pm 0.01 $ & $ 162.85\pm 0.10$ & $ 0.78\pm 0.04$ & $  9.94\pm 0.09 $\\
NGC~4649  & yes & $r^{1/n}+$exp.  & $18.18\pm 0.09$ & $ 14.88\pm 1.00$ & $1.63\pm 0.05$ & $0.85\pm 0.01$ & $  7.98\pm  0.31$ & $19.03\pm 0.11$ & $ 54.15\pm  2.33$ & $0.78\pm 0.01 $ & $  13.78\pm 0.01$ & $ 0.30\pm 0.01$ & $  9.56\pm 0.08 $\\
NGC~4697  & yes & $r^{1/n}$       & $21.82\pm 0.01$ & $128.68\pm 0.12$ & $4.96\pm 0.01$ & $0.69\pm 0.01$ & $ 63.86\pm  0.01$ & $ ...         $ & $ ...           $ & $...          $ & $ ...           $ & $ 1.00        $ & $  8.34\pm 0.01 $\\
NGC~5576  & yes & $r^{1/n}$       & $22.42\pm 0.07$ & $ 77.58\pm 2.70$ & $8.71\pm 0.17$ & $0.69\pm 0.01$ & $177.21\pm  0.02$ & $ ...         $ & $ ...           $ & $...          $ & $ ...           $ & $ 1.00        $ & $  9.62\pm 0.03 $\\
NGC~5845  & yes & $r^{1/n}$       & $17.63\pm 0.27$ & $  4.06\pm 0.72$ & $3.45\pm 0.26$ & $0.66\pm 0.01$ & $ 50.79\pm  0.04$ & $ ...         $ & $ ...           $ & $...          $ & $ ...           $ & $ 1.00        $ & $ 11.78\pm 0.18 $\\

\label{tab:gasp2d_phot}
\end{longtable}
\end{tiny}


\begin{scriptsize}
\begin{minipage}{22cm}
{\sc Notes.} ---
Col.(1): Galaxy name.
Col.(2): Result of the two dimensional decomposition of the galaxy surface
brightness distribution: yes = successful fit; no = unsuccessful fit;
Col.(3): Comments about the fit:
$r^{1/n}$ = the galaxy has been successfully fitted with a \sersic
law;
$r^{1/n}+$exp. = the galaxy has been successfully fitted with a
\sersic bulge and exponential disc;
bad = inadequate one or two-component fit;
Freeman II disc = the disc seems to follow a Freeman Type II law;
embedded disc = the bulge and disc components have been fitted with a \sersic
and
exponential law, respectively. But, the bulge is dominating the light
contribution at both small and large radii;
bar = a strong bar is present in addition to the bulge and disc
components;
light excess = the light excess measured at large radii can not be
parametrised with an exponential function. A cut-off profile
\citep[e.g.,][]{Oemler1976} is required.

Col.(4): Effective surface brightness of the bulge.
Col.(5): Effective radius of the bulge.
Col.(6): Shape parameter of the bulge.
Col.(7): Axial ratio of the bulge isophotes.
Col.(8): Position angle of the bulge major-axis.
Col.(9): Central surface brightness of the disc.
Col.(10): Scale length of the disc.
Col.(11): Axial ratio of the disc isophotes.
Col.(12): Position angle of the disc major-axis.
Col.(13): Bulge-to-total luminosity ratio.
Col.(14): Total magnitude of the bulge.
\end{minipage}
\end{scriptsize}
\end{center}
\end{landscape}

\clearpage


\begin{table*}
\begin{scriptsize}
\begin{center}
\caption{Fitting parameters and correlation coefficients for the different relations}
\begin{tabular}{l l c c r c c c c c}
\hline
\hline
\noalign{\smallskip}
\multicolumn{1}{c}{$x$} &
\multicolumn{1}{c}{Sample} &
\multicolumn{1}{c}{$N$} &
\multicolumn{1}{c}{$\rho$} &
\multicolumn{1}{c}{$P$}  &
\multicolumn{1}{c}{$\alpha$}  &
\multicolumn{1}{c}{$\beta$}  &
\multicolumn{1}{c}{$\epsilon$}  &
\multicolumn{1}{c}{$\epsilon_{int}$}  &
\multicolumn{1}{c}{$x_0$}  \\
\hline
\sigmae       & B          & 49 & $0.85$ & $<0.1\%$ & 8.19 $\pm$ 0.07 & 4.17 $\pm$ 0.32 & 0.41 $\pm$0.06& 0.36 $\pm$0.07& 200 \kms\\ 
              & A$+$B      &143 & $0.69$ & $<0.1\%$ & 7.99 $\pm$ 0.06 & 4.42 $\pm$ 0.30 & 0.44 $\pm$0.05& ...           & \\
              & comparison & 30 & $0.64$ & $<0.1\%$ & 7.88 $\pm$ 0.20 & 4.97 $\pm$ 1.22 & 0.52 $\pm$0.10& ...           & \\
\hline                                                                                                        
\Lbui         & B          & 19 & $0.26$ & $  26\%$ & 8.60 $\pm$ 0.30 & 0.47 $\pm$ 0.31 & 0.56 $\pm$0.10& 0.58 $\pm$0.11& $10^{11}$ \lsun\\ 
              & A$+$B      & 57 & $0.34$ & $   1\%$ & 8.17 $\pm$ 0.22 & 0.79 $\pm$ 0.24 & 0.81 $\pm$0.13& ...           & \\
              & comparison & 30 & $0.22$ & $  24\%$ & 7.24 $\pm$ 0.65 & 0.45 $\pm$ 0.53 & 0.85 $\pm$0.32& ...           & \\
 \hline                                                                                                       
\Mbuvir       & B          & 19 & $0.46$ & $ 4.6\%$ & 8.25 $\pm$ 0.13 & 0.79 $\pm$ 0.26 & 0.47 $\pm$0.06& 0.46 $\pm$0.07& $10^{11}$ \msun\\ 
              & A$+$B      & 57 & $0.52$ & $<0.1\%$ & 7.84 $\pm$ 0.12 & 0.91 $\pm$ 0.16 & 0.61 $\pm$0.08& ...           & \\
              & comparison & 30 & $0.44$ & $ 1.8\%$ & 7.79 $\pm$ 0.39 & 1.23 $\pm$ 0.53 & 0.77 $\pm$0.20& ...           & \\
\hline                                                                                                       
\n            & B          & 19 & $-0.1$ & $  38\%$ & 8.25 $\pm$ 0.18 & 0.06 $\pm$ 0.80 & 0.60 $\pm$0.12& 0.61 $\pm$0.13& 3\\ 
              & A$+$B      & 57 & $0.24$ & $ 7.6\%$ & 7.39 $\pm$ 0.19 & 1.25 $\pm$ 0.60 & 0.93 $\pm$0.17& ...           & \\
              & comparison & 30 & $0.30$ & $  11\%$ & 6.96 $\pm$ 0.36 & 1.84 $\pm$ 0.89 & 0.97 $\pm$0.28& ...           & \\
 \hline                                                                                                       
\muemean      & B          & 19 & $-0.0$ & $  46\%$ & 8.21 $\pm$ 0.19 & 1.78 $\pm$ 3.40 & 0.60 $\pm$0.10& 0.62 $\pm$0.12& 19 \mas\\ 
              & A$+$B      & 57 & $0.16$ & $  23\%$ & 7.47 $\pm$ 0.20 & 4.91 $\pm$ 4.00 & 0.96 $\pm$0.16& ...           & \\
              & comparison & 30 & $0.11$ & $  53\%$ & 6.90 $\pm$ 0.44 & 7.31 $\pm$ 7.75 & 1.08 $\pm$0.35& ...           & \\
\hline                                                                                                       
\ligal        & B          & 28 & $0.44$ & $   2\%$ & 8.69 $\pm$ 0.22 & 1.14 $\pm$ 0.40 & 0.53 $\pm$0.10& 0.52 $\pm$0.11& $10^{11}$ \msun\\ 
              & A$+$B      & 90 & $0.34$ & $0.15\%$ & 7.92 $\pm$ 0.23 & 1.33 $\pm$ 0.35 & 0.91 $\pm$0.12& ...           & \\
              & comparison & 30 & $0.26$ & $  17\%$ & 7.41 $\pm$ 0.55 & 1.22 $\pm$ 0.99 & 1.06 $\pm$0.21& ...           & \\
\hline                                                                                                       
\mstar\       & B          & 28 & $0.51$ & $0.59\%$ & 8.21 $\pm$ 0.10 & 1.43 $\pm$ 0.36 & 0.49 $\pm$0.09& 0.42 $\pm$0.14& $10^{11}$ \msun \\ 
              & A$+$B      & 90 & $0.43$ & $<0.1\%$ & 7.47 $\pm$ 0.15 & 1.42 $\pm$ 0.28 & 0.81 $\pm$0.11& ...           & \\
              & comparison & 30 & $0.40$ & $   3\%$ & 7.17 $\pm$ 0.31 & 1.55 $\pm$ 0.75 & 0.88 $\pm$0.14& ...           & \\
\hline                                                                                                       
\vc\          & B          & 23 & $0.72$ & $<0.1\%$ & 7.82 $\pm$ 0.15 & 3.29 $\pm$ 0.61 & 0.54 $\pm$0.08& 0.51 $\pm$0.09& 200 \kms\\ 
              & A$+$B      & 88 & $0.38$ & $<0.1\%$ & 6.91 $\pm$ 0.16 & 4.02 $\pm$ 0.87 & 0.84 $\pm$0.11& ...           & \\
              & comparison & 30 & $0.37$ & $ 4.5\%$ & 6.82 $\pm$ 0.31 & 3.59 $\pm$ 1.57 & 0.84 $\pm$0.19& ...           & \\
\hline                                                                                                       
\megal\       & B          & 24 & $0.55$ & $ 0.3\%$ & 8.17 $\pm$ 0.13 & 0.83 $\pm$ 0.27 & 0.54 $\pm$0.11& 0.53 $\pm$0.12& $10^{11}$ \msun \\ 
              & A$+$B      & 51 & $0.49$ & $<0.1\%$ & 7.86 $\pm$ 0.09 & 1.00 $\pm$ 0.17 & 0.49 $\pm$0.07& ...           & \\
\hline                                                                                                       
\mdyn\        & B          &  6 & $0.24$ & $  35\%$ & 7.62 $\pm$ 0.19 & 0.32 $\pm$ 0.87 & 0.46 $\pm$0.16& 0.53 $\pm$0.02& $10^{11}$ \msun \\ 
              & A$+$B      & 47 & $0.13$ & $  36\%$ & 6.10 $\pm$ 0.39 & 0.55 $\pm$ 0.59 & 0.96 $\pm$0.31& ...           & \\
\hline
\noalign{\smallskip}
\end{tabular}

\begin{minipage}{\textwidth}
{\sc Notes.} ---
For the variable $x$ a correlation of the form $\log(M_\bullet/{\rm
  M}_{\odot})=\alpha +\beta \log(x/x_0)$ is assumed for the $N$ data
points.  For Samples A$+$B and the comparison sample, the total
scatter $\epsilon$ is defined as the root-mean square deviation in
$\log (M_{\bullet}/{\rm M}_{\odot})$ from the fitted relation assuming
no measurement errors. For sample B, the total scatter takes
measurement errors in both variables into account.
\end{minipage}
\end{center}
\label{tab:fit_single}
\end{scriptsize}
\end{table*}

\clearpage

\begin{landscape}
\begin{table*}
\begin{scriptsize}
\begin{center}
\caption{Fitting parameters and correlation coefficients for the different linear combinations}
\begin{tabular}{c c l c c r c c c c c c c}
\hline
\hline
\noalign{\smallskip}
\multicolumn{1}{c}{$x$} &
\multicolumn{1}{c}{$y$} &
\multicolumn{1}{c}{Sample} &
\multicolumn{1}{c}{$N$} &
\multicolumn{1}{c}{$\rho$} &
\multicolumn{1}{c}{$P$}  &
\multicolumn{1}{c}{$\gamma$}  &
\multicolumn{1}{c}{$\alpha$} &
\multicolumn{1}{c}{$\beta$}  &
\multicolumn{1}{c}{$\epsilon$}  &
\multicolumn{1}{c}{$\epsilon_{\rm int}$}  &
\multicolumn{1}{c}{$x_0$}  &
\multicolumn{1}{c}{$y_0$}  \\
\noalign{\smallskip}
\hline
\sigmae & \reb     & B          & 19 & 0.81 & $<0.1\%$ & 8.34$\pm$0.13 & 3.93$\pm$0.68 & $0.32$$\pm$0.21 & 0.37 $\pm$0.06& 0.36 $\pm$0.07& 200 \kms        & 5 kpc\\ 
        &          & A$+$B      & 57 & 0.70 & $<0.1\%$ & 8.13$\pm$0.11 & 4.37$\pm$0.56 & $0.27$$\pm$0.14 & 0.39 $\pm$0.04& ...           &                 &      \\
        &          & comparison & 30 & 0.68 & $<0.1\%$ & 5.77$\pm$2.69 & 4.95$\pm$1.98 & $0.90$$\pm$0.53 & 0.37 $\pm$0.09& ...           &                 &      \\
\hline                                                                                                                        
\Lbui   & \reb     & B          & 19 & 0.11 & $  66\%$ & 8.62$\pm$0.30 & 0.41$\pm$0.40 & $0.19$$\pm$0.44 & 0.60 $\pm$0.10& 0.63 $\pm$0.11& $10^{11}$ \lsun & 5 kpc\\ 
        &          & A$+$B      & 57 & 0.33 & $ 1.2\%$ & 8.16$\pm$0.25 & 0.72$\pm$0.57 & $0.01$$\pm$0.60 & 0.81 $\pm$0.13& ...           &                & \\
        &          & comparison & 30 & 0.20 & $  28\%$ & 5.18$\pm$4.23 & 0.07$\pm$0.70 & $0.71$$\pm$1.14 & 0.88 $\pm$0.26& ...           &                & \\
\hline                                                                                                                        
\sigmae & \Lbui    & B          & 19 & 0.79 & $<0.1\%$ & 8.33$\pm$0.19 & 3.88$\pm$0.70 & $0.15$$\pm$0.20 & 0.39 $\pm$0.06& 0.38 $\pm$0.07& 200\kms        & $10^{11}$ \lsun\\         
        &          & A$+$B      & 57 & 0.68 & $<0.1\%$ & 8.11$\pm$0.17 & 4.42$\pm$0.63 & $0.14$$\pm$0.18 & 0.41 $\pm$0.05& ...           &                & \\
        &          & comparison & 30 & 0.63 & $<0.1\%$ & 7.83$\pm$0.54 & 4.97$\pm$2.48 &$-0.05$$\pm$0.30 & 0.52 $\pm$0.10& ...           &                & \\
\hline                                                                                                                        
\Mbuvir & \reb     & B          & 19 & 0.80 & $<0.1\%$ & 7.54$\pm$0.15 & 2.08$\pm$0.34 &$-1.82$$\pm$0.33 & 0.37 $\pm$0.06& 0.29 $\pm$0.08 & $10^{11}$ \msun & 5 kpc\\   
        &          & A$+$B      & 57 & 0.70 & $<0.1\%$ & 7.33$\pm$0.16 & 2.19$\pm$0.28 &$-1.92$$\pm$0.33 & 0.39 $\pm$0.04& ...           &                 & \\ 
        &          & comparison & 30 & 0.49 & $ 0.9\%$ &11.92$\pm$1.92 & 2.07$\pm$0.91 &$-1.46$$\pm$0.83 & 0.62 $\pm$0.13& ...           &                 & \\
\hline                                                                                                                        
\sigmae & \Mbuvir  &  B         & 19 & 0.81 & $<0.1\%$ & 8.22$\pm$0.10 & 3.29$\pm$0.61 & $0.32$$\pm$0.22 & 0.37 $\pm$0.06& 0.36 $\pm$0.07& 200 \kms        & $10^{11}$ \msun\\
        &          & A$+$B      & 57 & 0.70 & $<0.1\%$ & 8.03$\pm$0.09 & 3.83$\pm$0.65 & $0.27$$\pm$0.14 & 0.39 $\pm$0.04& ...           &                 & \\
        &          & comparison & 30 & 0.66 & $<0.1\%$ & 8.08$\pm$0.28 & 4.09$\pm$1.84 & $0.46$$\pm$0.44 & 0.45 $\pm$0.10& ...           &                 & \\
 \hline                                                                                                                        
\sigmae & \reg     & B          & 28 & 0.83 & $<0.1\%$ & 8.27$\pm$0.10 & 4.03$\pm$0.47 & $0.24$$\pm$0.23 & 0.36 $\pm$0.05& 0.34 $\pm$0.06& 200 \kms        & 5 kpc\\ 
        &          & A$+$B      & 90 & 0.71 & $<0.1\%$ & 8.02$\pm$0.09 & 4.81$\pm$0.40 & $0.03$$\pm$0.21 & 0.40 $\pm$0.04& ...           &                 & \\
        &          & comparison & 30 & 0.64 & $<0.1\%$ & 7.90$\pm$0.32 & 4.99$\pm$1.97 & $0.09$$\pm$0.72 & 0.52 $\pm$0.10& ...           &                 & \\
\hline                                                                                                                        
\li     & \reg     & B          & 28 & 0.53 & $ 0.4\%$ & 8.74$\pm$0.29 & 1.59$\pm$0.76 &$-0.72$$\pm$0.74 & 0.56 $\pm$0.12& 0.57 $\pm$0.14& $10^{11}$ \lsun & 5 kpc\\ 
        &          & A$+$B      & 90 & 0.57 & $<0.1\%$ & 8.45$\pm$0.25 & 3.15$\pm$0.57 &$-2.76$$\pm$0.64 & 0.77 $\pm$0.10& ...           &                 & \\
        &          & comparison & 30 & 0.58 & $ 0.2\%$ & 7.73$\pm$0.75 & 3.06$\pm$1.79 &$-3.69$$\pm$2.03 & 0.97 $\pm$0.20& ...           &                 & \\
\hline                                                                                                                        
\sigmae & \li      & B          & 28 & 0.84 & $<0.1\%$ & 8.27$\pm$0.20 & 4.02$\pm$0.81 & $0.12$$\pm$0.37 & 0.37 $\pm$0.05& 0.35 $\pm$0.0& 200 \kms        & $10^{11}$ \lsun \\ 
        &          & A$+$B      & 90 & 0.70 & $<0.1\%$ & 8.04$\pm$0.13 & 4.74$\pm$0.51 & $0.05$$\pm$0.23 & 0.41 $\pm$0.04& ...           &                 & \\
        &          & comparison & 30 & 0.64 & $<0.1\%$ & 7.90$\pm$0.36 & 4.95$\pm$1.99 & $0.06$$\pm$0.52 & 0.52 $\pm$0.11& ...           &                 & \\
\hline                                                                                                                        
\mstar  & \reg     & B          & 28 & 0.62 & $<0.1\%$ & 7.53$\pm$0.27 & 3.85$\pm$0.84 &$-3.19$$\pm$1.03 & 0.68 $\pm$0.16& 0.58 $\pm$0.15& $10^{11}$ \msun & 5 kpc\\ 
        &          & A$+$B      & 90 & 0.59 & $<0.1\%$ & 7.29$\pm$0.15 & 2.74$\pm$0.41 &$-2.33$$\pm$0.51 & 0.66 $\pm$0.09& ...           &                 & \\
        &          & comparison & 30 & 0.63 & $<0.1\%$ & 6.74$\pm$0.51 & 3.07$\pm$1.65 &$-3.51$$\pm$1.82 & 0.78 $\pm$0.19& ...           &                 & \\
 \hline                                                                                                                        
\sigmae & \mstar   & B          & 28 & 0.83 & $<0.1\%$ & 8.21$\pm$0.08 & 3.59$\pm$1.23 & $0.32$$\pm$0.55 & 0.37 $\pm$0.05& 0.34 $\pm$0.05& 200 \kms        & $10^{11}$ \msun\\ 
        &          & A$+$B      & 90 & 0.70 & $<0.1\%$ & 8.02$\pm$0.07 & 4.68$\pm$0.56 & $0.08$$\pm$0.24 & 0.41 $\pm$0.58& ...           &                 & \\
        &          & comparison & 30 & 0.68 & $<0.1\%$ & 7.93$\pm$0.21 & 4.28$\pm$1.79 & $0.61$$\pm$0.54 & 0.48 $\pm$0.11& ...           &                 & \\
\hline
\noalign{\smallskip}
\label{tab:fit_combination}
\end{tabular}

\begin{minipage}{\textwidth}

{\sc Notes.} ---
For the variable $(x,y)$ a correlation of the form
$\log(M_\bullet/{\rm M}_{\odot})=\gamma + \alpha \log(x/x_0) +\beta
\log(y/y_0)$ is assumed for the $N$ data points.  For Samples A$+$B
and the comparison sample, the total scatter $\epsilon$ is defined as
the root-mean square deviation in $\log (M_{\bullet}/{\rm M}_{\odot})$
from the fitted relation assuming zero measurement errors. For sample
B, the total scatter takes measurement errors for all the variables
into account.

\end{minipage}
\end{center}
\end{scriptsize}
\end{table*}
\end{landscape}

\bsp

\label{lastpage}

\end{document}